\documentclass[11pt,english]{article}
\pdfoutput=1
\usepackage{jheppub}
\usepackage{lscape}
\usepackage{amsmath}
\usepackage{slashed}
\usepackage{mathtools}

\usepackage{tikz-cd}
\usetikzlibrary{shapes,arrows,shadows}
\usepackage{amsmath,bm,times}
\newcommand{\aes}{\mathrm{Cov} (\hat H, \hat S )} % Vector command
\newcommand{\aee}{\mathrm{Var} ( \hat H )} % Vector command
\usepackage{cleveref}
\usepackage{tcolorbox}
\makeatletter
\newcommand{\@fpheader}{}
\makeatother

\usepackage[english]{babel}

\usepackage{amsfonts}
\usepackage{amssymb}
\usepackage{amsmath}
\usepackage{graphicx}

\begin{document}

\title{The Hagedorn Temperature as a Nonequilibrium Dynamical Bottleneck in String Thermodynamics}
\author[a]{Cesar Damian,} 
\author[b]{Oscar Loaiza-Brito}

\affiliation[a]{Departamento de Ingenier\'ia Mec\'anica, Universidad de Guanajuato, \\
Carretera Salamanca-Valle de Santiago km 3.5+1.8 Comunidad de Palo Blanco, Salamanca, Mexico}
\affiliation[b]{Departamento de F\'isica, Universidad de Guanajuato, \\
Loma del Bosque No. 103 Col. Lomas del Campestre C.P 37150 Leon, Guanajuato, Mexico.}

\emailAdd{cesar.damian@ugto.mx,oloaiza@fisica.ugto.mx}

\abstract{We revisit the Hagedorn regime of string theory from a
nonequilibrium perspective using Steepest-Entropy-Ascent Quantum
Thermodynamics (SEAQT), which formulates thermodynamic evolution directly on
the state manifold without requiring a globally well-defined canonical
ensemble. Treating the inverse temperature as an instantaneous,
state-dependent quantity, we derive its exact scalar evolution equation,
governed in the commuting limit by higher-order fluctuation moments. In an
open-system extension we prove that the inverse-temperature mobility equals
unity identically on the quasi-canonical family: a divergent Hagedorn variance
is necessary but not sufficient for slowing-down, and a reservoir coupled
democratically to the whole spectrum drives the system across the Hagedorn
scale unimpeded. Slowing-down requires instead a hierarchy of injection
rates, in which direct production of highly excited strings by the reservoir
is slow compared with a cascade through light quasi-particle modes. The
mobility then collapses to a floor set by the ratio of the two rates, the
intensive variable is pinned near the Hagedorn scale, and energy condenses
into the long-string sector at a finite rate; the pinning is a long-lived
transient of computable duration. Promoting the Hagedorn scale to a
moduli-dependent quantity along emergent-string limits yields a finite
critical distance $\Delta_*$ at which the thermodynamic inertia diverges as
$(\Delta_*-\Delta)^{a-3}$ for $1<a<3$, and beyond which pinning can activate.
Since Kaluza--Klein towers admit no such divergence, this provides a
thermodynamic discriminant between emergent-string and decompactification
limits.}

\maketitle

\section{Introduction}

A gas of weakly interacting strings exhibits a thermodynamic behavior unlike that of ordinary particle systems: because the string density of states grows exponentially with energy, the canonical partition function develops a Hagedorn singularity and the standard equilibrium description ceases to be reliable near a critical temperature $T_H$ \cite{Hagedorn:1965statistical,Atick:1988si,Bowick:1989us}. From the microcanonical perspective, the same exponential growth implies that increasing the total energy drives the system toward highly excited long-string configurations rather than ordinary further heating \cite{Giddings:1989cx,Deo:1989jn}.\\

More precisely, the asymptotic density of states takes the schematic form
\begin{equation}
\Omega(E)\sim E^{-a}e^{\beta_H E},
\end{equation}
where $\beta_H=1/T_H$ and the exponent $a$ is model dependent, reflecting the number of spacetime dimensions, compactification data, and conservation constraints. This behavior originates from the rapid proliferation of oscillatory string modes at high excitation levels, a universal feature of string spectra in the weakly interacting regime \cite{Bowick:1989us,Giddings:1989cx,Deo:1989jn}. A direct consequence is that the canonical partition function
\begin{equation}
Z(\beta)=\int dE\,\Omega(E)e^{-\beta E}
\end{equation}
converges only for $\beta>\beta_H$. As $\beta\to\beta_H^{+}$, the canonical description becomes dominated by large fluctuations and highly excited string states. In this regime, additional energy is preferentially absorbed into long-string configurations that carry a macroscopic fraction of the total energy, while the remaining degrees of freedom remain distributed near the Hagedorn scale \cite{Giddings:1989cx,Deo:1989pp}.\\

Later developments clarified that near-Hagedorn physics can also be described in terms of thermal winding modes and the thermal scalar, providing a more geometric and background-sensitive interpretation of the approach to the critical regime \cite{sathiapalan1987vortices,mertens2014near,Kruczenski:2006random,Mertens:2016thermal}. At the same time, the precise interpretation of the transition remains model dependent: depending on the string background and on the effective description adopted, the transition may be described differently, and even its order need not be universal \cite{Atick:1988si,dienes2005re,brustein2021effective}.\\

These features make the Hagedorn regime a natural setting in which to seek a genuinely nonequilibrium and time-resolved thermodynamic description. Such a need is sharpened by cosmological applications, where Hagedorn phases have long played an important role in string gas scenarios for the very early universe \cite{Brandenberger:1988aj,Brandenberger:2011et}. It is also sharpened by more recent quantum-gravity arguments suggesting that infinite-distance limits in theory space are accompanied by towers of light states and, in many cases, by emergent weakly coupled strings \cite{palti2019swampland,Ooguri:2006in,lee2022emergent}. Together, these observations suggest that the Hagedorn regime may provide a useful thermodynamic window into broader questions about the breakdown of effective descriptions in string theory and quantum gravity.\\

In this work, we adopt the steepest-entropy-ascent quantum thermodynamic
(SEAQT) framework, which formulates nonequilibrium thermodynamic evolution
directly on the state manifold without requiring the existence of a globally
well-defined canonical partition function
\cite{Beretta:2014se,li2016generalized}. This framework has been applied in a
broad variety of nonequilibrium settings, including chemically reactive
systems, heat and mass diffusion, electron--phonon transport, decoherence in
microscopic quantum systems, quantum-information protocols, microstructure
evolution, polymer systems, and adsorption processes
\cite{LiVonSpakovsky2016Chem,li2016generalized,LiVonSpakovskyHin2018,CanoAndradeBerettaVonSpakovsky2015,MontanezBarreraDamianVonSpakovskyCano2020,MontanezBarreraVonSpakovskyDamianCano2022,McDonaldVonSpakovskyReynolds2022,SaldanaRoblesDamianReynoldsVonSpakovsky2025Adsorption}. More recently,
SEAQT has also been used as an effective nonequilibrium model for the
instability and decoherence of coherent states of gravitons in string theory,
providing a bridge between entropy-driven quantum dynamics and questions of
de Sitter stability in the Swampland program
\cite{DamianLoaizaBrito2025}.

This makes it possible to study the approach to the Hagedorn regime as a dynamical nonequilibrium process rather than solely as a singular feature of equilibrium thermodynamics.   Our aim is to investigate whether, and under which conditions, the Hagedorn temperature acts not only as a limiting or critical scale of the equilibrium description but also as a dynamical bottleneck of nonequilibrium string evolution.

\section{Construction of the density of states in string theory}
\label{sec:density_states_string}

A precise discussion of the Hagedorn regime is most naturally formulated in the
microcanonical ensemble, where the central object is the density of states
\begin{equation}
    \Omega(E) = \mathrm{Tr}\,\delta(E-\widehat H)\, .
    \label{eq:Omega_micro_def}
\end{equation}
This quantity counts the number of physical string states with total energy $E$.
For weakly interacting strings it is useful to separate the problem into two steps:
first, one constructs the \emph{single-string} density of states; second, one uses it
to build the \emph{multi-string} density of states of the full gas. This is the route
followed in the microcanonical analyses of high-temperature string thermodynamics
and it is precisely this construction that makes the Hagedorn behavior transparent
\cite{Bowick:1989us,Giddings:1989cx}. At the same time, the same asymptotic
structure can be recovered from the analytic structure of the canonical partition
function, whose singularity at $\beta=\beta_H$ is interpreted, in the Euclidean
formalism, as the point at which a thermal winding mode becomes massless and then
tachyonic \cite{Atick:1988si}.

\subsection{Microcanonical and canonical viewpoints}

The microcanonical density of states and the canonical partition function are related
by the Laplace transform
\begin{equation}
    Z(\beta)=\mathrm{Tr}\,e^{-\beta \widehat H}
    =\int_0^\infty dE\, \Omega(E)e^{-\beta E}\, ,
    \label{eq:Z_Laplace_Omega}
\end{equation}
with inverse transform
\begin{equation}
    \Omega(E)=\frac{1}{2\pi i}\int_{\beta_0-i\infty}^{\beta_0+i\infty}
    d\beta\, e^{\beta E} Z(\beta)\, ,
    \qquad \beta_0>\beta_H .
    \label{eq:Omega_inverse_Laplace}
\end{equation}
The canonical description is reliable only when the integral defining $Z(\beta)$
converges. Therefore, if the large-energy asymptotics of the density of states takes
the form
\begin{equation}
    \Omega(E)\sim E^{-a} e^{\beta_H E}\, ,
    \qquad E\to\infty ,
    \label{eq:Omega_Hagedorn_generic}
\end{equation}
then $\beta_H$ is the abscissa of convergence of the canonical partition function.
In that sense the Hagedorn inverse temperature is already encoded in the
microcanonical growth law. The canonical ensemble can approach $\beta_H$ from
$\beta>\beta_H$, but the exponential proliferation of states prevents an analytic
continuation as an ordinary thermal gas beyond that point. In the Euclidean
worldsheet description this same obstruction appears as the onset of a thermal
winding instability \cite{Atick:1988si}.\\

For the purposes of the present work, Eq.~\eqref{eq:Omega_Hagedorn_generic} is the crucial structural result: the exponential factor determines the Hagedorn scale, whereas the power-law prefactor is model dependent and encodes dimensionality, compactification data, and global conservation laws.

\subsection{\label{sec:single_string}Single-string spectrum and oscillator degeneracy}

To make the origin of Eq.~\eqref{eq:Omega_Hagedorn_generic} explicit, consider a closed string compactified on a torus, with some spatial directions non-compact and some compact with radii $R_i$. Let $k$ denote the momentum in the non-compact directions, and let $(n_i,w_i)$ be the discrete momentum and winding quantum numbers in the compact directions. For definiteness, we choose units with $\alpha'=1/2$.

The physical closed-string states are built from left- and right-moving oscillators. Their excitation levels, $N_L$ and $N_R$, are constrained both by the mass-shell condition and by level matching. In the toroidal case these constraints may be written as
\begin{align}
    8N_L &=
    E^2-k^2-\sum_i\left(\frac{n_i}{R_i}+2R_i w_i\right)^2 ,
    \label{eq:NL_constraint}
    \\
    8N_R &=
    E^2-k^2-\sum_i\left(\frac{n_i}{R_i}-2R_i w_i\right)^2 ,
    \label{eq:NR_constraint}
\end{align}
together with
\begin{equation}
    N_L-N_R=-\sum_i n_i w_i\, .
    \label{eq:level_matching}
\end{equation}
The important point is that the total energy is distributed among center-of-mass motion, momentum and winding in the compact directions, and oscillator
excitations. At very large $E$, the overwhelmingly large contribution to the density of states comes from the enormous degeneracy of oscillator states.\\

The left- and right-moving oscillator degeneracies are obtained from the generating functions of the corresponding worldsheet conformal field theory. Denoting the oscillator partition functions schematically by $f_L(q)$ and $f_R(q)$, one extracts the asymptotic level densities by inverse contour integration,
\begin{equation}
    d_L(N_L)=\frac{1}{2\pi i}\oint \frac{dq}{q^{N_L+1}} f_L(q), 
    \qquad
    d_R(N_R)=\frac{1}{2\pi i}\oint \frac{dq}{q^{N_R+1}} f_R(q),
    \label{eq:oscillator_inverse_contour}
\end{equation}
and evaluates them by the saddle-point method for large level number. One obtains the universal Cardy-type behavior \cite{Cardy:1986operator}
\begin{equation}
    d_L(N_L)\sim N_L^{-\gamma_L} e^{4\pi\sqrt{a_L N_L}},
    \qquad
    d_R(N_R)\sim N_R^{-\gamma_R} e^{4\pi\sqrt{a_R N_R}},
    \label{eq:oscillator_asymptotics}
\end{equation}
where the constants $a_L$, $a_R$, $\gamma_L$, and $\gamma_R$ depend on the
left/right worldsheet sectors and hence on the specific string theory under
consideration.\\

Substituting Eqs.~\eqref{eq:NL_constraint}--\eqref{eq:level_matching} into
Eq.~\eqref{eq:oscillator_asymptotics} and summing over momentum and winding
quantum numbers yields the single-string density of states. It is convenient to introduce chemical potentials conjugate to the conserved quantum numbers in order to keep track of the constraints throughout the calculation. Thus one defines
\begin{equation}
    \omega_1(E;\kappa,\mu,\nu)
    =
    \mathrm{Tr}\!\left[
        \delta(E-\widehat H)\,
        e^{2\pi i(\kappa\cdot k+\mu\cdot w+\nu\cdot n)}
    \right],
    \label{eq:single_string_density_chem}
\end{equation}
where the trace is over the single-string Hilbert space. The physical density with fixed total charges is recovered later by integrating the chemical potentials over one period.\\

After replacing the sums over large momentum and winding numbers by integrals
and evaluating the result by saddle point, one finds the asymptotic single-string growth law
\begin{equation}
    \omega_1(E;\kappa,\mu,\nu)
    \sim
    V\,E^{-p_1}
    e^{\beta_H(\kappa,\mu,\nu)\,E},
    \label{eq:single_string_asymptotic_general}
\end{equation}
for large E, where $V$ is the non-compact spatial volume, $p_1$ is an algebraic exponent, and $\beta_H(\kappa,\mu,\nu)$ is a generalized inverse Hagedorn temperature that depends on the chemical potentials. Setting the chemical potentials to zero gives the physical Hagedorn scale,
\begin{equation}
    \beta_H \equiv \beta_H(0,0,0).
    \label{eq:betaH_zero_chem}
\end{equation}
The precise value of $\beta_H$ depends on the underlying theory and conventions, but the essential point is universal: the exponential growth of oscillator degeneracy drives an exponential growth of the single-string density of states.

\subsection{\label{sec:multi_strings}From the single-string density to the multi-string density}

The full string gas contains an arbitrary number of strings, so the quantity relevant for thermodynamics is the multi-string density of states $\Omega(E)$, not the single-string density $\omega_1(E)$. The standard construction begins from the canonical partition function. For weakly interacting strings, one may write
\begin{equation}
    \ln Z(\beta)
    =
    \sum_{r=1}^{\infty}
    \frac{1}{r}
    \left[
        f_B(r\beta)+(-1)^{r+1}f_F(r\beta)
    \right],
    \label{eq:logZ_bf}
\end{equation}
where $f_B$ and $f_F$ are the Laplace transforms of the bosonic and fermionic
single-string densities,
\begin{equation}
    f_{B,F}(\beta)
    =
    \int_0^\infty dE\, \omega_{1,B/F}(E)e^{-\beta E}.
    \label{eq:fBF_def}
\end{equation}
Eq.~\eqref{eq:logZ_bf} is the string analogue of the standard relation between the grand canonical partition function and the one-particle partition function. The full multi-string density is then recovered from the inverse Laplace transform, given in Eq.~\eqref{eq:Omega_inverse_Laplace}.\\

When quantum-statistical corrections are subleading, the high-energy asymptotics is well captured by the Maxwell--Boltzmann approximation. In that approximation one expands the gas into sectors with fixed string number $n$ and convolves the single-string densities
\begin{equation}
    \Omega(E)
    \simeq
    \sum_{n=0}^{\infty}\frac{1}{n!}
    \int_0^\infty \prod_{i=1}^n dE_i\,
    \omega_1(E_i)\,
    \delta\!\left(E-\sum_{i=1}^n E_i\right).
    \label{eq:Omega_MB_convolution}
\end{equation}
This representation is especially useful for understanding which configurations dominate the asymptotic density of states.\\

For systems with at least one non-compact spatial direction, the convolution
integral is dominated at large $E$ by configurations in which one string carries a macroscopic fraction of the total energy while the remaining strings populate a thermal background. In other words, the large-energy asymptotics of the multi-string density inherits the same exponential factor as the single-string sector,
\begin{equation}
    \Omega(E)\sim E^{-a}e^{\beta_H E},
    \qquad E\to\infty ,
    \label{eq:Omega_asymptotic_final}
\end{equation}
but with a modified algebraic power $a$ that encodes the convolution measure,
compactification data, and the projection onto sectors of fixed total momentum and winding \cite{Bowick:1989us,Giddings:1989cx}. This dominance of one highly excited string is the microcanonical origin of the long-string picture of the Hagedorn regime.

\subsection{Role of conservation laws and compactification}
\label{sec:roleof}
An important refinement of the argument above is that the asymptotic power-law prefactor in Eq.~\eqref{eq:Omega_asymptotic_final} is not universal. It is sensitive to the global conservation laws imposed on the ensemble and, in compact backgrounds, to the way the winding sectors are treated.\\
 
If one works at fixed charge, the constraints are implemented by keeping the
chemical potentials $(\kappa,\mu,\nu)$ throughout the calculation and
projecting onto the desired charge sector only at the end,
\begin{equation}
\Omega_{\rm phys}(E)=\int_{-1/2}^{1/2} d\mu\, d\nu\; \Omega(E;\mu,\nu).
\label{eq:projection}
\end{equation}
Carrying out this projection onto the zero-winding sector of a toroidal
compactification yields the strongly suppressed prefactors obtained
in~\cite{Bowick:1989us},
\begin{equation}
\Omega^{w=0}_{\rm II}(E)\sim\frac{e^{\beta_H E}}{E^{10}},
\qquad
\Omega^{w=0}_{\rm het}(E)\sim\frac{e^{\beta_H E}}{E^{18}} .
\label{eq:projected}
\end{equation}
These are the densities of states of an ensemble held at fixed winding number.\\
 
In a generic thermodynamic setting the winding sectors are not fixed but summed
over, which amounts to setting the chemical potentials to zero. It is the
projection that generates the strong
algebraic suppression of Eq.~\eqref{eq:projected}. Once the sum is performed the prefactor is much weaker. At sufficiently high energy a general argument based on the worldsheet conformal field theory gives
\begin{equation}
a\;\longrightarrow\;1 ,
\label{eq:a-asymptotic}
\end{equation}
as shown in Appendix~A of~\cite{Brandenberger:1988aj}. The physical origin of the difference is transparent: a typical highly excited string has enough energy to wrap the compact directions, so the sum over winding modes is effectively saturated and contributes its own energy dependence to the
counting. The exponent is therefore not a constant of the background but a slowly varying function of the energy,
\begin{equation}
a=a(E),
\end{equation}
controlled by the effective number of compact directions that a string of energy $E$ is able to wrap, and interpolating toward the asymptotic value given in Eq.~\eqref{eq:a-asymptotic}. This structure was analysed in~\cite{Abel:1999rq}, and it is reproduced by the random-walk description of
highly excited strings~\cite{Kruczenski:2006random,mertens2014near}.\\
 
For the purposes of the present work only one property of $a(E)$ is required: that it varies slowly over the energy window which dominates the tail moments of Sec.~\ref{sec:hagedornpinnin}, so that the local form $\Omega\sim E^{-a}e^{\beta_H E}$ may be used with a constant effective exponent evaluated in that window. All statements below involving a numerical value of $a$ are to be read in this
sense.\\

Two consequences may be anticipated here and are developed in Sec.~\ref{sec:hagedornpinnin}. First, the asymptotic value $a=1$ lies inside the divergent regime derived there: it separates the branch $a<1$, for which $\mathrm{Var}_S(H_S)\sim\delta^{-2}$, from the branch $1<a<3$, for which $\mathrm{Var}_S(H_S)\sim\delta^{a-3}$, and itself gives $\mathrm{Var}_S(H_S)\sim\delta^{-2}/|\ln\delta|$. The correct treatment of the winding sum therefore enlarges the class of backgrounds carrying a divergent thermodynamic inertia. Second, the non-divergent branch $a>3$ requires non-compact directions: with $d$ non-compact spatial directions the dominant sector is the single long string, with effective exponent $a_{\rm eff}=1+d/2$, so that $a_{\rm eff}>3$ only for $d\ge5$.

\subsection{Thermodynamic consequences}

Once Eq.~\eqref{eq:Omega_asymptotic_final} is established, the microcanonical
entropy is
\begin{equation}
    S(E)=\ln \Omega(E)
    =\beta_H E-a\ln E+\mathcal{O}(1),
    \qquad E\to\infty .
    \label{eq:entropy_asymptotic}
\end{equation}
Therefore the microcanonical inverse temperature is
\begin{equation}
    \beta_{\rm micro}(E)
    =
    \frac{\partial S}{\partial E}
    =
    \beta_H-\frac{a}{E}
    +\mathcal{O}(E^{-2}) .
    \label{eq:beta_micro_asymptotic}
\end{equation}

This expression shows that, as the energy is increased, the inverse temperature approaches $\beta_H$ from below. Equivalently, the temperature approaches the Hagedorn temperature from above. In the weakly interacting microcanonical picture, the additional energy is funneled predominantly into the excitation of one long string, while the remainder of the gas stays distributed at the Hagedorn scale \cite{Giddings:1989cx,Bowick:1989us}.\\

From the canonical perspective, the same asymptotic law explains the singular role
of $\beta_H$: substituting Eq.~\eqref{eq:Omega_asymptotic_final} into
Eq.~\eqref{eq:Z_Laplace_Omega} shows that the exponential growth of states pins
the convergence boundary at $\beta=\beta_H$. Thus the Hagedorn phenomenon is
not an arbitrary feature of a particular ensemble, but a direct consequence of the
microscopic construction of the string density of states.\\

As we discuss in Sec.~\ref{sec:hagedornpinnin}, the value of $a$ controls the strength
of the nonequilibrium bottleneck. Which value applies is fixed by the
background and by the sector that dominates the absorption of energy, and, as
explained in Sec.~\ref{sec:roleof}, in compact backgrounds it is an
energy-dependent quantity once the winding sectors are summed. In the presence of at least one non-compact direction the dominant
sector is the single long string, with $a_{\rm eff}=1+d/2$. Broad Hagedorn
tails with $a<3$ produce a divergent thermodynamic inertia, while steeper tails
yield a finite, though possibly large, response.

\section{Nonequilibrium thermodynamics framework}
\label{sec:nonequilibrium}

Before proceeding it is useful to delimit which parts of this section are
standard and which are new. The SEAQT equation of motion and its geometric
formulation as a constrained steepest-entropy-ascent gradient flow on the state
manifold, together with the determination of the Lagrange multipliers by the
conservation constraints, are due to Beretta~\cite{Beretta:1984qw,Beretta:1985er,Beretta:2014se,Beretta:2020th};
the material up to and including the equation of motion obtained in
Sec.~\ref{sec:special_case}, and the gradient computations recorded in
Appendix~\ref{app:phi}, are reproduced here only to fix notation and to make
the paper self-contained. What is new begins with Sec.~\ref{sec:derivation}:
the exact scalar evolution equation for the nonequilibrium inverse temperature
and its expansion in powers of $\beta$ in Sec.~\ref{sec:structure_nonequilibrium}
are derived here for the first time to our knowledge. These include in
particular the treatment of the genuinely noncommuting contribution through the
Fr\'echet derivative of the matrix logarithm, and the diagonal terms
$\mathrm{Var}_{\rm d}$ and $\mathrm{Cov}_{\rm d}$, which have no analogue in
the existing SEAQT literature. The commuting-limit result of
Sec.~\ref{sec:structure_nonequilibrium}, showing that the evolution of $\beta$
is governed entirely by third-order mixed fluctuation moments, and the
discussion of Sec.~\ref{sec:Hagedorn_attractor} are likewise new, as is the
whole of Sec.~\ref{sec:open_system_Hagedorn} apart from the block-metric
kinematics recalled in Sec.~\ref{sec:composite}.\\

In the SEAQT framework, the nonequilibrium dynamics is formulated on an underlying energy eigenstructure, or on a suitable coarse-grained pseudo-eigenstructure, and the density of states provides the essential microscopic information needed to build that structure. The density of states controls the diagonal coarse-grained sector of the
dynamics. The genuinely noncommuting corrections, however, require additional
microscopic information about the operator structure and coherences of the
state, and cannot be reconstructed from $\Omega(E)$ alone. For this reason, once the spectral structure is
specified in terms of the density of states, the SEAQT formalism can translate the microscopic string spectrum into a dynamical thermodynamic flow. In the present context, this is precisely why the asymptotic Hagedorn form of the string density of states is sufficient to control the behavior of the effective inverse temperature $\beta(t)$ near the critical regime.\\

In order to construct the SEAQT dynamics, let $\mathcal L$ be a Hilbert space, endowed with an inner product $(\cdot|\cdot)$, whose elements $\hat{\gamma} \in \mathcal L$ represent the states of the system within the SEAQT formulation. In the quantum setting, the physical state is represented by a density operator $\rho$. Instead of working directly with $\hat \rho$, it is convenient to introduce a generalized square-root representation by defining
\begin{equation}
\hat \rho = \hat{\gamma} \hat{\gamma}^\dagger.
\end{equation}
This representation ensures by construction the non-negativity and Hermiticity of $\hat \rho$. Moreover, it guarantees that the functional derivatives of relevant thermodynamic quantities belong to the same space $\mathcal L$, which is essential for the geometric formulation of the dynamics. The representation is not unique, since $\hat{\gamma} \to \hat{\gamma} \hat U$ with $\hat U$ unitary leaves $\hat \rho$ invariant; however, this non-uniqueness does not affect the formulation.\\

The time evolution of $\hat{\gamma}$ is assumed to satisfy
\begin{equation}
\frac{d\hat{\gamma}}{dt} + \frac{i}{\hbar} \hat H \hat{\gamma} = \hat \Pi_{\gamma},
\end{equation}
together with its adjoint equation
\begin{equation}
\frac{d\hat{\gamma}^\dagger}{dt} - \frac{i}{\hbar} \hat{\gamma}^\dagger \hat H = \hat \Pi_\gamma^\dagger,
\end{equation}
where $H$ is the Hamiltonian operator and $\Pi_\gamma$ represents the dissipative contribution to the dynamics. These equations imply the evolution equation for the density operator
\begin{equation}
\frac{d\hat \rho}{dt} + \frac{i}{\hbar}[H,\hat \rho] = \hat \Pi_\gamma \hat{\gamma}^\dagger + \hat{\gamma} \hat \Pi_\gamma^\dagger.
\end{equation}

Consider now a set of conserved functionals
\begin{equation}
\widetilde C_i : \mathcal L \to \mathbb{R}, \qquad i=1,\dots,k,
\end{equation}
and define the mapping
\begin{equation}
\mathbf C : \mathcal L \to \mathbb{R}^k, \qquad
\mathbf C(\hat{\gamma}) = \big(\widetilde C_1(\hat{\gamma}),\dots,\widetilde C_k(\hat{\gamma})\big).
\end{equation}

Under suitable regularity conditions (functional independence of the $\widetilde C_i$), the map $\mathbf C$ is locally a submersion and induces a foliation of $\mathcal L$ into level sets
\begin{equation}
\mathcal M_{\mathbf c}
=
\left\{
\hat{\gamma} \in \mathcal L : \widetilde C_i(\hat{\gamma})=c_i,\; i=1,\dots,k
\right\}.
\end{equation}

Each $\mathcal M_{\mathbf c}$ represents the set of physically admissible states with fixed values of the conserved quantities. Thus, at any point $\hat{\gamma} \in \mathcal M_{\mathbf c}$, the tangent space is given by
\begin{equation}
T_\gamma \mathcal M_{\mathbf c}
=
\left\{
\delta\hat{\gamma} \in \mathcal L :
D\widetilde C_i(\hat{\gamma})[\delta\hat{\gamma}]=0,\; \forall i
\right\},
\end{equation}
where $D\widetilde C_i(\hat{\gamma})[\delta\hat{\gamma}]$ denotes the Fréchet derivative of $\widetilde C_i$ at $\hat{\gamma}$ in the direction $\delta\hat{\gamma}$, defined as
\begin{equation}
D\widetilde C_i(\hat{\gamma})[\delta\hat{\gamma}]
=
\lim_{\epsilon \to 0}
\frac{\widetilde C_i(\hat{\gamma} + \epsilon \delta\hat{\gamma}) - \widetilde C_i(\hat{\gamma})}{\epsilon}.
\end{equation}

By the Riesz representation theorem, there exist vectors
\begin{equation}
|\Psi_i) = \frac{\delta \widetilde C_i}{\delta \hat{\gamma}} \in \mathcal L
\end{equation}
such that
\begin{equation}
D\widetilde C_i(\hat{\gamma})[\delta\hat{\gamma}]
=
(\Psi_i|\delta\hat{\gamma}).
\end{equation}

Hence,
\begin{equation}
T_\gamma \mathcal M_{\mathbf c}
=
\left\{
\delta\hat{\gamma} \in \mathcal L :
(\Psi_i|\delta\hat{\gamma})=0,\; \forall i
\right\}.
\end{equation}

The normal space is therefore
\begin{equation}
N_\gamma \mathcal M_{\mathbf c}
=
\mathrm{span}\{|\Psi_1),\dots,|\Psi_k)\}.
\end{equation}

Let $\widehat G$ be a symmetric positive-definite operator on $\mathcal L$, defining a metric via
\begin{equation}
\langle u, v \rangle_G := (u|\widehat G|v).
\end{equation}
This metric encodes the dissipative structure of the system and determines how distances and thus ``steepest'' directions are measured in state space. Let $\widetilde S(\hat{\gamma})$ be the entropy functional, and define
\begin{equation}
|\Phi) = \frac{\delta \widetilde S}{\delta \hat{\gamma}}.
\end{equation}

The dissipative evolution is defined as the vector $\Pi_\gamma \in T_\gamma \mathcal M_{\mathbf c}$ that maximizes the entropy production under a fixed norm induced by $\widehat G$. The resulting evolution equation is
\begin{equation} \label{eq:pi_gamma}
\hat \Pi_\gamma
=
\widehat L
\left(
\frac{1}{k_B}|\Phi) - \sum_{i=1}^k \zeta_i |\Psi_i)
\right),
\qquad
\widehat L = \frac{1}{\tau_D}\widehat G^{-1},
\end{equation}
where the Lagrange multipliers $\zeta_i$ are determined by the conservation constraints
\begin{equation} \label{eq:constraints}
(\Psi_j|\Pi_\gamma)=0, \qquad \forall j.
\end{equation}

The SEAQT dynamics can be interpreted as a constrained gradient flow of the entropy on the manifold $\mathcal M_{\mathbf c}$ with respect to the metric induced by $\widehat G$. Thus, $\mathcal L$ is the ambient Hilbert space, $\mathbf C$ induces a foliation into invariant manifolds $\mathcal M_{\mathbf c}$, the vectors $|\Psi_i)$ span the normal space, and the evolution vector $\hat \Pi_\gamma$ lies in the tangent space. In the special case where the only conserved quantities are the energy and the normalization of the density operator, the SEAQT evolution takes the form
\begin{equation}
\hat \Pi_\gamma
=
\widehat L
\left(
\frac{1}{k_B}|\Phi)-\zeta_1 |\Psi_I)-\zeta_2 |\Psi_H)
\right),
\end{equation}
where $\zeta_2$ and $\zeta_1$ are the Lagrange multipliers associated with energy conservation and normalization, respectively. These multipliers are determined by imposing that the dissipative evolution remains tangent to the constraint manifold, namely,
\begin{equation}
(\Psi_H|\Pi_\gamma)=0,
\qquad
(\Psi_I|\Pi_\gamma)=0.
\end{equation}
Substituting into the above conditions yields
\begin{align}
(\Psi_H|\widehat L|\Phi)
-k_B\zeta_2(\Psi_H|\widehat L|\Psi_H)
-k_B\zeta_1(\Psi_H|\widehat L|\Psi_I)&=0,
\\
(\Psi_I|\widehat L|\Phi)
-k_B\zeta_2(\Psi_I|\widehat L|\Psi_H)
-k_B\zeta_1(\Psi_I|\widehat L|\Psi_I)&=0.
\end{align}
Solving the second equation for $\zeta_1$ gives

\begin{equation}
\label{eq:beta1_general}
k_B\zeta_1
=
\frac{
(\Psi_I|\widehat L|\Phi)}{(\Psi_I|\widehat L|\Psi_I)}-k_B\zeta_2 \frac{(\Psi_I|\widehat L|\Psi_H)}{(\Psi_I|\widehat L|\Psi_I)}
\end{equation}
Substituting this expression into the first equation yields
\begin{equation} \label{eq:beta2_general}
k_B\zeta_2
=
\frac{
(\Psi_H|\widehat L|\Phi)(\Psi_I|\widehat L|\Psi_I)
-
(\Psi_I|\widehat L|\Phi)(\Psi_H|\widehat L|\Psi_I)
}{
(\Psi_H|\widehat L|\Psi_H)(\Psi_I|\widehat L|\Psi_I)
-
(\Psi_H|\widehat L|\Psi_I)^2
}.
\end{equation}
Therefore, the Lagrange multipliers are completely determined by the metric-weighted inner products among the entropy gradient and the gradients associated with the conserved quantities.
\subsection{Special case for \texorpdfstring{$\widehat G$}{G}}
\label{sec:special_case}
Now, let us consider the simplest case where
\begin{equation}
\widehat G = \widehat I,
\end{equation}
where $\widehat I$ is the identity superoperator. Physically, this corresponds to an isotropic dissipation in the state space, where all directions relax with the same characteristic time scale $\tau_D$. Thus, using\footnote{A detailed derivation is shown in Appendix~\ref{app:phi}.}
\begin{equation} \label{eq:basishi}
|\Psi_H)=|2 \hat H\hat{\gamma}), \qquad |\Psi_I)=|2\hat{\gamma}),
\end{equation}
and the real inner product, the relevant inner products become
\begin{align*}
(\Psi_H|\widehat L|\Phi)
&=
\frac{4}{\tau_D}
\left(
\langle \widehat H\widehat S\rangle
-
k_B\langle\widehat H\rangle
\right),
\\
(\Psi_I|\widehat L|\Psi_I)
&=
\frac{4}{\tau_D},
\\
(\Psi_H|\widehat L|\Psi_I)
&=
\frac{4}{\tau_D}\langle\widehat H\rangle,
\\
(\Psi_I|\widehat L|\Phi)
&=
\frac{4}{\tau_D}
\left(
\langle\widehat S\rangle-k_B
\right),
\\
(\Psi_H|\widehat L|\Psi_H)
&=
\frac{4}{\tau_D}
\langle \widehat H^2\rangle .
\end{align*}
where
\begin{equation}
\langle A \rangle = \mathrm{Tr}(\hat \rho A), \qquad \hat S = - k_B\ln \hat \rho.
\end{equation}

Substituting into the general expression for the Lagrange multiplier yields
\begin{equation}
k_B\zeta_2
=
\frac{\langle \hat H \hat S \rangle - \langle \hat H \rangle \langle \hat S \rangle}
{\langle \hat H^2 \rangle - \langle \hat H \rangle^2}.
\end{equation}
Thus, $\zeta_2$ has units of inverse energy and is identified with the
canonical inverse temperature,
\begin{equation} \label{eq:beta}
\zeta_2 = \frac{1}{k_B}\frac{\aes}{\aee } = \beta.
\end{equation}
Similarly, the Lagrange multiplier $\zeta_1$ is
\begin{equation} \label{eq:beta1_sh}
k_B\zeta_1 = \langle \hat S \rangle - k_B\beta \langle \hat H \rangle - k_B.
\end{equation}

Substituting these into the SEAQT dissipative term yields
\begin{equation}
\Pi_\gamma
=
-\frac{2}{\tau_D}
\left(
 \ln \hat \rho +\hat I+ \beta \widehat H + \zeta_1 \hat I
\right)\hat{\gamma}.
\end{equation}

Defining the free energy operator
\begin{equation} \label{eq:definion_F}
\hat F = \hat H - \theta \hat S, \qquad \theta = \frac{1}{k_B\beta},
\end{equation}
one obtains
\begin{equation} \label{eq:pi_gamma2}
\Pi_\gamma
=
-\frac{2 \beta}{\tau_D}
\left(
\hat F - \langle \hat F \rangle
\right)\hat{\gamma}.
\end{equation}

Therefore, the evolution equation for $\hat \rho$ becomes
\begin{equation} \label{eq:eqofmotion}
\frac{d\hat \rho}{dt} + \frac{i}{\hbar}[\hat H,\hat \rho]
=
-\frac{2  \beta}{\tau_D}
\{ \Delta F , \hat \rho \},
\end{equation}
where $\Delta \hat F = \hat F - \langle \hat F \rangle$ and $\{\hat A ,\hat B\}=\hat A \hat B+\hat B\hat A$. 
\subsection{Derivation of the temperature dynamics}
\label{sec:derivation}
To examine whether the Hagedorn temperature can act as a dynamical bottleneck, we consider the rate of change of the nonequilibrium inverse temperature given by Eq.~\eqref{eq:beta}. This reads
\begin{equation} \label{eq:dbetadt}
k_B\frac{d\beta}{dt}
=
\frac{1}{\aee}
\left(
\frac{d\aes}{dt}
-
k_B\beta \frac{d\aee}{dt}
\right),
\end{equation}
or equivalently
\begin{equation}
\label{eq:Aee_dbeta_general}
k_B\aee\frac{d\beta}{dt}
=
-
k_B\beta \frac{d\aee}{dt}
+
\frac{d\aes}{dt} \,.
\end{equation}

Thus, the rate of change of $\beta$ is governed by the interplay between the covariance between energy and entropy, $\aes$, and the variance of the energy, $\aee$. In particular, $\aes$ encodes how entropy gradients project along energy fluctuations, while $\aee$ measures the spread of the energy distribution. Therefore, as long as energy and entropy are not perfectly aligned, $\beta$ evolves dynamically, approaching a stationary condition corresponding to thermodynamic equilibrium.\\

We now evaluate the first term on the right-hand side. Since
\begin{equation}
\aee=\langle (\Delta \hat H)^2\rangle,
\end{equation}
we first compute the energy balance
\begin{equation}
\frac{d}{dt}\langle \hat H\rangle
=
\mathrm{Tr}\!\left(\hat H \frac{d\hat\rho}{dt}\right).
\end{equation}
By direct substitution of Eq.~\eqref{eq:eqofmotion}, we obtain
\begin{equation}
\frac{d}{dt}\langle \hat H\rangle
=
-\frac{i}{\hbar}\mathrm{Tr}\!\left(\hat H[\hat H,\hat\rho]\right)
-
\frac{2\beta}{\tau_D}
\mathrm{Tr}\!\left(
\hat H \{ \Delta \hat F,\hat\rho\}
\right).
\end{equation}

The unitary contribution vanishes by cyclicity of the trace,
\begin{equation}
\mathrm{Tr}\!\left(\hat H[\hat H,\hat\rho]\right)=0,
\end{equation}
so only the dissipative term contributes. Using cyclicity again, the latter can be rewritten as
\begin{align}
\mathrm{Tr}\!\left(
\hat H \{ \Delta \hat F,\hat\rho\}
\right)
&=
\mathrm{Tr}\!\left(
\hat\rho \{ \hat H,\Delta \hat F\}
\right),
\end{align}
and therefore
\begin{equation}
\frac{d}{dt}\langle \hat H\rangle
=
-\frac{2\beta}{\tau_D}
\left\langle
\{ \hat H,\Delta \hat F\}
\right\rangle
=
-\frac{4\beta}{\tau_D}\mathrm{Cov}(\hat H,\hat F).
\end{equation}

Substituting Eq.~\eqref{eq:definion_F}, we notice that
\begin{equation} \label{eq:deltaF}
\Delta \hat F = \Delta \hat H - \theta\Delta \hat S,
\end{equation}
and using the definition of $\beta$ in Eq.~\eqref{eq:beta}, one finds
\begin{equation}
\mathrm{Cov}(\hat H,\hat F)
=
\aee - \theta\aes
=0.
\end{equation}
due to Eq.~(\ref{eq:beta}). Hence,
\begin{equation} \label{eq:dHdt_zero}
\frac{d}{dt}\langle \hat H\rangle = 0.
\end{equation}

This result reflects the fact that energy is a conserved constraint within the SEAQT framework: the dissipative dynamics redistributes populations while preserving the mean energy. Since $\langle \hat H\rangle$ is constant, $\Delta \hat H$ is time-independent, thus the rate of change of the energy variance can be written as
\begin{equation}
\frac{d\aee}{dt}
=
\mathrm{Tr}\!\left( (\Delta \hat H)^2 \frac{d\hat\rho}{dt} \right).
\end{equation}
Substituting again Eq.~\eqref{eq:eqofmotion}, we obtain
\begin{equation}
\frac{d\aee}{dt}
=
-\frac{i}{\hbar}\mathrm{Tr}\!\left( (\Delta \hat H)^2 [\hat H,\hat\rho] \right)
-
\frac{2\beta}{\tau_D}
\mathrm{Tr}\!\left( (\Delta \hat H)^2 \{\Delta \hat F,\hat\rho\} \right).
\end{equation}

As before, the unitary contribution vanishes because $(\Delta \hat H)^2$ commutes with $\hat H$, and therefore
\begin{equation}
\label{eq:dAee_dt_compact}
\frac{d\aee}{dt}
=
-
\frac{2\beta}{\tau_D}
\left\langle
\{ (\Delta \hat H)^2,\Delta \hat F\}
\right\rangle .
\end{equation}

For the second term on the right-hand side of Eq.~\eqref{eq:dbetadt}, we note that the covariance between the Hamiltonian and entropy operators can be written as the expectation value of their joint fluctuations, namely
\begin{equation} \label{eq:covHF}
\aes
=
\frac{1}{2}
\left\langle
\{ \Delta \hat H,\Delta \hat S\}
\right\rangle .
\end{equation}
Taking the time derivative,
\begin{equation}
\frac{d\aes}{dt}
=
\frac{1}{2}
\frac{d}{dt}
\mathrm{Tr}\!\left(
\hat\rho
\{ \Delta \hat H,\Delta \hat S\}
\right)
\end{equation}
where after imposing Eq.~\eqref{eq:dHdt_zero}, we obtain,
\begin{equation}
\label{eq:dAes_dt_before_frechet}
\frac{d\aes}{dt}
=
\frac{1}{2}
\mathrm{Tr}\!\left(
\frac{d\hat\rho}{dt}
\{ \Delta \hat H,\Delta \hat S\}
\right)
+
\frac{1}{2}
\mathrm{Tr} \left(
\hat \rho \left\{
\Delta \hat H,\frac{d\hat S}{dt}
\right\}
\right) .
\end{equation}

Substituting Eq.~\eqref{eq:eqofmotion} into the first term, we obtain
\begin{align}
\frac{1}{2}
\mathrm{Tr}\!\left(
\frac{d\hat\rho}{dt}
\{ \Delta \hat H,\Delta \hat S\}
\right)
&=
-\frac{i}{2\hbar}
\mathrm{Tr}\!\left(
[\hat H,\hat\rho]\{ \Delta \hat H,\Delta \hat S\}
\right)
\nonumber\\
&\quad
-
\frac{\beta}{\tau_D}
\mathrm{Tr}\!\left(
\{\Delta \hat F,\hat\rho\}
\{ \Delta \hat H,\Delta \hat S\}
\right).
\end{align}

The unitary contribution as we will see cancels with the unitary contribution of the rate of change of the second term on Eq.~\eqref{eq:dAes_dt_before_frechet}, and thus, does not contribute to the scalar $\beta$-dynamics. Thus, the rate of change of the covariance can be written as
\begin{equation}
\label{eq:dAes_dt_compact}
\frac{d\aes}{dt}
=-\frac{i}{2\hbar}
\mathrm{Tr}\!\left(
[\hat H,\hat\rho]\{ \Delta \hat H,\Delta \hat S\}
\right)
-
\frac{\beta}{\tau_D}
\left\langle
\left\{
\{ \Delta \hat H,\Delta \hat S\},
\Delta \hat F
\right\}
\right\rangle
+
\frac{1}{2}
\mathrm{Tr} \left(
\hat \rho \left\{
\Delta \hat H,\frac{d\hat S}{dt}
\right\}
\right) .
\end{equation}

Substituting Eqs.~\eqref{eq:dAee_dt_compact} and \eqref{eq:dAes_dt_compact} into Eq.~\eqref{eq:Aee_dbeta_general}, we obtain
\begin{align}
\label{eq:Aee_dbeta_intermediate}
k_B\aee\frac{d\beta}{dt} &= 
\frac{2k_B\beta^2}{\tau_D}
\left\langle
\{ (\Delta \hat H)^2,\Delta \hat F\}
\right\rangle  
-\frac{i}{2\hbar}
\mathrm{Tr}\!\left(
[\hat H,\hat\rho]\{ \Delta \hat H,\Delta \hat S\}
\right) \\ \nonumber
&
-
\frac{\beta}{\tau_D}
\left\langle
\left\{
\{ \Delta \hat H,\Delta \hat S\},
\Delta \hat F
\right\}
\right\rangle
+
\frac{1}{2}
\mathrm{Tr} \left(
\hat \rho \left\{
\Delta \hat H,\frac{d\hat S}{dt}
\right\}
\right) 
\nonumber
\end{align}

At this point, it is convenient to use the identity
\begin{equation}
\label{eq:DeltaS_beta_relation}
\Delta \hat S = k_B\beta(\Delta \hat H - \Delta \hat F),
\end{equation}
which follows directly from Eq.~\eqref{eq:deltaF}. Then
\begin{align}
\{ \Delta \hat H,\Delta \hat S\}
&=
k_B\beta \{ \Delta \hat H,\Delta \hat H-\Delta \hat F\}
\nonumber\\
&=
2k_B\beta (\Delta \hat H)^2
-
k_B\beta \{ \Delta \hat H,\Delta \hat F\}.
\end{align}

Substituting into the third term of Eq.~\eqref{eq:Aee_dbeta_intermediate}, we obtain
\begin{align} \label{eq:deltahdeltasdeltaf}
-
\frac{\beta}{\tau_D}
\left\langle
\left\{
\{ \Delta \hat H,\Delta \hat S\},
\Delta \hat F
\right\}
\right\rangle
&=
-
\frac{2k_B\beta^2}{\tau_D}
\left\langle
\left\{
(\Delta \hat H)^2,\Delta \hat F
\right\}
\right\rangle
\nonumber\\
&\quad
+
\frac{k_B\beta^2}{\tau_D}
\left\langle
\left\{
\{ \Delta \hat H,\Delta \hat F\},
\Delta \hat F
\right\}
\right\rangle .
\end{align}

The first term on the right-hand side of Eq.~\eqref{eq:deltahdeltasdeltaf} cancels exactly the first term on the right-hand side of  Eq.~\eqref{eq:Aee_dbeta_intermediate} 
\begin{equation}
\label{eq:Aee_dbeta_with_dSdt}
k_ B\aee\frac{d\beta}{dt}
=-\frac{i}{2\hbar}
\mathrm{Tr}\!\left(
[\hat H,\hat\rho]\{ \Delta \hat H,\Delta \hat S\}
\right)+
\frac{k_B\beta^2}{\tau_D}
\left\langle
\left\{
\{ \Delta \hat H,\Delta \hat F\},
\Delta \hat F
\right\}
\right\rangle
+
\frac{1}{2}
\mathrm{Tr} \left(
\hat \rho \left\{
\Delta \hat H,\frac{d\hat S}{dt}
\right\}
\right)  .
\end{equation}

The remaining term involves the time derivative of the operator functional
\begin{equation}
\hat S(\hat\rho) = -k_B\ln \hat\rho.
\end{equation}
Since $\hat\rho$ and $d\hat\rho/dt$ do not generally commute,  the rate of change of the operator is  the Fréchet derivative of the logarithm \cite{Al:2013computing},
\begin{equation}
\label{eq:Frechet_log_integral}
D(\ln \hat\rho)[\hat X]
=
\int_0^\infty
(\hat\rho+\lambda \hat I)^{-1}
\hat X
(\hat\rho+\lambda \hat I)^{-1}
\,d\lambda .
\end{equation}
Therefore,
\begin{equation}
\label{eq:dSdt_frechet}
\frac{d\hat S}{dt}
=
-
k_BD(\ln \hat\rho)\!\left[\frac{d\hat\rho}{dt}\right].
\end{equation}

Let the instantaneous spectral decomposition of $\hat\rho$ be
\begin{equation}
\hat\rho = \sum_n p_n |n\rangle \langle n|.
\end{equation}
Then the matrix elements of the Fréchet derivative are
\begin{equation}
\label{eq:Frechet_log_matrix_elements}
\left[D(\ln \hat\rho)[\hat X]\right]_{nm}
=
\begin{cases}
\dfrac{\ln p_n - \ln p_m}{p_n-p_m}\,X_{nm}, & n\neq m,\\[2ex]
\dfrac{1}{p_n}\,X_{nn}, & n=m.
\end{cases}
\end{equation}

The off-diagonal elements carry the genuinely noncommuting correction.
The diagonal elements give
$(d\hat S/dt)_{nn} = -k_B\, p_n^{-1}\,(d\hat\rho/dt)_{nn}$;
inserting the dissipative part of Eq.~\eqref{eq:eqofmotion}, whose diagonal matrix
elements are $\dot\rho_{nn} = -(4\beta/\tau_D)\, p_n \Delta F_{nn}$,
one obtains
\begin{equation} \label{eq:dsdtnn}
\left(\frac{d\hat S}{dt}\right)_{nn}
= \frac{4 k_B \beta}{\tau_D}\,\Delta F_{nn} .
\end{equation} 

Substituting into the remaining term yields
\begin{align}
\label{eq:Frechet_quantum_term}
\frac{1}{2}
\mathrm{Tr} \left(
\hat \rho \left\{
\Delta \hat H,\frac{d\hat S}{dt}
\right\}
\right) 
 &= \frac{i}{2\hbar}
\mathrm{Tr}\!\left( \hat \rho
\{ \Delta \hat H,[\hat S,\hat H]\}
\right)   \\ \nonumber
&+\frac{k_B\beta}{\tau_D}
\sum_{n\neq m}
\frac{(p_n+p_m)^2}{p_n-p_m}
\left(\ln \frac{p_n}{p_m} \right)
\Delta H_{nm}
\Delta F_{mn} +  \frac{4 k_B \beta}{\tau_D} \sum_n p_n\, \Delta H_{nn}\,\Delta F_{nn}.
\end{align}

Notice that for the reversible contribution, we have that 
\begin{equation} \label{eq:comm_iden}
-\frac{i}{2\hbar} \mathrm{Tr} \left( [ \hat H , \hat \rho ] \{ \Delta \hat H, \Delta \hat S\} \right) = -\frac{i}{2 \hbar } \mathrm{Tr} \left( \hat \rho\{\Delta \hat H, [ \hat S, \hat H ] \} \right)
\end{equation}
Thus the reversible contribution cancels out in the rate of change of $\beta$. Thus, by substituting Eq.~\eqref{eq:Frechet_quantum_term} into Eq.~\eqref{eq:Aee_dbeta_with_dSdt}, we finally obtain
\begin{align} \label{eq:beta-exact}
k_B\aee\frac{d\beta}{dt}
&=
\frac{k_B\beta^2}{\tau_D}
\left\langle
\left\{
\{ \Delta \hat H,\Delta \hat F\},
\Delta \hat F
\right\}
\right\rangle \\ \nonumber
&+
\frac{k_B\beta}{\tau_D}
\sum_{n\neq m}
\frac{(p_n+p_m)^2}{p_n-p_m}
\Delta H_{nm}\Delta F_{mn}
\ln \frac{p_n}{p_m}+\frac{4 k_B \beta}{\tau_D} \sum_n p_n\, \Delta H_{nn}\,\Delta F_{nn}.
\end{align}

Expanding $\Delta F_{nn} = \Delta H_{nn} - \theta\,\Delta S_{nn}$, the
diagonal contribution reads
\begin{equation}
\frac{4 k_B \beta}{\tau_D} \sum_n p_n\, \Delta H_{nn}\,\Delta F_{nn}=\frac{4 k_B \beta}{\tau_D}\,\mathrm{Var}_{\rm d}(\hat H)
- \frac{4}{\tau_D}\,\mathrm{Cov}_{\rm d}(\hat H,\hat S),
\end{equation}
where $\mathrm{Var}_{\rm d} (\hat H) = \sum_n p_n (\Delta H_{nn})^2$ and , $\mathrm{Cov}^{\rm d} (\hat H , \hat S) = \sum_n p_n \Delta H_{nn}\Delta S_{nn}$, thus  contributing $(4k_B/\tau_D)\mathrm{Var}_{\rm d}$ to $C_1$ and
$-(4/\tau_D)\mathrm{Cov}^{\rm d}$ to $C_0$ of the expansion in powers of
$\beta$ given in Sec.~\ref{sec:structure_nonequilibrium}. Two properties are worth noting. First, by
$\mathrm{Cov}(\hat H,\hat F)=0$ the diagonal sum equals minus the
coherence part of the $H$--$F$ covariance, so it vanishes whenever
$\hat\rho$ and $\hat H$ commute. Second, in the commuting limit
$\mathrm{Var}_{\rm d} = \mathrm{Var}(\hat H)$ and
$\mathrm{Cov}^{\rm d} = \mathrm{Cov}(\hat H,\hat S) =
k_B\beta\,\mathrm{Var}(\hat H)$ by the definition given in Eq.~\eqref{eq:beta} of $\beta$,
so the two diagonal terms cancel identically.\\

\subsection{Structure of the nonequilibrium temperature evolution and the classical limit}
\label{sec:structure_nonequilibrium}

To make explicit the thermodynamic forces governing the evolution of $\beta$, it is
convenient to expand the free-energy fluctuation,
\begin{equation}
    \Delta \hat F = \Delta \hat H - \theta\Delta \hat S,
\end{equation}
inside the nested anticommutator appearing in the first term of the evolution
equation. In this way, the $\beta$-dynamics is written entirely in terms of energy
and entropy fluctuations.

For notational brevity, let us define
\begin{equation}
    A := \Delta \hat H,
    \qquad
    B := \Delta \hat S,
    \qquad
    \Delta \hat F = A-\theta B.
\end{equation}
The first anticommutator is
\begin{equation}
    \{A,\Delta \hat F\}
    =
    \{A,A-\theta B\}
    =
    2A^2-\theta\{A,B\}.
\end{equation}
Substituting this into the nested anticommutator gives
\begin{align}
    \bigl\{\{A,\Delta \hat F\},\Delta \hat F\bigr\}
    &=
    \left\{2A^2-\theta\{A,B\},\,A-\theta B\right\}
    \nonumber\\
    &=
    2\{A^2,A\}
    -2\theta\{A^2,B\}
    -\theta\{\{A,B\},A\}
    +\theta^2\{\{A,B\},B\}.
\end{align}
Since $A$ commutes with $A^2$, one has $\{A^2,A\}=2A^3$, and therefore
\begin{equation}
    \bigl\{\{\Delta \hat H,\Delta \hat F\},\Delta \hat F\bigr\}
    =
    4(\Delta \hat H)^3
    -2\theta\{(\Delta \hat H)^2,\Delta \hat S\}
    -\theta\bigl\{\{\Delta \hat H,\Delta \hat S\},\Delta \hat H\bigr\}
    +\theta^2\bigl\{\{\Delta \hat H,\Delta \hat S\},\Delta \hat S\bigr\}.
    \label{eq:nested_anticommutator_expanded}
\end{equation}

Substituting Eq.~\eqref{eq:nested_anticommutator_expanded} into the evolution
equation for $\beta$, we obtain
\begin{align} \label{eq:beta_expanded}
k_B\mathrm{Var}(\hat H)\frac{d\beta}{dt}
=&\,
\frac{4k_B\beta^2}{\tau_D}
\left\langle(\Delta\hat H)^3\right\rangle
-
\frac{2\beta}{\tau_D}
\left\langle\left\{(\Delta\hat H)^2,\Delta\hat S\right\}\right\rangle
\nonumber\\
&-
\frac{\beta}{\tau_D}
\left\langle
\left\{
\{\Delta\hat H,\Delta\hat S\},\Delta\hat H
\right\}
\right\rangle
+
\frac{1}{k_B\tau_D}
\left\langle
\left\{
\{\Delta\hat H,\Delta\hat S\},\Delta\hat S
\right\}
\right\rangle
\nonumber\\
&+
\frac{k_B\beta}{\tau_D}
\sum_{n\neq m}
\frac{(p_n+p_m)^2}{p_n-p_m}
\ln\left(\frac{p_n}{p_m}\right)
\Delta H_{nm}\Delta F_{mn} \nonumber \\
&+ \frac{4k_B\beta}{\tau_D}\,\mathrm{Var}_{\rm d}(\hat H)
\;-\;\frac{4}{\tau_D}\,\mathrm{Cov}_{\rm d}(\hat H,\hat S).
\end{align}

The last term contains the genuinely noncommuting Fr\'echet contribution. Expanding
$\Delta F_{mn}=\Delta H_{mn}-\theta\Delta S_{mn}$ yields
\begin{align} \label{eq:deltahdeltaf}
&
\frac{k_B\beta}{\tau_D}
\sum_{n\neq m}
\frac{(p_n+p_m)^2}{p_n-p_m}
\ln\left(\frac{p_n}{p_m}\right)
\Delta H_{nm}\Delta F_{mn}
\nonumber\\
&=
\frac{k_B\beta}{\tau_D}
\sum_{n\neq m}
\frac{(p_n+p_m)^2}{p_n-p_m}
\ln\left(\frac{p_n}{p_m}\right)
|\Delta H_{nm}|^2
\nonumber\\
&\quad
-
\frac{1}{\tau_D}
\sum_{n\neq m}
\frac{(p_n+p_m)^2}{p_n-p_m}
\ln\left(\frac{p_n}{p_m}\right)
\Delta H_{nm}\Delta S_{mn}.
\end{align}

It is therefore natural to collect the result according to powers of $\beta$ and
write the evolution equation in the form
\begin{equation}
   k_B \aee \frac{d\beta}{dt}
    =
     C_2[\hat\rho]\,\beta^2
    + C_1[\hat\rho]\,\beta
    + C_0[\hat\rho] \equiv  \mathcal{J}_{\mathrm{int}}[\rho_S],
    \label{eq:beta_quad}
\end{equation}
where the coefficients are state-dependent functionals
\begin{align} \label{eq:fluctuationH}
C_2[\widehat\rho]
&=
\frac{4 k_B}{\tau_D}
\left\langle
(\Delta\widehat H)^3
\right\rangle ,
\end{align}
\begin{align}
C_1[\widehat\rho]
&=
-\frac{2}{\tau_D}
\left\langle
\left\{
(\Delta\widehat H)^2,\Delta\widehat S
\right\}
\right\rangle
-
\frac{1}{\tau_D}
\left\langle
\left\{
\left\{
\Delta\widehat H,\Delta\widehat S
\right\},
\Delta\widehat H
\right\}
\right\rangle
\\  \nonumber
&\quad
+
\frac{k_B}{\tau_D}
\sum_{n\neq m}
\frac{(p_n+p_m)^2}{p_n-p_m}
\ln\left(\frac{p_n}{p_m}\right)
|\Delta H_{nm}|^2  \\ \nonumber
&+\frac{4k_B}{\tau_D}\,\mathrm{Var}_{\rm d}(\hat H) \,, 
\end{align}
and
\begin{align}  \label{eq:betac0}
C_0[\widehat\rho]
&=
\frac{1}{k_B\tau_D}
\left\langle
\left\{
\left\{
\Delta\widehat H,\Delta\widehat S
\right\},
\Delta\widehat S
\right\}
\right\rangle
\nonumber \\
&\quad
-
\frac{1}{\tau_D}
\sum_{n\neq m}
\frac{(p_n+p_m)^2}{p_n-p_m}
\ln\left(\frac{p_n}{p_m}\right)
\Delta H_{nm}\Delta S_{mn} -\,\frac{4}{\tau_D}\,\mathrm{Cov}_{\rm d}(\hat H,\hat S).
\end{align}

In the commuting limit, $[\hat\rho,\hat H]=0$, all relevant operators are
simultaneously diagonalizable. Hence
\begin{equation} \label{eq:deltahnm}
    \Delta H_{nm}=\Delta S_{nm}=0,
    \qquad n\neq m,
\end{equation}
so the Fr\'echet contributions vanish identically. Moreover, all anticommutators
reduce to ordinary products. One then finds
\begin{equation}
    k_B\aee \frac{d\beta}{dt}
    =
    \frac{4k_B\beta^2}{\tau_D}\,\langle (\Delta H)^3\rangle
    -\frac{8\beta}{\tau_D}\,\langle (\Delta H)^2\Delta S\rangle
    +\frac{4}{k_B\tau_D}\,\langle \Delta H(\Delta S)^2\rangle.
    \label{eq:beta_commuting_limit_expanded}
\end{equation}
This may be written more compactly as
\begin{equation}
    \aee \frac{d\beta}{dt}
    =
    \frac{4\beta^2}{\tau_D}\,\langle \Delta H(\Delta F)^2\rangle,
    \label{eq:beta_commuting_limit_compact}
\end{equation}
which provides a useful check of the preceding expansion. In this limit, the
evolution of $\beta$ is governed entirely by third-order mixed fluctuation moments.\\

Near equilibrium, where the distribution is only weakly asymmetric and higher-order
cumulants are small, the right-hand side of
Eq.~\eqref{eq:beta_commuting_limit_compact} is correspondingly suppressed, and
the evolution of $\beta$ becomes slow. By contrast, in strongly non-Gaussian
regimes, such as those associated with broad, highly skewed energy distributions
near the Hagedorn sector, the higher-order moments become dominant and govern
the nonequilibrium thermodynamic trajectory.

\subsection{The Hagedorn Temperature as a Dynamical Bottleneck}
\label{sec:Hagedorn_attractor}

The discussion of Sec.~\ref{sec:density_states_string} shows that the Hagedorn regime is characterized by the asymptotic growth
\begin{equation}
\Omega(E)\sim E^{-a}e^{\beta_H E},
\label{eq:HagedornGrowth34}
\end{equation}
so that, in the microcanonical description, the inverse temperature approaches the constant value
$\beta_H$ as the energy increases. In the canonical description, the same exponential growth identifies
$\beta_H$ as the convergence boundary of the partition function. These two facts establish the
Hagedorn scale as a distinguished thermodynamic threshold, but by themselves they do not provide a
dynamical law for how a nonequilibrium string system approaches that regime.

The SEAQT framework supplies such a law. Since the effective inverse temperature is defined as
\begin{equation}
\beta(t)=\frac{1}{k_B}\frac{\mathrm{Cov}(\hat H,\hat S)}{\mathrm{Var}(\hat H)},
\label{eq:beta_def_34}
\end{equation}
its evolution is governed directly by the fluctuation structure of the nonequilibrium state. In
Sec.~\ref{sec:structure_nonequilibrium}, we showed that the exact scalar dynamics can be written schematically as a quadratic
polynomial in $\beta$, with coefficients determined by the instantaneous density matrix. In the absence
of quantum effects, this evolution reduces to the classical-like contribution given in Eq.~\eqref{eq:beta_commuting_limit_compact},
making explicit that the dynamics of $\beta$ are governed by higher-order fluctuation moments, while
their overall rate is suppressed by the inverse energy variance.\\

This structure is suggestive in the Hagedorn regime. Because the density of states
grows exponentially, the nonequilibrium distribution can develop a broad
high-energy tail, the energy variance becomes large, and the scalar evolution of
$\beta$ carries the prefactor $1/\mathrm{Var}(\hat H)$. It is tempting to read
this prefactor directly as a dynamical bottleneck, with $\beta_H$ an effective
attractor in sufficiently broad Hagedorn sectors.

That reading is, however, premature, and one of the main results of this paper
is that it is incorrect as stated. The prefactor $1/\mathrm{Var}(\hat H)$ cannot
be assessed in isolation, because the energy variance plays two roles at once in
the dynamics of $\beta$. It is the thermodynamic inertia, measuring the energy
that must be supplied to displace the intensive variable; but it also controls
the rate at which the state exchanges energy, since the energy flux driven by a
mismatch of intensive variables is itself proportional to the variance. A
divergent variance produces slowing-down only if these two roles are played by
different degrees of freedom; when they are not, the divergence cancels exactly.
This is what happens when a reservoir drives a quasi-canonical Hagedorn state
directly, as shown in Sec.~\ref{sec:relationtoisolated}. The intrinsic
fluctuation contribution of the isolated dynamics vanishes identically on that
family, a canonical state being stationary under the isolated flow, while the
reservoir contribution grows in exact proportion to the variance. The
inverse-temperature mobility therefore equals unity, and the state crosses the
Hagedorn scale at an unsuppressed rate however broad the tail. This cancellation
does not depend on the density of states, and in particular the Hagedorn
divergence drops out of it entirely. A divergent Hagedorn variance is thus
necessary but not sufficient, and what the present section establishes is the
weaker and correct statement that the Hagedorn regime supplies the divergent
inertia. The bottleneck arises not from the injection of energy as such but from
its route: it appears when energy reaches the Hagedorn sector only indirectly,
through degrees of freedom of finite variance, so that the rate of energy uptake
is set by a finite variance while the cost of displacing the intensive variable
is set by the divergent one. Sections~\ref{sec:relationtoisolated}--\ref{sec:direct}
establish this, and quantify how the effect degrades when a direct channel into
the Hagedorn sector is also present.

\section{Open-system extension and Hagedorn pinning}
\label{sec:open_system_Hagedorn}

The preceding analysis was carried out for an isolated system, for which the mean
energy is conserved and the effective inverse-temperature dynamics is governed by
Eq.~\eqref{eq:beta_quad}. In order to study genuine heating toward the Hagedorn regime, it is natural to consider an open-system extension in which the string system $S$ is
coupled to a large environment or reservoir $R$. The simplest SEAQT realization of
this idea is to treat the composite $S+R$ as an isolated total system while allowing
energy exchange between the two sectors.

\subsection{Composite state space and block-diagonal dissipative metric}
\label{sec:composite}

At the microscopic level, the composite system is described on
\begin{equation}
    \mathcal H_{SR}=\mathcal H_S\otimes\mathcal H_R.
\end{equation}
This is an effective decomposition of the SEAQT state-space directions, or dissipative sectors, associated with the system and the reservoir.\\

We assume that the directions relevant to the constrained projection
decompose into a normalization direction and orthogonal system and
reservoir sectors,
\begin{equation}
\mathcal{L} \supset \mathrm{span}\{\hat\gamma\}\oplus\mathcal{L}_S\oplus\mathcal{L}_R,
 \label{eq:L_split}
\end{equation}
which for a weakly coupled composite in product form is not an additional
assumption but a consequence of the product structure of the state, as
shown in Appendix~\ref{app:block}, Eq.~\eqref{eq:app-decomposition}, and that the relevant observables are approximately block separated,
\begin{equation}
    \hat H \simeq \hat H_S \oplus \hat H_R,
    \qquad
    \hat S \simeq \hat S_S \oplus \hat S_R,
    \label{eq:H_S_split}
\end{equation}
which is appropriate in the weak-coupling regime, where interaction terms are
subleading in the dissipative projection. The SEAQT dissipative operator is then
taken to be block diagonal,

\begin{equation}
    \hat{L}=\tau_0^{-1}\,\hat{P}_0\;\oplus\;\tau_S^{-1}\hat{I}_S\;\oplus\;\tau_R^{-1}\hat{I}_R,
    \label{eq:L_open}
\end{equation}
where $\tau_S$ and $\tau_R$ are the characteristic relaxation times of the system
and reservoir sectors and $\tau_0$ is the rate assigned to the normalization direction. As shown in Appendix~\ref{app:block}, $\beta_{\rm eff}$ is independent of $\tau_0$ so this rate is never needed in what follows. Physically, the reservoir-dominated limit
corresponds to a sector that both relaxes internally much faster than the system and
has sufficiently large thermodynamic weight to remain effectively canonical during
the exchange process.\\

A concrete realization of such a reservoir in string theory is a gas of open
strings in contact with the closed-string gas that plays the role of the
system $S$. In the out-of-equilibrium analysis
of~\cite{Frey:2023xep,Frey:2024jqy}, the open-string gas carries the dominant
share of the energy and equilibrates the closed-string sector considerably
more efficiently than the closed strings equilibrate among themselves. Both
features correspond naturally to the reservoir-dominated regime discussed
below: the dominant energy share to a large thermodynamic weight of the
reservoir sector, and the efficient equilibration to a short relaxation time
for the exchange between the two sectors. Two caveats are in order. First,
open and closed strings of the same background share the same Hagedorn scale,
so in this example the reservoir cannot itself sit well above $T_H$; a
reservoir hotter than the Hagedorn scale of the subsystem requires the two
scales to be set independently, as happens in a warped throat or along the
moduli-space trajectories of Sec.~\ref{sec:SDC}. Second, the open strings
interact with the entire closed-string spectrum, including its highly excited
states. In the terminology developed in
Secs.~\ref{sec:relationtoisolated}--\ref{sec:direct} this is a direct drive of
the Hagedorn sector, which carries unit mobility; we return to its
consequences in Sec.~\ref{sec:direct}.\\

The general SEAQT expression for the Lagrange multiplier $\beta$ is metric
dependent. Under the present decomposition, the weighted inner products split into
system and reservoir contributions. As shown in detail in Appendix~\ref{app:block}, for a weakly coupled composite in (approximate) product form the metric-weighted inner products separate exactly into system and reservoir contributions, and the effective inverse temperature of the composite takes the form
\begin{equation}
    \beta_{\mathrm{eff}}
    \simeq \frac{1}{k_B}
    \frac{
        \tau_S^{-1}\,\mathrm{Cov}_S(\hat H_S,\hat S_S)
        +
        \tau_R^{-1}\,\mathrm{Cov}_R(\hat H_R,\hat S_R)
    }{
        \tau_S^{-1}\,\mathrm{Var}_S(\hat H_S)
        +
        \tau_R^{-1}\,\mathrm{Var}_R(\hat H_R)
    }.
    \label{eq:beta_eff_open}
\end{equation}
If each sector is close to an internal canonical or hypoequilibrium form \cite{li2016generalized}, then
\begin{equation}
    \mathrm{Cov}_X(\hat H_X,\hat S_X)
    \simeq
    k_B\beta_X\,\mathrm{Var}_X(\hat H_X),
    \qquad X=S,R,
    \label{eq:sector_canonical_relation}
\end{equation}
and therefore
\begin{equation}
    \beta_{\mathrm{eff}}
    \simeq
    \frac{
        \tau_S^{-1}\beta_S\,\mathrm{Var}_S(\hat H_S)
        +
        \tau_R^{-1}\beta_R\,\mathrm{Var}_R(\hat H_R)
    }{
        \tau_S^{-1}\mathrm{Var}_S(\hat H_S)
        +
        \tau_R^{-1}\mathrm{Var}_R(\hat H_R)
    }.
    \label{eq:beta_eff_weighted_average}
\end{equation}
Thus, in the reservoir-dominated regime
\begin{equation}
    \tau_R^{-1}\mathrm{Var}_R(\hat H_R)
    \gg
    \tau_S^{-1}\mathrm{Var}_S(\hat H_S),
    \label{eq:reservoir_dominance}
\end{equation}
one finds
\begin{equation}
    \beta_{\mathrm{eff}} \simeq \beta_R .
    \label{eq:beta_eff_to_betaR}
\end{equation}
This is the open-system analogue of the weighted-temperature result obtained in the
hypoequilibrium formalism for interacting subspaces.

\subsection{Reduced open-system dynamics of the subsystem}

In the limit given in Eq.~\eqref{eq:beta_eff_to_betaR}, the dissipative dynamics of the system
sector is governed by the reservoir inverse temperature. In consequence, setting $\beta_{\rm eff}\rightarrow \beta_R$ the exact equation of motion becomes
\begin{equation}
    \frac{d\hat\rho_S}{dt}
    + \frac{i}{\hbar}[\hat H_S,\hat\rho_S]
    =
    -\frac{2\beta_R}{\tau_S}\,
    \left\{
        \Delta \hat F_S^{(R)},
        \hat\rho_S
    \right\},
    \label{eq:rho_open_reduced}
\end{equation}
where
\begin{equation}
    \hat F_S^{(R)}
    :=
    \hat H_S - (k_B\beta_R)^{-1}\hat S_S,
    \qquad
    \Delta \hat F_S^{(R)}
    :=
    \hat F_S^{(R)}-\langle \hat F_S^{(R)}\rangle_S .
    \label{eq:F_R_defined}
\end{equation}
Eq.~\eqref{eq:rho_open_reduced} is of the same SEAQT form as the isolated
equation, except that the intensive parameter entering the dissipator is now fixed by
the reservoir rather than by the instantaneous internal constraint structure of the
subsystem.\\

The essential point is that the subsystem energy is no longer conserved. Indeed,
taking the trace of $\hat H_S$ against Eq.~\eqref{eq:rho_open_reduced} gives
\begin{align}
    \frac{d}{dt}\langle \hat H_S\rangle
    &=
    -\frac{4\beta_R}{\tau_S}
    \,\mathrm{Cov}_S(\hat H_S,\hat F_S^{(R)})
    \nonumber\\
    &=
    -\frac{4\beta_R}{\tau_S}
    \left[
        \mathrm{Var}_S(\hat H_S)
        -(k_B\beta_R)^{-1}\mathrm{Cov}_S(\hat H_S,\hat S_S)
    \right].
    \label{eq:E_open_balance_1}
\end{align}
Introducing the subsystem nonequilibrium inverse temperature
\begin{equation}
    \beta_S
    :=
    \frac{1}{k_B}\frac{\mathrm{Cov}_S(\hat H_S,\hat S_S)}
    {\mathrm{Var}_S(\hat H_S)},
    \label{eq:beta_S_definition}
\end{equation}
this becomes
\begin{equation}
    \frac{d}{dt}\langle \hat H_S\rangle
    =
    -\frac{4}{\tau_S}\,
    \mathrm{Var}_S(\hat H_S)\,
    (\beta_R-\beta_S).
    \label{eq:E_open_balance_2}
\end{equation}
Hence, if the reservoir is hotter than the system, $\beta_R<\beta_S$, the subsystem absorbs energy; if it is colder,
$\beta_R>\beta_S$, the subsystem loses energy. Notice that the total energy of the composite $S+R$ remains conserved, but the subsystem energy does not.

\subsection{Relation to the isolated \texorpdfstring{$\beta$}{beta}-dynamics}
\label{sec:relationtoisolated}

For the isolated system, the inverse-temperature dynamics derived previously takes the form given in Eq.~\eqref{eq:beta_quad} written for the subsystem sector. In the open-system setting, the reservoir introduces an additional external driving. At the level of the $\beta_S$ dynamics, the corresponding equation acquires an additional term $\mathcal{J}_R[\hat\rho_S,\beta_R]$ giving
\begin{equation}
    k_B\mathrm{Var}_S(\hat H_S)\,\frac{d\beta_S}{dt}
    =
     \mathcal{J}_{\mathrm{int}}[\rho_S]
    -
    \mathcal{J}_R[\hat\rho_S,\beta_R].
    \label{eq:beta_open_schematic}
\end{equation}
The reservoir contribution $\mathcal{J}_R$ can in fact be computed
exactly, without committing to a closure. The computation rests on the
identity
\begin{equation}
\beta_R\,\Delta\hat F^{(R)}_S
\;=\;
\beta_S\,\Delta\hat F_S
\;+\;
\big(\beta_R - \beta_S\big)\,\Delta\hat H_S ,
\qquad
\hat F_S := \hat H_S - \frac{1}{k_B \beta_S}\,\hat S_S ,
\label{eq:split-identity}
\end{equation}
which follows from the fact that $\beta_R \hat F^{(R)}_S$ and
$\beta_S \hat F_S$ differ only in their energy term, both fluctuations
being taken with respect to the same state $\hat\rho_S$. Substituting
Eq.~\eqref{eq:split-identity} into the reduced dynamics given in Eq.~\eqref{eq:rho_open_reduced}, the
dissipator separates exactly into the isolated steepest-entropy-ascent
flow evaluated at $\beta_S$, plus a drive along pure energy
fluctuations,
\begin{equation}
\frac{d\hat\rho_S}{dt} + \frac{i}{\hbar}\,[\hat H_S,\hat\rho_S]
=
-\frac{2\beta_S}{\tau_S}\,\{\Delta\hat F_S,\hat\rho_S\}
\;-\;
\frac{2(\beta_R-\beta_S)}{\tau_S}\,\{\Delta\hat H_S,\hat\rho_S\} .
\label{eq:dissipator-split}
\end{equation}
The first term is the isolated dissipator of Eq.~\eqref{eq:eqofmotion} with
$\tau_D \to \tau_S$; the second vanishes when $\beta_S = \beta_R$ and
reverses sign across that point, so that the structure anticipated
below Eq.~\eqref{eq:beta_open_schematic} is manifest already at the level of the equation of
motion. Since the scalar dynamics of $\beta_S$ is, at fixed state, a
linear functional of $d\hat\rho_S/dt$, the two terms in Eq.~
(\ref{eq:dissipator-split}) contribute additively: the first reproduces
the fluctuation polynomial of Eq.~\eqref{eq:beta_quad}, and the second yields the
reservoir contribution in closed form (the derivation is given in the
Appendix~\ref{app:reservoir}),
\begin{equation}
\mathcal{J}_R[\hat\rho_S,\beta_R]
=
\frac{4k_B\,\mathrm{Var}_S(\hat H_S)(\beta_S-\beta_R)}{\tau_S} \mu_R \left[ \hat \rho_S\right]
\label{eq:JR-exact}
\end{equation}
where the dimensionless drive mobility is
\begin{equation}
\mu_R[\hat\rho_S] :=
\frac{\beta_S \big\langle \{\{\Delta\hat H_S,\Delta\hat F_S\},
\Delta\hat H_S\}\big\rangle + Q_H + 4\,\mathrm{Var}_{\rm d}(\hat H_S)}
{4\,\mathrm{Var}_S(\hat H_S)} ,
\label{eq:mobility}
\end{equation}
and
\begin{equation}
Q_H[\hat\rho_S] := \sum_{n\neq m}
\frac{(p_n+p_m)^2}{p_n-p_m}\,
\ln\!\frac{p_n}{p_m}\;
\big|\Delta H_{nm}\big|^2 \;\geq\; 0 ,
\qquad
\mathrm{Var}_{\rm d}(\hat H_S) := \sum_n p_n \big(\Delta H_{nn}\big)^2 ,
\label{eq:QH-Vard}
\end{equation}
in the instantaneous eigenbasis of $\hat\rho_S$. The quantity $Q_H$ is
nonnegative term by term, since $(p_n - p_m)$ and $\ln(p_n/p_m)$ always
carry the same sign, and $\mathrm{Var}_{\rm d}$ is the diagonal
(population) part of the energy variance,
$\mathrm{Var}_S(\hat H_S) = \mathrm{Var}_{\rm d}(\hat H_S)
+ \sum_{n\neq m} p_n |\Delta H_{nm}|^2$, the remainder being carried by
the coherences.\\

First,  $\mathcal{J}_R$ vanishes at $\beta_S = \beta_R$ and reverses sign across
that point, as required by the energy balance Eq.~\eqref{eq:E_open_balance_2}. Second, in the
hypoequilibrium limit, where $\hat\rho_S$ is canonical within its
eigenstructure, one has $\Delta\hat F_S = 0$, $Q_H = 0$ and
$\mathrm{Var}_{\rm d} = \mathrm{Var}_S$, so that
\begin{equation}
\mathcal{J}_R \;\longrightarrow\;
\frac{4 k_B}{\tau_S}\,\mathrm{Var}_S(\hat H_S)\,
\big(\beta_S - \beta_R\big) ,
\label{eq:JR-hypo}
\end{equation}
while the intrinsic polynomial vanishes identically (the canonical
state is stationary for the isolated flow); Eq.~\eqref{eq:beta_open_schematic} then collapses to
$d\beta_S/dt = (4/\tau_S)(\beta_R - \beta_S)$, which is precisely the
quasi-canonical closure result in Section~\ref{sec:hagedornpinnin}.\\

The bottleneck condition formulated in Sec.~\ref{sec:hagedornpinnin} below is
then the statement that $\mu_R \to 0$ (together with the analogous suppression
of the intrinsic term). In the hypoequilibrium family $\mu_R \equiv 1$
identically, which explains, now at the level of the exact dynamics, why the
one-parameter closure of Sec.~\ref{sec:hagedornpinnin} cannot exhibit
slowing-down. This leads to a no-go statement: a reservoir that drives a
canonically prepared subsystem directly produces no slowing-down of
$\beta_S$, however broad the Hagedorn tail, because the divergent inertia
$\mathrm{Var}_S(\hat H_S)$ is cancelled by an equally divergent energy flux.
The canonical family is moreover invariant under such a drive
(Sec.~\ref{sec:twosector}), so that the statement holds for the exact dynamics
and not only for the closure. A divergent variance is therefore necessary but
not sufficient for a Hagedorn bottleneck: the suppression of the
inverse-temperature mobility requires the fluctuation structure of the state
to depart from the canonical form, so that the numerator of
Eq.~(\ref{eq:mobility}) grows more slowly than $\mathrm{Var}_S(\hat H_S)$.
This is exactly what the two-sector construction of Sec.~\ref{sec:twosector}
implements, where the drive reaches the Hagedorn sector only through the
quasi-particle band and the effective numerator is controlled by the band
variance.\\

Finally, the fully explicit form of Eq.~\eqref{eq:beta_open_schematic} can be assembled.
Combining Eqs.~\eqref{eq:beta_quad}--\eqref{eq:betac0} (with $\tau_D \to \tau_S$) and Eq.~(\ref{eq:JR-exact}), or equivalently by direct computation with the
dissipator given in Eq.~\eqref{eq:rho_open_reduced} (see Appendix~\ref{app:reservoir}), one finds
\begin{align}
k_B \mathrm{Var}_S(\hat H_S)\,\frac{d\beta_S}{dt}
={}& \frac{4 k_B\,\beta_S\beta_R}{\tau_S}\,
\big\langle (\Delta\hat H_S)^3 \big\rangle
- \frac{2\beta_S}{\tau_S}\,
\big\langle \{(\Delta\hat H_S)^2, \Delta\hat S_S\} \big\rangle
- \frac{\beta_R}{\tau_S}\,
\big\langle \{\{\Delta\hat H_S, \Delta\hat S_S\}, \Delta\hat H_S\}
\big\rangle
\nonumber\\
&+ \frac{1}{k_B \tau_S}\,
\big\langle \{\{\Delta\hat H_S, \Delta\hat S_S\}, \Delta\hat S_S\}
\big\rangle
+ \frac{k_B \beta_R}{\tau_S}\, Q_H
- \frac{1}{\tau_S}\, Q_{HS}
\nonumber\\
&+ \frac{4 k_B \beta_R}{\tau_S}\, \mathrm{Var}_{\rm d}(\hat H_S)
- \frac{4}{\tau_S}\, \mathrm{Cov}^{\rm d}(\hat H_S, \hat S_S) ,
\label{eq:415-explicit}
\end{align}
with
\begin{equation}
Q_{HS} := \sum_{n\neq m}
\frac{(p_n+p_m)^2}{p_n-p_m}\,\ln\!\frac{p_n}{p_m}\;
\Delta H_{nm}\,\Delta S_{mn} ,
\qquad
\mathrm{Cov}^{\rm d}(\hat H_S,\hat S_S)
:= \sum_n p_n\, \Delta H_{nn}\, \Delta S_{nn} .
\label{eq:QHS-Covd}
\end{equation}
Eq.~\eqref{eq:415-explicit} is the open-system analogue of
Eq.~\eqref{eq:beta_expanded}: the isolated expression is recovered term by term upon
setting $\beta_R \to \beta_S$ and $\tau_S \to \tau_D$. Note that,
because the dissipator no longer depends on the instantaneous
$\beta_S$, the right-hand side of (\ref{eq:415-explicit}) is
\emph{affine} in $\beta_S$ at fixed state: the quadratic structure of
the isolated dynamics given in Eq.~\eqref{eq:fluctuationH} originates entirely from the
self-consistency of the instantaneous Lagrange multiplier, and is
removed by the reservoir. As a check, evaluating
(\ref{eq:415-explicit}) on a hypoequilibrium state, where
$\Delta\hat S = k_B \beta_S \Delta\hat H$ and $\hat\rho_S$ is diagonal,
all four third-moment terms cancel pairwise, $Q_H = Q_{HS} = 0$, and
the diagonal terms give
$(4k_B/\tau_S)\mathrm{Var}_S(\beta_R - \beta_S)$, reproducing the
quasi-canonical closure result of Sec.~\ref{sec:hagedornpinnin}.

\subsection{Hagedorn pinning as a mobility-induced bottleneck}
\label{sec:hagedornpinnin}
The open-system setting is the natural arena in which to discuss the
Hagedorn bottleneck mechanism, since the subsystem can exchange energy
with a reservoir and can therefore be driven toward the Hagedorn regime.
We consider a diagonal coarse-grained subsystem whose asymptotic density
of states is
\begin{equation}
    \Omega_S(E)
    =
    A E^{-a}e^{\beta_H E}\Theta(E-E_0),
    \label{eq:Omega_S_Hag}
\end{equation}
where $a>0$ is the model-dependent algebraic exponent and $E_0$ is an
infrared cutoff. Here $\beta_H$ has units of inverse energy, as in the
canonical string partition function.\\

A useful point of clarification is the distinction between shell
probabilities and microscopic probabilities. Let $P_S(E;t)$ denote the
probability density that the subsystem occupies the energy shell
$[E,E+dE]$. This quantity already includes the degeneracy of that energy
shell. The probability per microscopic state inside the same shell is
therefore
\begin{equation}
    q_S(E;t)
    =
    \frac{P_S(E;t)}{\Omega_S(E)}.
    \label{eq:q_micro_from_shell}
\end{equation}
Accordingly, the appropriate coarse-grained Gibbs entropy is
\begin{equation}
    S_S[P]
    =
    -k_B
    \int_{E_0}^{\infty}
    dE\,
    P_S(E;t)
    \ln\left(
    \frac{P_S(E;t)}{\Omega_S(E)}
    \right).
    \label{eq:coarse_grained_Gibbs_entropy}
\end{equation}
Thus, the logarithm entering the entropy is not the logarithm of the
energy-shell probability $P_S(E;t)$, but rather the logarithm of the
microscopic probability $q_S(E;t)$.

As a local quasi-canonical closure, assume that at each instant the
microscopic probability has the diagonal form
\begin{equation}
    q_S(E;t)
    =
    \frac{1}{Z_S(\lambda)}
    e^{-\lambda(t)E},
    \qquad
    \lambda(t)>\beta_H .
    \label{eq:q_quasi_canonical}
\end{equation}
The parameter $\lambda(t)$ has units of inverse energy and therefore
plays the role of a local canonical inverse temperature. Equivalently, the
shell probability is
\begin{equation}
    P_S(E;t)
    =
    \frac{1}{Z_S(\lambda)}
    \Omega_S(E)e^{-\lambda(t)E}.
    \label{eq:P_shell_quasi_canonical}
\end{equation}
Defining
\begin{equation}
    \delta(t)
    :=
    \lambda(t)-\beta_H>0,
    \label{eq:delta_lambda_betaH}
\end{equation}
one obtains
\begin{equation}
    P_S(E;t)
    =
    \frac{A}{Z_S(\delta)}
    E^{-a}e^{-\delta(t)E}\Theta(E-E_0).
    \label{eq:P_shell_delta}
\end{equation}

The normalization and raw energy moments are determined by
\begin{equation}
    I_n(\delta)
    :=
    \int_{E_0}^{\infty}
    dE\,
    E^{n-a}e^{-\delta E}
    =
    \delta^{a-n-1}
    \Gamma(n+1-a,\delta E_0),
    \label{eq:I_n_delta}
\end{equation}
so that
\begin{equation}
    Z_S(\delta)=A I_0(\delta),
    \qquad
    \langle E^n\rangle_S
    =
    \frac{I_n(\delta)}{I_0(\delta)} .
    \label{eq:moments_In}
\end{equation}
In particular,
\begin{equation}
    \langle H_S\rangle
    =
    \frac{I_1}{I_0},
    \qquad
    \mathrm{Var}_S(H_S)
    =
    \frac{I_2}{I_0}
    -
    \left(
    \frac{I_1}{I_0}
    \right)^2 .
    \label{eq:mean_variance_HS}
\end{equation}

The entropy eigenvalue associated with a microscopic state of energy $E$
is
\begin{equation}
    s_S(E)
    =
    -k_B\ln q_S(E;t)
    =
    k_B\left[
    \lambda(t)E+\ln Z_S(\lambda)
    \right].
    \label{eq:entropy_eigenvalue_microstate}
\end{equation}
Consequently,
\begin{equation}
    \mathrm{Cov}_S(H_S,S_S)
    =
    k_B\lambda(t)\,
    \mathrm{Var}_S(H_S).
    \label{eq:Cov_HS_SS_lambda}
\end{equation}
With the convention used in Sec.~\ref{sec:nonequilibrium}, the nonequilibrium inverse
temperature is the canonical inverse temperature,
\begin{equation}
    \beta_S
    :=
    \frac{1}{k_B}
    \frac{\mathrm{Cov}_S(H_S,S_S)}
    {\mathrm{Var}_S(H_S)} .
    \label{eq:betaS_canonical_definition}
\end{equation}
Therefore,
\begin{equation}
    \beta_S
    =
    \lambda(t)
    =
    \beta_H+\delta(t).
    \label{eq:betaS_lambda_delta}
\end{equation}
This result shows that the algebraic factor $E^{-a}$ belongs to the
density of states, not to the microscopic probability $q_S(E;t)$. Hence
it does not generate an additional term proportional to
$\mathrm{Cov}(E,\ln E)$ in the SEAQT definition of $\beta_S$. The
algebraic factor affects the moments and variances through the shell
measure, but it does not alter the local canonical identification
$\beta_S=\lambda$.\\

With the reduced SEAQT convention used in Eq.~\eqref{eq:rho_open_reduced}, the
subsystem energy balance then takes the form given in Eq.~\eqref{eq:E_open_balance_2}, and thus the quasi-canonical ansatz also makes $\langle H_S\rangle$ a function of
$\lambda$. Since
\begin{equation}
    \langle H_S\rangle_\lambda = -\partial_\lambda\ln Z_S(\lambda),
\end{equation}

one has
\begin{equation}
\partial_\lambda \langle H_S\rangle_\lambda
=
-\operatorname{Var}_S(H_S).
\end{equation}

Thus, combining this identity with the energy balance by means of the chain rule gives
\begin{equation}
\frac{d\lambda}{dt} = \frac{4}{\tau_S} \left(\beta_{R}-\lambda\right).
\end{equation}
Equivalently, in terms of $\delta=\lambda-\beta_H$,
\begin{equation}
    \frac{d\delta}{dt}
    =
    \frac{4}{\tau_S}
    \left(
    \beta_R-\beta_H-\delta
    \right).
    \label{eq:delta_dyn}
\end{equation}

Eq.~\eqref{eq:delta_dyn} is useful as a local
quasi-canonical closure, but it should not be interpreted as the origin
of Hagedorn slowing-down. Indeed, in this strict one-parameter closure,
the factor $\mathrm{Var}_S(H_S)$ appearing in the energy balance cancels
against $d\langle H_S\rangle/d\lambda$. Therefore
Eq.~\eqref{eq:delta_dyn} describes how the canonical parameter
$\lambda$ is driven by the reservoir, but it does not by itself
demonstrate a mobility-induced suppression of the inverse-temperature
dynamics.\\

The Hagedorn bottleneck should instead be formulated at the level of the
general SEAQT scalar dynamics. In the open-system setting, and using the
same canonical inverse-temperature convention as in Sec.~\ref{sec:nonequilibrium}, the subsystem
inverse-temperature evolution has the structure
\begin{equation}
    k_B\mathrm{Var}_S(H_S)
    \frac{d\beta_S}{dt}
    =
    \mathcal{J}_{\mathrm{int}}[\rho_S]
    -
    \mathcal{J}_{R}[\rho_S,\beta_R],
    \label{eq:general_open_beta_dynamics}
\end{equation}
where $\mathcal{J}_{\mathrm{int}}$ denotes the polynomial SEAQT
fluctuation contribution and $\mathcal{J}_{R}$ denotes the
reservoir-induced driving. In the isolated limit,
$\mathcal{J}_{\mathrm{int}}$ reduces to the fluctuation polynomial obtained from the isolated $\beta_S$-dynamics. Eq.~\eqref{eq:general_open_beta_dynamics} makes the mobility
structure explicit,
\begin{equation}
    \frac{d\beta_S}{dt}
    =
    \frac{
    \mathcal{J}_{\mathrm{int}}[\rho_S]
    -
    \mathcal{J}_{R}[\rho_S,\beta_R]
    }
    {k_B\mathrm{Var}_S(H_S)} .
    \label{eq:beta_mobi}
\end{equation}
Thus a Hagedorn bottleneck occurs when the energy variance becomes large
while the numerator grows more slowly than $k_B\mathrm{Var}_S(H_S)$,
namely
\begin{equation}
    \mathrm{Var}_S(H_S)\rightarrow \infty,
    \qquad
    \mathcal{J}_{\mathrm{int}}[\rho_S]
    -
    \mathcal{J}_{R}[\rho_S,\beta_R]
    =
    o\!\left(k_B\mathrm{Var}_S(H_S)\right).
    \label{eq:bottleneck}
\end{equation}
Equivalently, since $k_B$ is constant, the condition may be written as
growth slower than $\mathrm{Var}_S(H_S)$. Under these conditions,
\begin{equation}
    \frac{d\beta_S}{dt}
    \rightarrow 0,
    \label{eq:beta_slowing}
\end{equation}
even if the reservoir continues to inject energy into the subsystem. This
is the sense in which the Hagedorn regime acts as a mobility-induced
bottleneck rather than as an ordinary equilibrium fixed point. The strength of this bottleneck depends on the algebraic exponent $a$.
From Eq.~\eqref{eq:I_n_delta}, the variance behaves for
$\delta\rightarrow 0^+$ as
\begin{equation}
    \mathrm{Var}_S(H_S)
    \sim
    \begin{cases}
    \delta^{-2},
    & a<1,\\[0.4em]
    \delta^{-2}/|\ln \delta|,
    & a=1,\\[0.4em]
    \delta^{a-3},
    & 1<a<3,\\[0.4em]
    -\ln\delta,
    & a=3,\\[0.4em]
    \mathrm{finite},
    & a>3,
    \end{cases}
    \label{eq:variance_asym}
\end{equation}

up to model-dependent constants. It is instructive to ask which values of the exponent $a$ are realized in
string theory. As set out in Sec.~\ref{sec:roleof}, the answer is fixed by
the background, in particular by the number $d$ of non-compact spatial
dimensions, and by which sector of the gas dominates the absorption of
energy near the Hagedorn scale.\\
 
When at least one spatial direction is non-compact, the convolution
of Eq.~\eqref{eq:Omega_MB_convolution} is dominated at large $E$ by configurations
containing a single string of macroscopic energy, so that the sector absorbing
the incoming energy is the single long string rather than the multi-string gas.
From the random-walk representation of highly excited
strings~\cite{Kruczenski:2006random}, a single closed string of energy $E$
propagating in $d$ non-compact spatial dimensions has
$\omega_1(E)\sim V E^{-(1+d/2)}e^{\beta_H E}$, so that the effective exponent
of the dominant Hagedorn sector is $a_{\rm eff}=1+d/2$. For $d=3$ this gives
$a_{\rm eff}=5/2$ and a power-law divergence
$\mathrm{Var}_S(H_S)\sim\delta^{-1/2}$; for $d=4$ the divergence is logarithmic; for
$d\ge5$ the variance remains finite.\\
 
When all spatial directions are compact, the relevant exponent depends on the
treatment of the winding sectors. Summing over them, as is appropriate to a
thermodynamic ensemble at fixed energy, the
asymptotic exponent approaches $a=1$, which is the marginal case of
Eq.~\eqref{eq:variance_asym}, with
$\mathrm{Var}_S(H_S)\sim\delta^{-2}/|\ln\delta|$; at intermediate energies $a(E)$
interpolates according to the effective number of compact directions available
to the typical string. The strongly suppressed values $a=10$ and $a=18$ of
Eq.~\eqref{eq:projected} apply to an ensemble at fixed winding and are not the
generic compactified result.\\

The thermodynamic inertia therefore diverges in all the cases considered
here except two: when the asymptotic geometry has $d\ge5$ non-compact
spatial directions, and when the charge sectors are held fixed by hand.
For $d=4$ the divergence is only logarithmic. It is the case
$1<a_{\rm eff}<3$, realized for $1\leq d\le3$, that is relevant to the
emergent-string limits of Sec.~\ref{sec:SDC}, where the light tower
propagates in a non-compact asymptotic geometry.\\
 
It is useful to place this criterion against the equilibrium notion of a
limiting temperature \cite{Abel:1999rq}. In the canonical description $T_H$ is limiting when
reaching it costs an infinite amount of energy. From Eq.~\eqref{eq:I_n_delta},
$\langle H_S\rangle = I_1/I_0$ diverges as $\delta\to0^+$ only for $a\le2$,
whereas the energy variance, which we have called the thermodynamic inertia,
diverges for $a<3$ (logarithmically at $a=3$). The window $2<a<3$ is therefore
one in which $T_H$ is reached at finite energy but with divergent fluctuations.
The two-sector dynamics of Sec.~\ref{sec:twosector} reproduces this classification: for $a\le2$
the approach to $\beta_H$ is asymptotic, while for $2<a<3$ the
boundary is reached in a finite time with vanishing rate. The equilibrium
criterion is nonetheless necessary but not sufficient once the question is
posed dynamically: Sec.~\ref{sec:relationtoisolated} shows that on the
quasi-canonical family the inverse-temperature mobility is identically unity,
so that a system with a divergent Hagedorn variance, driven isotropically,
crosses $\beta_H$ at an unsuppressed rate. Sustained pinning requires in
addition a suppression of the mobility, which is a statement about how the
energy is delivered and not about the density of states alone.

\subsection{A two-sector realization of the mobility-induced bottleneck}
\label{sec:twosector}

The results of Secs.~\ref{sec:relationtoisolated} and \ref{sec:hagedornpinnin} sharply constrain how a Hagedorn bottleneck can arise. The exact reservoir contribution given in Eq.~\eqref{eq:JR-exact} shows that the driven relaxation of the subsystem inverse temperature proceeds at the rate $(4\mu_R/\tau_S)(\beta_R - \beta_S)$, with the mobility $\mu_R$ of Eq.~\eqref{eq:mobility} identically equal to unity on the hypoequilibrium family. The obstruction is in fact stronger than a property of the closure. In the diagonal case, the reduced dynamics given in Eq.~\eqref{eq:rho_open_reduced} leaves the canonical family \emph{invariant}: for a microscopic probability
$q(E) \propto e^{-\lambda E}$ one has $\beta_R\,\Delta\hat F^{(R)}_S = (\beta_R - \lambda)\,\Delta\hat H_S$, which is tangent to the family, so that Eq.~\eqref{eq:delta_dyn} is not an approximation but the exact diagonal dynamics of canonically prepared states. A direct reservoir drive therefore can never generate the departure from canonical fluctuation structure that $\mu_R < 1$ requires: the suppression of the inverse-temperature mobility demands a different route for the energy. Which route a given reservoir realizes is a microscopic question, and we do not presuppose the answer. A reservoir of light quanta can populate the Hagedorn
sector in two ways: indirectly, by exciting light quasi-particle modes which then transfer energy to the long-string sector through internal exchange; and directly, through collisions energetic enough to produce highly excited strings. Only the first route can suppress the mobility, since the second is again a direct drive on a sector-canonical family and carries unit mobility by the argument just given. In this subsection we isolate the first route by considering the limiting case in which the reservoir couples to the quasi-particle band alone. This is an idealization; the direct channel is reinstated in Sec.~\ref{sec:direct}, where we determine the hierarchy of rates required for the bottleneck to survive it. The construction requires no new machinery: it suffices to apply the block-metric construction of Sec.~\ref{sec:composite} within the subsystem itself.\\

To this end, we regard the subsystem as composed of two weakly coupled subsystems in (approximate) product form: a quasi-particle band (sector $1$), with an ordinary density of states and finite energy variance $\mathrm{Var}_1(\hat H_1)$, and the Hagedorn tail (sector $2$), with shell density
$\Omega_2(E) = A E^{-a} e^{\beta_H E}\,\Theta(E-E_0)$ and
quasi-canonical parameter $\lambda_2 = \beta_H + \delta_2$,
$\delta_2 > 0$, as in Sec.~\ref{sec:hagedornpinnin}. As shown in Appendix~\ref{app:block}, the product structure induces an
orthogonal decomposition of the relevant state-space directions,
Eq.~\eqref{eq:app-decomposition} applied within the subsystem,
\begin{equation}
\mathcal{L}_S \;\supset\;
\mathrm{span}\{\hat\gamma_S\}\oplus\mathcal{L}_1\oplus\mathcal{L}_2 ,
\label{eq:twosector-split}
\end{equation}
and the covariance and variance entering the definition in
Eq.~\eqref{eq:beta_S_definition} separate exactly into sector contributions, so that
the subsystem nonequilibrium inverse temperature is the variance-weighted
mean of the sector temperatures,
\begin{equation}
\beta_S \;=\; \frac{\beta_1 \mathrm{Var}_1(\hat H_1) + \beta_2 \mathrm{Var}_2(\hat H_2)}{\mathrm{Var}_1(\hat H_1) + \mathrm{Var}_2(\hat H_2)} ,
\label{eq:betaS-weighted}
\end{equation}
with $\beta_i$ the internal sector inverse temperatures and, under the
sector-canonical form of Sec.~\ref{sec:hagedornpinnin}, $\beta_2 = \lambda_2$.

Equipping the internal dynamics of $S$ with the block-diagonal dissipative
operator, restricted to the two sectors,
$\hat{L}_S=\tau_0^{-1}\hat{P}_{S0}\oplus\tau_1^{-1}\hat{I}_1
\oplus\tau_2^{-1}\hat{I}_2$, with $\hat{P}_{S0}$ the projector onto
$\mathrm{span}\{\hat\gamma_S\}$, the common Lagrange multiplier of the
composite is the mobility-weighted mean of Eq.~\eqref{eq:app-beta-eff},
\begin{equation}
\beta_{\rm eff}
\;=\;
\frac{\tau_1^{-1} \mathrm{Var}_1(\hat H_1)\, \beta_1 + \tau_2^{-1} \mathrm{Var}_2(\hat H_2)\, \beta_2}
     {\tau_1^{-1} \mathrm{Var}_1(\hat H_1) + \tau_2^{-1} \mathrm{Var}_2(\hat H_2)},
\label{eq:beta-eff-internal}
\end{equation}
and each sector obeys the energy balance,
\begin{align}
\frac{d}{dt}\langle \hat H_1 \rangle \;&=\; -\frac{4}{\tau_1}\, \mathrm{Var}_1(\hat H_1) \left(\beta_{\rm eff} - \beta_1\right)-\frac{4}{\tau_{R1}}\, \mathrm{Var}_1(\hat H_1) \left(\beta_{\rm R} - \beta_1\right), \nonumber \\
\frac{d}{dt}\langle \hat H_2 \rangle \;&=\; -\frac{4}{\tau_2}\, \mathrm{Var}_2(\hat H_2) \left(\beta_{\rm eff} - \beta_2\right) \,.
\label{eq:sector-balance}
\end{align}

The coupling to the external reservoir is then assigned to the quasi-particle band only, adding to the right-hand side of Eq.~\eqref{eq:sector-balance} for $d/dt \langle H_1\rangle$ with $\tau_{R1}$ the system--reservoir relaxation time: energy enters the Hagedorn sector only by cascading through the light modes. Using the sector compressibilities $\partial_{\beta_i}\langle \hat H_i\rangle = -\mathrm{Var}_i(\hat H_i)$, the sector
equations read
\begin{align} \label{eq:beta2-dot}
\frac{d\beta_1}{dt}
&= \frac{4}{\tau_1}\left(\beta_{\rm eff} - \beta_1\right)
 + \frac{4}{\tau_{R1}}\left(\beta_R - \beta_1\right), \\[2pt] 
\frac{d\beta_2}{dt}
&= \frac{4}{\tau_2}\left(\beta_{\rm eff} - \beta_2\right)
 = \frac{4}{\tau_2}\,
   \frac{\tau_1^{-1} \mathrm{Var}_1(\hat H_1)}{\tau_1^{-1} \mathrm{Var}_1(\hat H_1) + \tau_2^{-1} \mathrm{Var}_2(\hat H_2)}
   \left(\beta_1 - \beta_2\right). \nonumber
\end{align}
Eq.~\eqref{eq:beta2-dot} is the central result of this
subsection. In contrast with the direct-drive setting, the variance of
the Hagedorn sector no longer cancels: the energy influx into sector
$2$ is controlled by the \emph{band} variance $\mathrm{Var}_1(\hat H_1)$, which remains
finite, while displacing $\beta_2$ costs an amount of energy set by
the \emph{tail} variance $\mathrm{Var}_2(\hat H_2)$. In the regime
$\tau_2^{-1} \mathrm{Var}_2(\hat H_2) \gg \tau_1^{-1} \mathrm{Var}_1(\hat H_1)$ one finds
\begin{equation}
\frac{d\beta_2}{dt}
\;\simeq\;
\frac{4}{\tau_1}\,\frac{\mathrm{Var}_1(\hat H_1)}{\mathrm{Var}_2(\hat H_2)}\,(\beta_1 - \beta_2)
\;\longrightarrow\; 0 ,
\qquad
\frac{d}{dt}\langle \hat H_2\rangle
\;\simeq\;
\frac{4}{\tau_1}\, \mathrm{Var}_1(\hat H_1)\, (\beta_2 - \beta_1)
\;=\; \mathcal{O}(1),
\label{eq:suppression}
\end{equation}
so that the intensive variable of the Hagedorn sector freezes while
the energy flux into it remains finite and nonvanishing. Since
$\mathrm{Var}_2(\hat H_2) \to \infty$ also forces $\beta_S \to \beta_2$ in Eq.~\eqref{eq:betaS-weighted}, with
$\beta_S - \beta_2 = \mathcal{O}(\mathrm{Var}_1(\hat H_1)/\mathrm{Var}_2(\hat H_2))$, the subsystem inverse
temperature inherits the same suppression. This realizes explicitly
the bottleneck condition given in Eq.~\eqref{eq:bottleneck}: the numerator of the general scalar dynamics given in Eq.~\eqref{eq:beta_mobi} is here of order $k_B \mathrm{Var}_1(\hat H_1) (\beta_1 - \beta_2)$, while the denominator $k_B \mathrm{Var}_S(H_S) \simeq k_B \mathrm{Var}_2(\hat H_2)$ diverges.\\

The approach to the Hagedorn boundary can be made fully explicit. When
the reservoir is hotter than the Hagedorn scale, $\beta_R < \beta_H$,
the band relaxes to the stationary value obtained from Eq.~
\eqref{eq:beta2-dot} with $\beta_{\rm eff} \to \beta_2 \to \beta_H$,
\begin{equation}
\beta_1^{*}
\;=\;
\frac{\tau_1^{-1}\,\beta_H + \tau_{R1}^{-1}\,\beta_R}
     {\tau_1^{-1} + \tau_{R1}^{-1}}
\;\in\; \left(\beta_R,\, \beta_H\right),
\label{eq:beta1-star}
\end{equation}
so that the light modes equilibrate strictly between the reservoir and
the Hagedorn temperature, at a value fixed by the mobility ratio,  an observable prediction of the mechanism. Using $\mathrm{Var}_2(\hat H_2) \simeq C_a\,\delta_2^{a-3}$ from Eq.~\eqref{eq:variance_asym}, valid for $1<a<3$,
the suppressed dynamics\footnote{For $a<1$ the variance instead behaves as $\mathrm{Var}_S\sim\delta^{-2}$, so
that Eq.~\eqref{eq:delta-regimes} is replaced by
$\dot\delta_2\simeq-\mathcal{K}'\delta_2^{2}$ with algebraic freezing
$\delta_2\sim(\mathcal{K}'t)^{-1}$; at $a=1$ the additional $|\ln\delta|$ of
Eq.~\eqref{eq:variance_asym} modifies the rate logarithmically. Both cases are
relevant: as discussed in Sec.~\ref{sec:roleof}, $a=1$ is the asymptotic
value in compact backgrounds once the winding sectors are summed.} given in Eq.~\eqref{eq:suppression} becomes
\begin{equation} 
\frac{d\delta_2}{dt}
\;\simeq\;
-\,\mathcal{K}\, \delta_2^{\,3-a},
\qquad
\mathcal{K} := \frac{4 \mathrm{Var}_1(\hat H_1) (\beta_H - \beta_1^{*})}{\tau_1\, C_a} > 0 ,
\label{eq:delta-flow}
\end{equation}
whose solutions exhibit three regimes:
\begin{equation}
\delta_2(t) \;\sim\;
\begin{cases}
\big[(2-a)\,\mathcal{K}\, t\big]^{-1/(2-a)}, & 1<a < 2,
\quad \text{(algebraic freezing)}\\[4pt]
\delta_2(0)\, e^{-\mathcal{K} t}, & a = 2,\\[4pt]
\big[\delta_2(0)^{a-2} - (a-2)\mathcal{K} t\big]^{1/(a-2)}, & 2 < a < 3,
\quad \text{(boundary reached at finite $t_*$)} .
\end{cases}
\label{eq:delta-regimes}
\end{equation}
In all three cases $\dot\beta_S \propto \delta_2^{\,3-a} \to 0$: the
subsystem inverse temperature is dynamically pinned at $\beta_H$ while
the Hagedorn sector continues to absorb energy at the finite rate given in Eq.~\eqref{eq:suppression}, which is the long-string condensation channel.
For $a > 3$, by contrast, $\mathrm{Var}_2(\hat H_2)$ remains finite as $\delta_2 \to 0^{+}$, Eq.~(\ref{eq:beta2-dot}) retains a finite rate, and the trajectory reaches the normalizability boundary $\lambda_2 = \beta_H$ at finite speed without any freezing of $\beta_S$; there the quasi-canonical family is simply exhausted and the description must be continued beyond it. The mobility-induced bottleneck is therefore a genuine dynamical phenomenon of broad Hagedorn tails, $a < 3$, and not a kinematical consequence of the boundary itself. Finally, if the
reservoir is colder than the Hagedorn scale, $\beta_R > \beta_H$, then $\beta_1 - \beta_2 > 0$ in Eq.~\eqref{eq:beta2-dot} drives $\delta_2$ to larger values, $\mathrm{Var}_2(\hat H_2)$ decreases, and the ordinary relaxation of the whole subsystem toward $\beta_R$ is recovered, consistently with the discussion at the end of Sec.~\ref{sec:hagedornpinnin}.

In the language of Sec.~\ref{sec:relationtoisolated}, the construction realizes the suppression of the inverse-temperature mobility explicitly. Defining the measured mobility through $\dot\beta_S =: (4\mu_{\rm eff}/\tau_{R1})(\beta_R - \beta_S)$, the leading behavior given in Eq.~\eqref{eq:suppression} gives
\begin{equation}
\mu_{\rm eff}
\;\simeq\;
\frac{\tau_{R1}}{\tau_1}\,
\frac{\mathrm{Var}_1(\hat H_1)}{\mathrm{Var}_2(\hat H_2)}\,
\frac{\beta_1 - \beta_2}{\beta_R - \beta_S}
\;\longrightarrow\; 0
\qquad (\delta_2 \to 0^{+},\; a<3),
\label{eq:mu-eff}
\end{equation}
in contrast with $\mu_R \equiv 1$ for the direct drive: the
energy-injection topology, not the Hagedorn density of states alone,
is what converts the divergent variance into a vanishing mobility. Two
remarks are in order. First,
Eqs.~(\ref{eq:beta2-dot}) are not merely a
closure: by the same invariance argument given at the beginning of
this subsection, applied sector by sector with $\beta_{\rm eff}$ (and,
for the band, $\beta_R$) in place of the drive parameter, the diagonal
block dynamics leaves each sector-canonical family invariant, so the
two-parameter system is the \emph{exact} reduced dynamics of
canonically prepared sectors; we have verified numerically that the
pinning behavior is robust when the tail is prepared in a
non-canonical (mixed) state. Second, coupling the reservoir to the band alone is the limiting case isolated here; its role is made transparent by Eq.~\eqref{eq:mu-eff} and the effect of relaxing it is quantified in Sec.~\ref{sec:direct}; and any injection channel that reaches the
Hagedorn sector only through degrees of freedom of finite variance
will produce the same suppression.

\begin{figure}[!htb]
\centering
a)\includegraphics[scale=0.29]{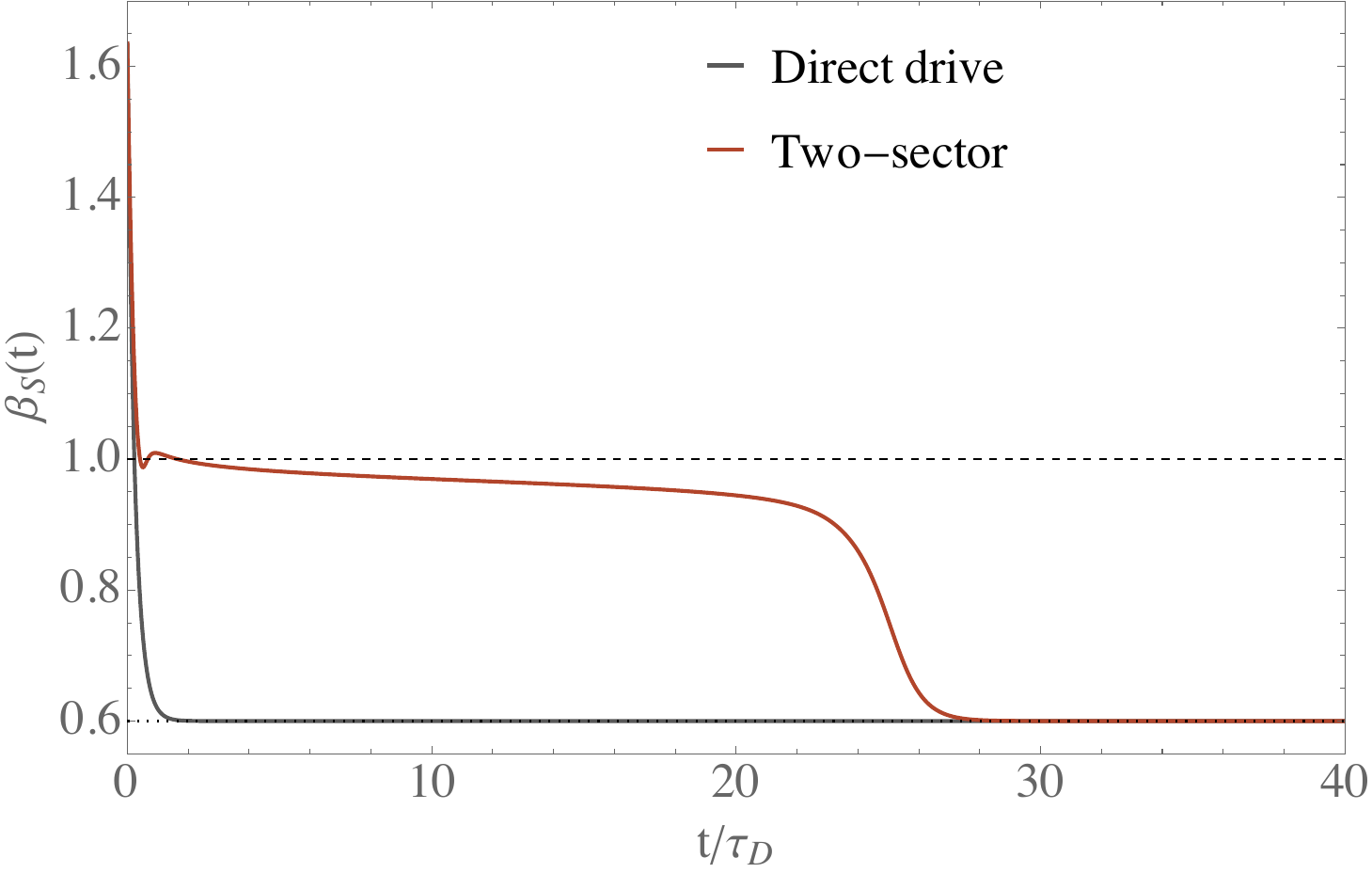}
b)\includegraphics[scale=0.29]{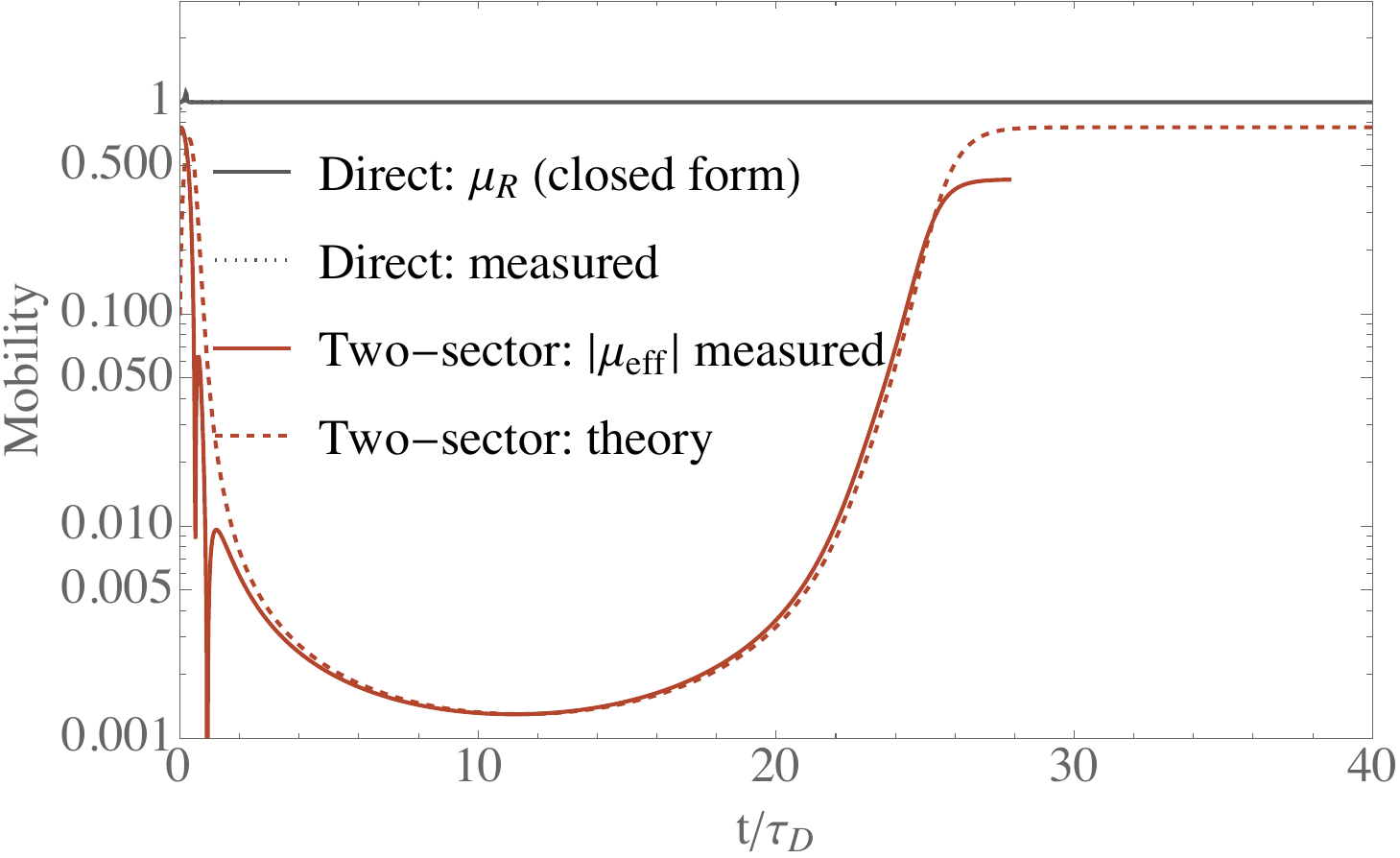}
c)\includegraphics[scale=0.29]{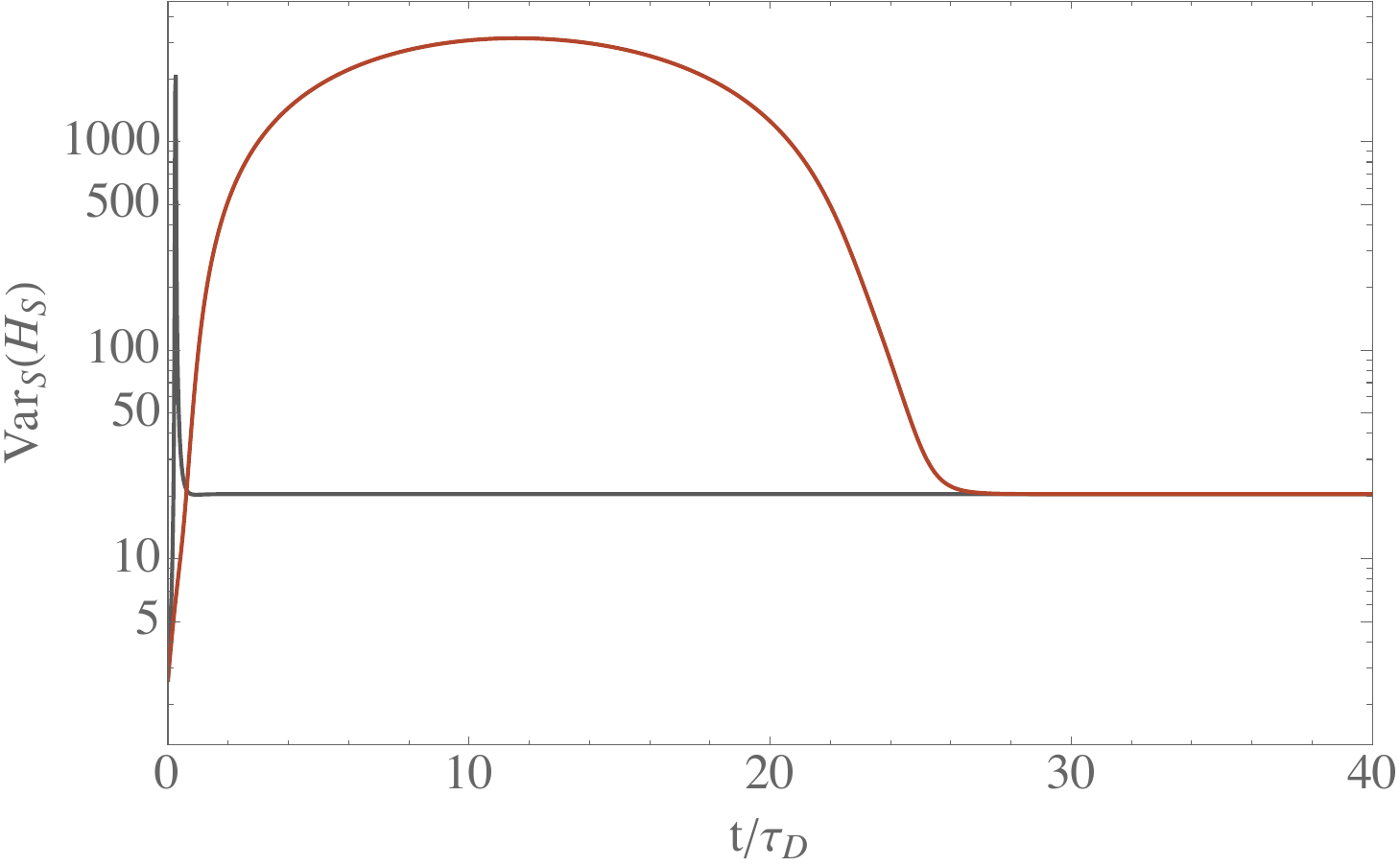}
\caption{Mobility collapse from the exact diagonal dynamics, without any
closure. The subsystem is a product of a quasi-particle band
($\omega \propto \varepsilon^{p-1}$, $p=5$) and a Hagedorn tail
($\Omega \propto E^{-3/2}e^{\beta_H E}$ on $[\beta_H^{-1},
150\,\beta_H^{-1}]$), evolved at the level of the full populations with
identical, deliberately non-canonical initial data (the tail is a
canonical mixture) and $\beta_R = 0.6\,\beta_H$.
(a)~Under the direct isotropic drive, Eq.~\eqref{eq:rho_open_reduced}, $\beta_S$ crosses the
Hagedorn scale unimpeded; under the two-sector dynamics of
Sec.~\ref{sec:twosector} it is pinned at $\beta_H$ for
$\sim 25\,\tau_D$, escaping only when the finite truncated Hagedorn
sector is exhausted (the plateau duration grows with $E_{\rm max}$,
i.e.\ with the available long-string phase space). Both runs reach the
same final state.
(b)~Inverse-temperature mobility: for the direct drive the closed-form
$\mu_R$ of Eq.~\eqref{eq:mobility} equals unity throughout, and the
exact affine equation~(\ref{eq:415-explicit}) is verified along the
trajectory to a median residual of $10^{-8}$; for the two-sector
dynamics the measured $|\mu_{\rm eff}|$ collapses by nearly three
orders of magnitude and tracks the prediction (\ref{eq:mu-eff}) to a
few percent (median ratio $1.01$) in the pinned regime. The brief early
sign change of $\mu_{\rm eff}$ is the transient weighted-mean shift of
Eq.~(\ref{eq:betaS-weighted}) as the tail variance grows.
(c)~Thermodynamic inertia $\mathrm{Var}_S(H_S)$.}
\label{fig:mobility}
\end{figure}

Figure~\ref{fig:mobility} tests the mechanism without invoking the
quasi-canonical closure at all. The subsystem is evolved at the level of
the full diagonal populations, with a quasi-particle band and a truncated
Hagedorn tail prepared in deliberately non-canonical initial data, under
two dynamics with identical initial conditions: the direct isotropic drive
of Eq.~\eqref{eq:rho_open_reduced} and the two-sector dynamics of the present
subsection. Under the direct drive $\beta_S$ crosses the Hagedorn scale
unimpeded and the closed-form mobility $\mu_R$ of Eq.~\eqref{eq:mobility}
equals unity throughout, with the affine Eq.~\eqref{eq:415-explicit}
satisfied along the trajectory to a median residual of $10^{-8}$; this is
a direct numerical confirmation of the no-go statement of
Sec.~\ref{sec:relationtoisolated}. Under the two-sector dynamics the measured
mobility collapses by nearly three orders of magnitude and tracks the
prediction given in Eq.~\eqref{eq:mu-eff} to a few percent in the pinned regime. Two
features are worth noting. First, the pinning is not permanent for a
truncated tail: the plateau persists for $\sim 25\,\tau_D$ and terminates
once the finite Hagedorn sector is exhausted, its duration growing with
$E_{\max}$, that is, with the available long-string phase space. The
divergent-inertia limit of Eq.~\eqref{eq:variance_asym} is thus recovered as
$E_{\max}\to\infty$, and at finite $E_{\max}$ the mechanism produces a
long but finite plateau. Second, both runs reach the same final state, so
the two-sector construction alters the route to equilibrium and not its
endpoint; the bottleneck is dynamical, as claimed.

\begin{figure}[!htb]
\centering
a) \includegraphics[scale=0.29]{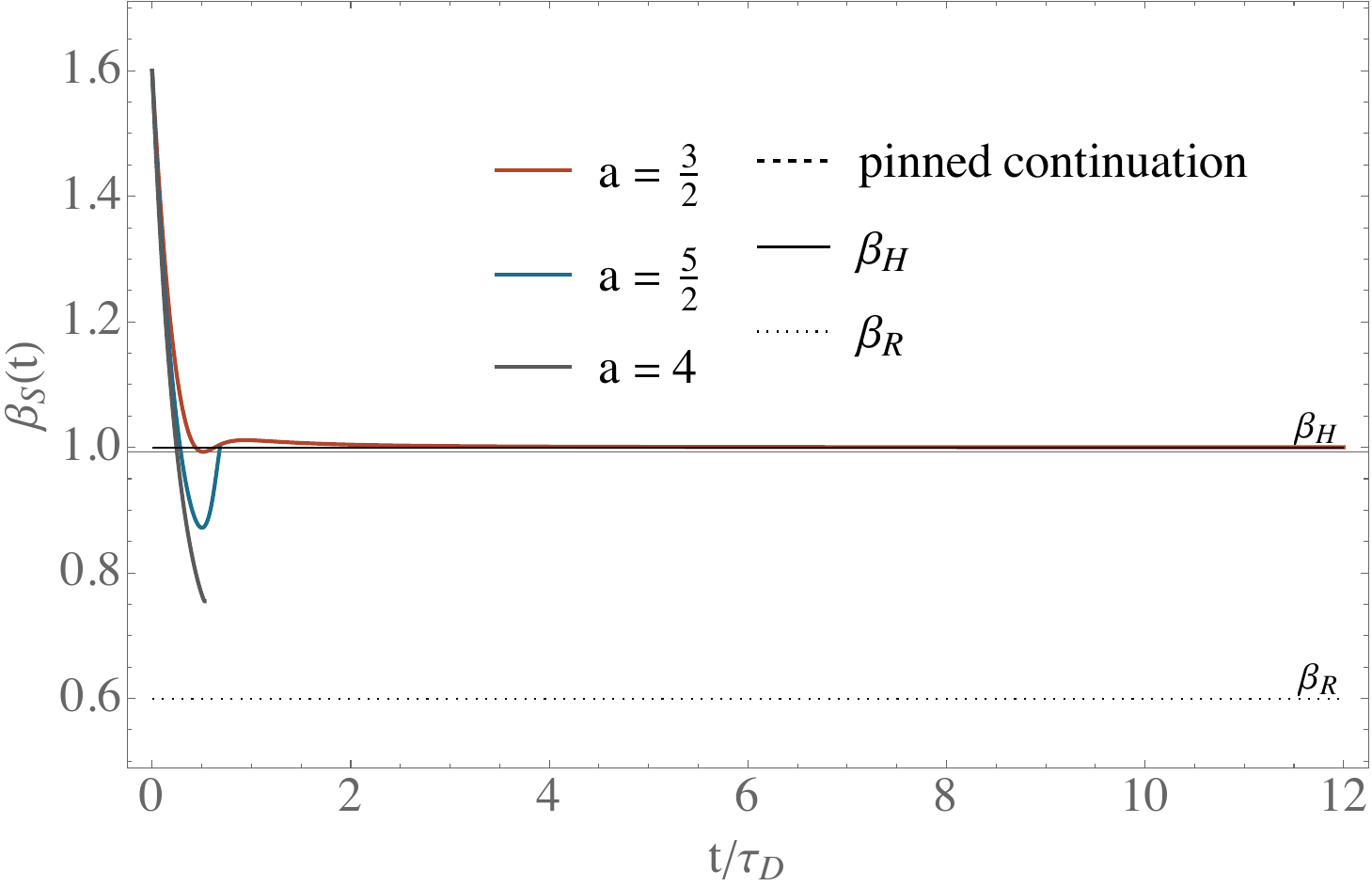} 
b)  \includegraphics[scale=0.29]{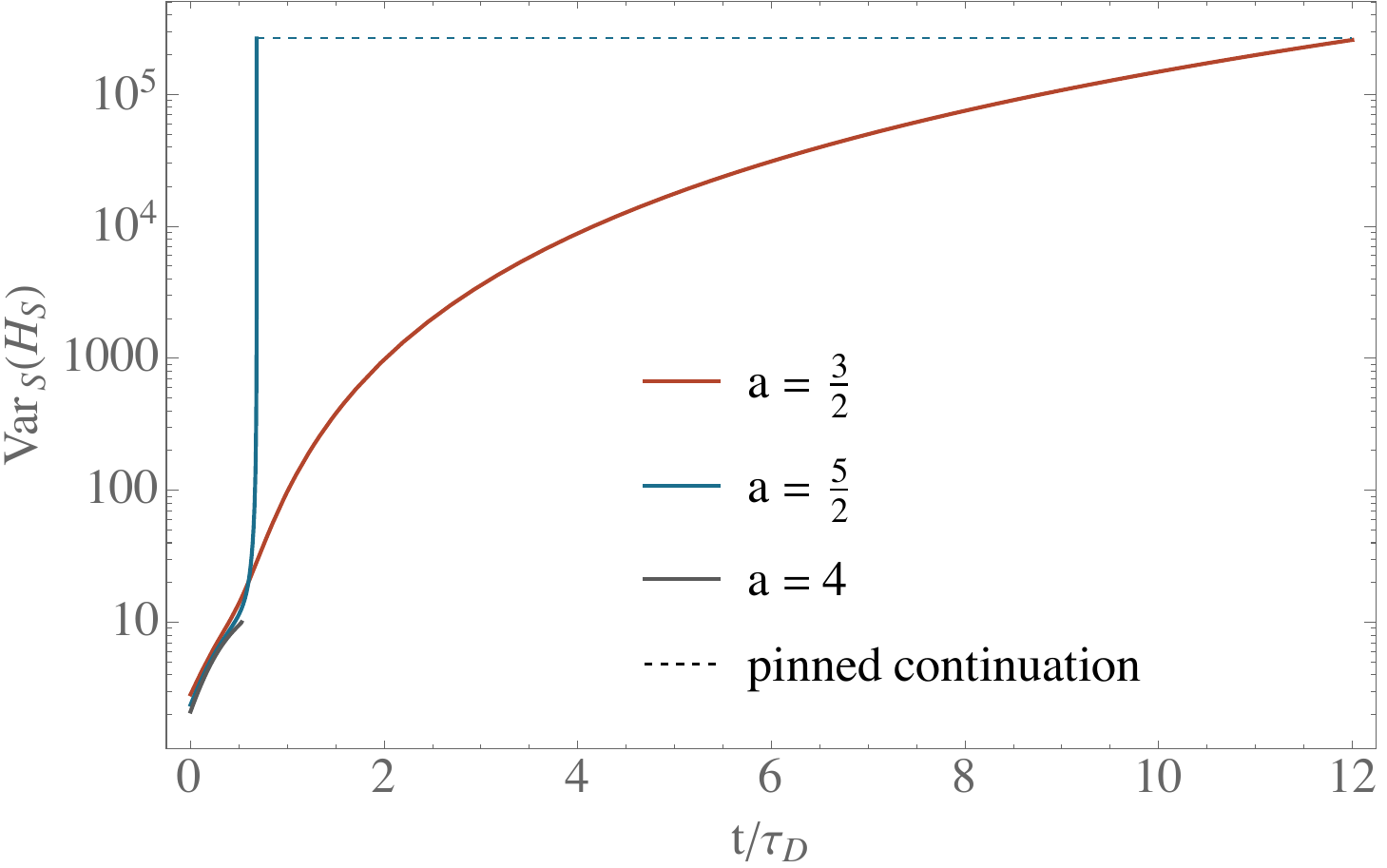}
\caption{Two-sector realization of the Hagedorn bottleneck,
Eqs.~\eqref{eq:beta2-dot}, for a reservoir hotter than
the Hagedorn scale ($\beta_R = 0.6\,\beta_H$), a quasi-particle band with
$\mathrm{Var}_1(\hat H_1) = p/\beta_1^2$, $p= 5$, Hagedorn tail with $E_0 = \beta_H^{-1}$,
equal relaxation times $\tau_1 = \tau_2 = \tau_{R1} = \tau_D$, and cold
initial condition $\beta_1(0) = \beta_2(0) = 1.6\,\beta_H$.
(a) Subsystem inverse temperature $\beta_S(t)$ from Eq.~\eqref{eq:betaS-weighted}.
For $a = 5/2$ the boundary $\delta_2 \to 0^+$ is reached at finite time (open
circle) with vanishing rate, $\dot\beta_S \sim \delta_2^{3-a} \to 0$, after
which the state is pinned at $\beta_H$ (dashed continuation); for $a = 4$ the
tail variance stays finite, no pinning of $\beta_S$ occurs, and the
quasi-canonical family is exhausted at finite rate while $\beta_S$ relaxes
toward $\beta_R$. In the pinned regime the band equilibrates at
$\beta_1^* = (\beta_H + \beta_R)/2 = 0.8\,\beta_H$, in exact agreement with
Eq.~\eqref{eq:beta1-star}. (b) Thermodynamic inertia $\mathrm{Var}_S(H_S)=\mathrm{Var}_1(\hat H_1)+\mathrm{Var}_2(\hat H_2)$ on a logarithmic scale: it diverges for $a<3$, here as $\delta_2^{a-3}$, and saturates at an $\mathcal{O}(1)$ value for $a=4$. For $a=3/2$, $\beta_S$ freezes algebraically onto $\beta_H$ without ever reaching the boundary, Eq.~\eqref{eq:delta-regimes}.}
\label{fig:twosector}
\end{figure}

Figure~\ref{fig:twosector} shows the numerical integration of
Eq.~(\ref{eq:beta2-dot}) with the exact
incomplete-Gamma moments of Eq.~\eqref{eq:I_n_delta}, for three representative
exponents. The asymptotic law given in Eq.~\eqref{eq:delta-regimes} is verified quantitatively: for $a = 3/2$ the predicted $\delta_2(t) \simeq [(2-a)\mathcal{K}t]^{-1/(2-a)}$ reproduces the integrated value at $t = 12\,\tau_D$ to within a few percent. The consequences of this mechanism along moduli-space trajectories, where the Hagedorn scale itself becomes moduli dependent, are developed in Sec.~\ref{sec:SDC}.

\subsection{Reinstating the direct injection channel}
\label{sec:direct}
 
The construction of Sec.~\ref{sec:twosector} assigns the reservoir coupling
entirely to the quasi-particle band. Physically this is an idealization: a
reservoir able to produce highly excited string states in collisions of light
quanta populates the Hagedorn sector directly as well. We now reinstate that
channel and determine when the bottleneck survives it.\\
 
Let $\tau_{R2}$ denote the relaxation time of the direct reservoir--tail
channel. The second of Eqs.~\eqref{eq:beta2-dot} becomes
\begin{equation}
\frac{d\beta_2}{dt}
=\frac{4}{\tau_2}\bigl(\beta_{\rm eff}-\beta_2\bigr)
+\frac{4}{\tau_{R2}}\bigl(\beta_R-\beta_2\bigr),
\label{eq:beta2-direct}
\end{equation}
the equation for $\beta_1$ being unchanged. Thus, the
added term is a direct drive on a sector-canonical family, so that the tail
variance cancels between the energy flux
$-(4/\tau_{R2})\mathrm{Var}_2(\hat H_2)(\beta_R-\beta_2)$ and the sector
compressibility $\partial_{\beta_2}\langle\hat H_2\rangle=-\mathrm{Var}_2(\hat H_2)$:
it carries unit mobility. The cascade term, by contrast, has its rate
controlled by $\mathrm{Var}_1(\hat H_1)/\mathrm{Var}_2(\hat H_2)$ and collapses as the tail
variance diverges.\\
 
The immediate consequence is that the mobility collapse saturates. Repeating
the computation leading to Eq.~\eqref{eq:mu-eff}, the measured mobility acquires
an additive floor,
\begin{equation}
\mu_{\rm eff}\;\simeq\;
\frac{\tau_{R1}}{\tau_1}\,
\frac{\mathrm{Var}_1(\hat H_1)}{\mathrm{Var}_2(\hat H_2)}\,
\frac{\beta_1-\beta_2}{\beta_R-\beta_S}
\;+\;\frac{\tau_{R1}}{\tau_{R2}}\,
\frac{\beta_R-\beta_2}{\beta_R-\beta_S}\,.
\label{eq:mu-floor}
\end{equation}
so that the attainable suppression is bounded below by $\tau_{R1}/\tau_{R2}$. It should be stressed that the hierarchy does not improve along the trajectory: since $\mathrm{Var}_2 (H_2) \rightarrow\infty$ as the Hagedorn scale is approached, the cascade rate vanishes while the direct rate does not, and the direct channel becomes relatively more important the deeper into the Hagedorn regime the system is driven. The bottleneck is in this sense intrinsically a transient, as it already was for a tail of finite extent.\\

Its lifetime can be estimated analytically in the quasi-canonical closure.
Using $\mathrm{Var}_2(\hat H_2)\simeq C_a\delta_2^{a-3}$ and expanding
Eq.~\eqref{eq:beta2-direct} for $\delta_2\ll\beta_H-\beta_R$,
\begin{equation}
\frac{d\delta_2}{dt}\simeq-\mathcal{K}\,\delta_2^{\,3-a}-g,
\qquad
g:=\frac{4}{\tau_{R2}}(\beta_H-\beta_R)>0,
\label{eq:delta-direct}
\end{equation}
with $\mathcal{K}$ as in Eq.~\eqref{eq:delta-flow}. The two terms balance at
$\delta_\times=(g/\mathcal{K})^{1/(3-a)}$, which sets the resolution of the
pinning: the intensive variable freezes at $\beta_H+O(\delta_\times)$, and below $\delta_\times$ the direct channel carries
the state across at finite speed. For $1<a<2$ this gives a plateau of duration\footnote{For $a<2$ the cascade alone never reaches the boundary: $\delta_2$
freezes algebraically, Eq.~\eqref{eq:delta-regimes} (and footnote~2 for
$a\le1$), so the plateau length is set entirely by the direct channel; for
$1<a<2$ it scales universally with $\tau_{R2}$ as in Eq.~\eqref{eq:tcross}.
For $a=2$ the approach is exponential and the plateau grows only
logarithmically, $t_\times\simeq\mathcal{K}^{-1}\ln(1+\mathcal{K}\delta_2(0)/g)$.
For $2<a<3$ the cascade already reaches $\delta_2=0$ at a finite time $t_*$
fixed by the initial data, Eq.~\eqref{eq:delta-regimes}, and the direct
channel only shortens that plateau.}
\begin{equation}
t_\times\simeq\frac{\delta_\times^{-(2-a)}}{(2-a)\mathcal{K}}
\;\propto\;\tau_{R2}^{\,(2-a)/(3-a)} ,
\label{eq:tcross}
\end{equation}
obtained by integrating the cascade-dominated regime of Eq.~\eqref{eq:delta-direct} down to $\delta_\times$; the final stage, $\delta_2<\delta_\times$, lasts $\delta_\times/g$ and has the same scaling.\\

\begin{figure}[!htb]
\centering
a)\includegraphics[width=0.4\textwidth]{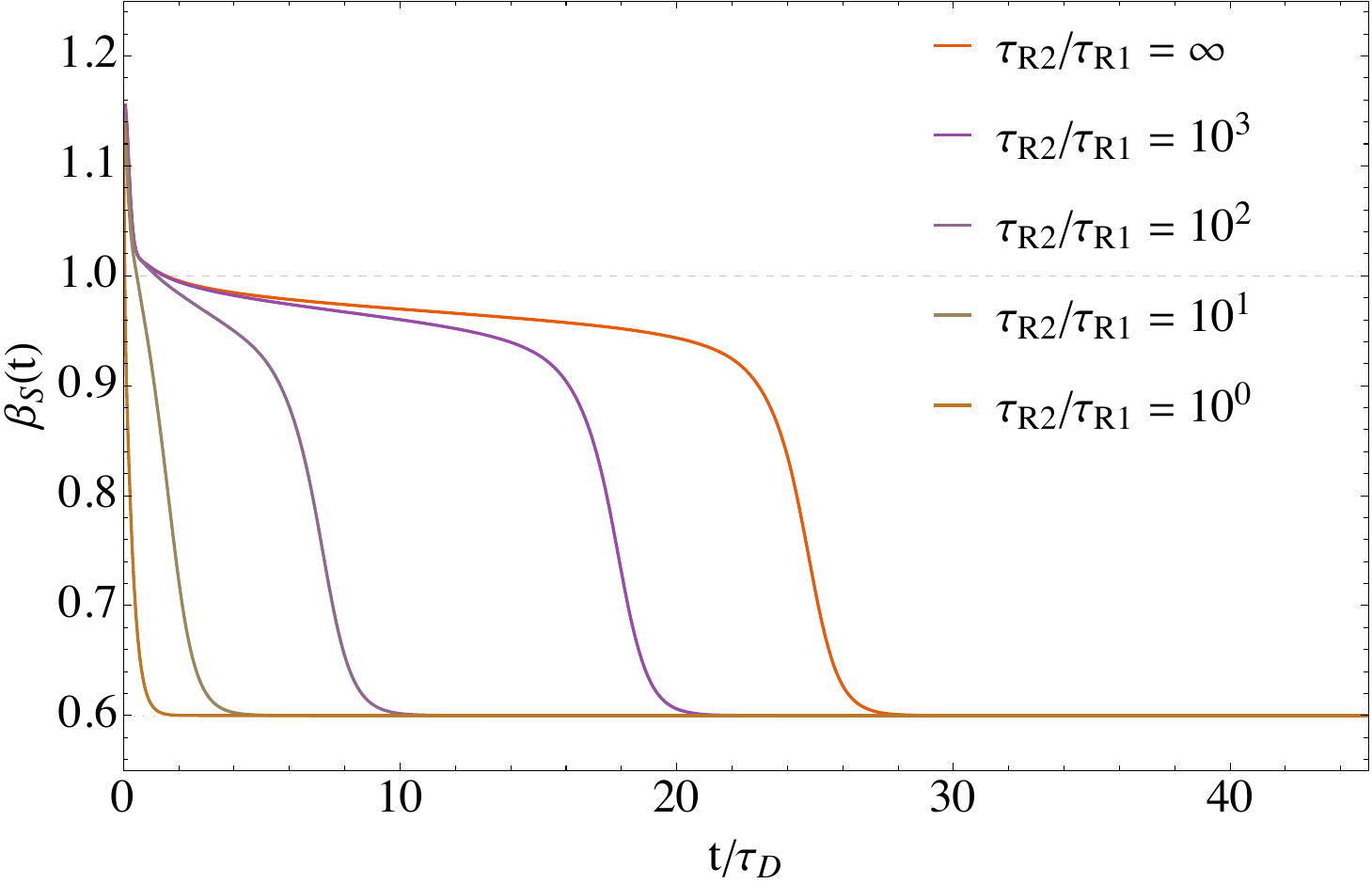}
b)\includegraphics[width=0.4\textwidth]{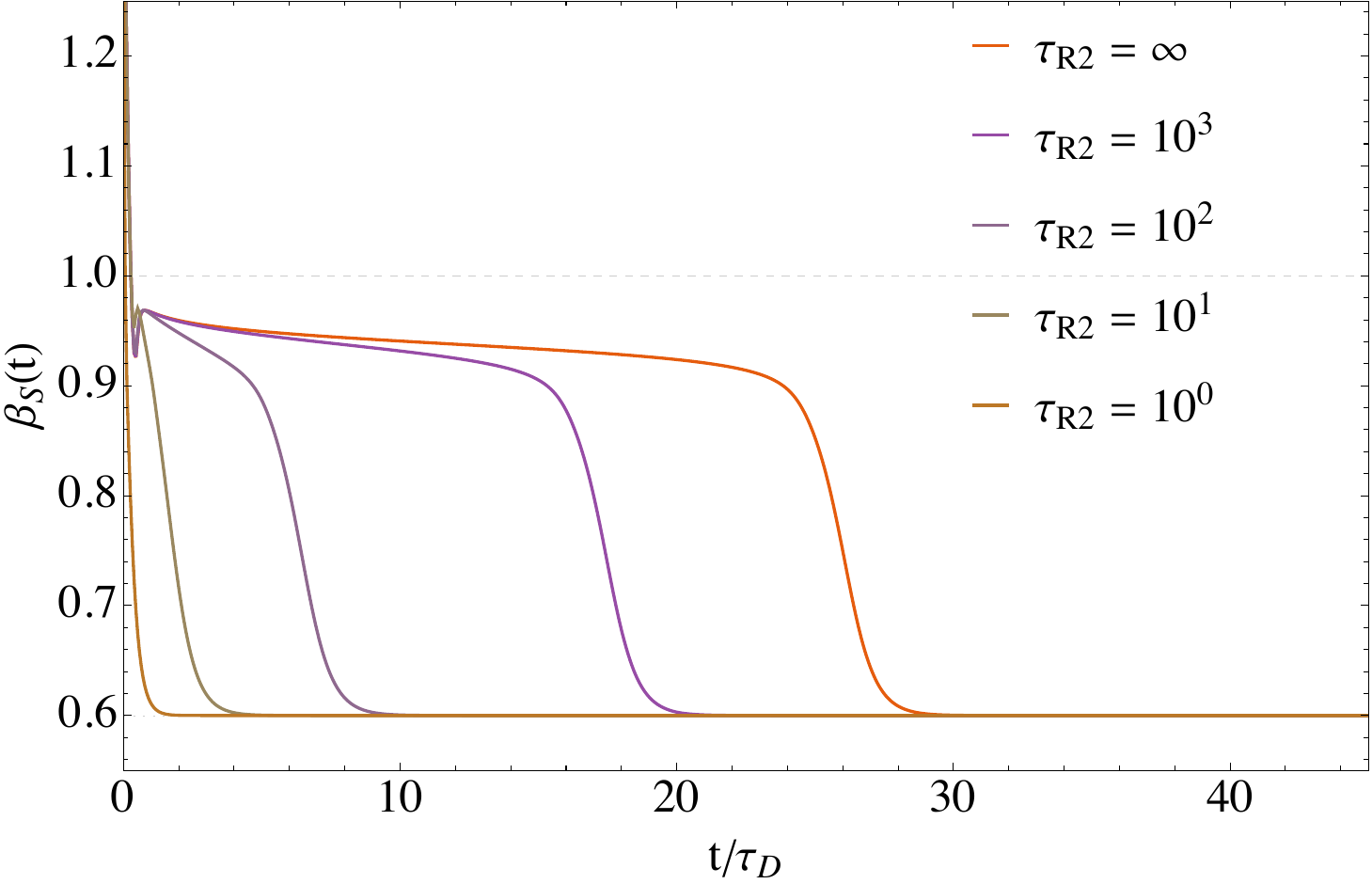}\\
c)\includegraphics[width=0.4\textwidth]{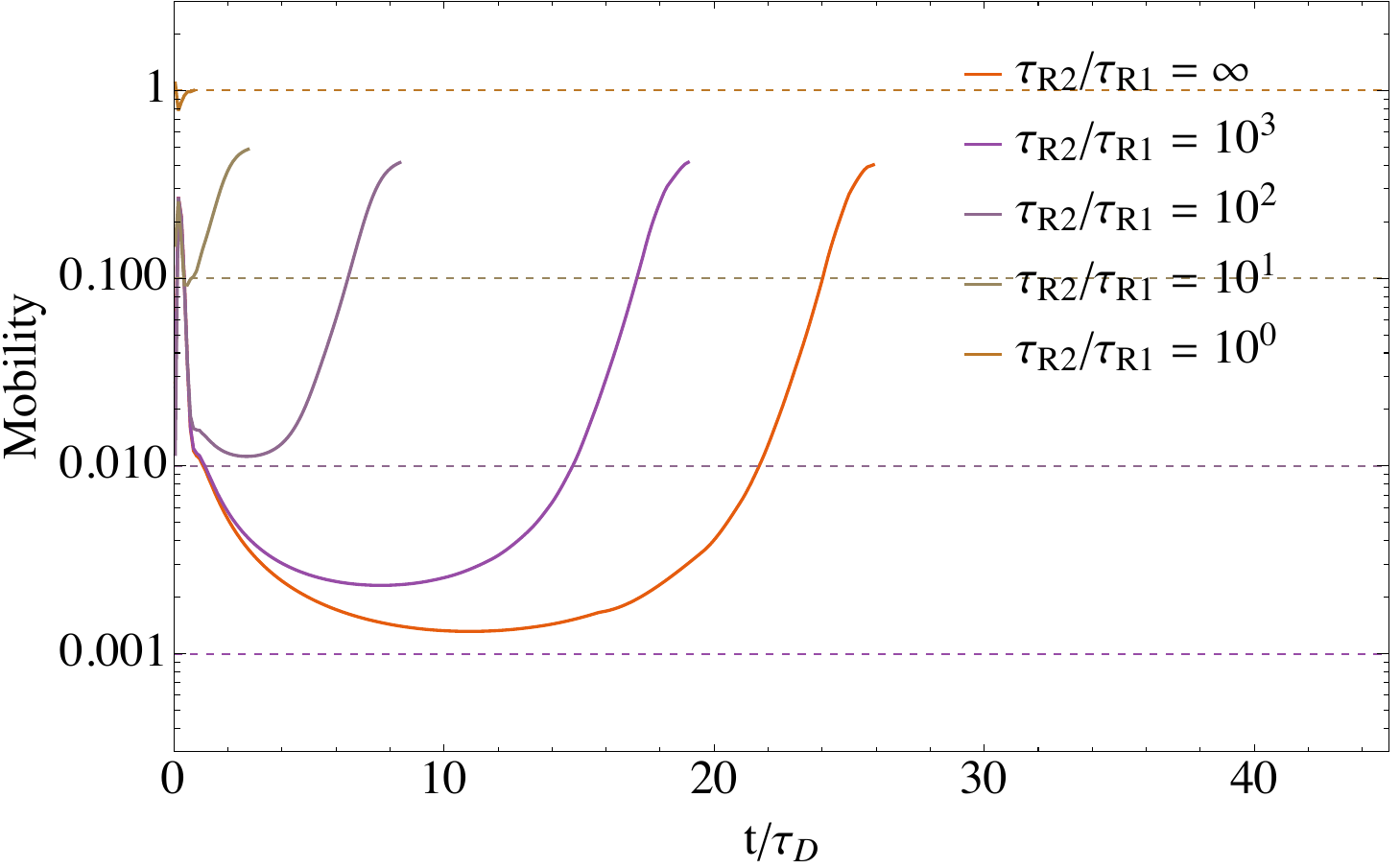}
d)\includegraphics[width=0.4\textwidth]{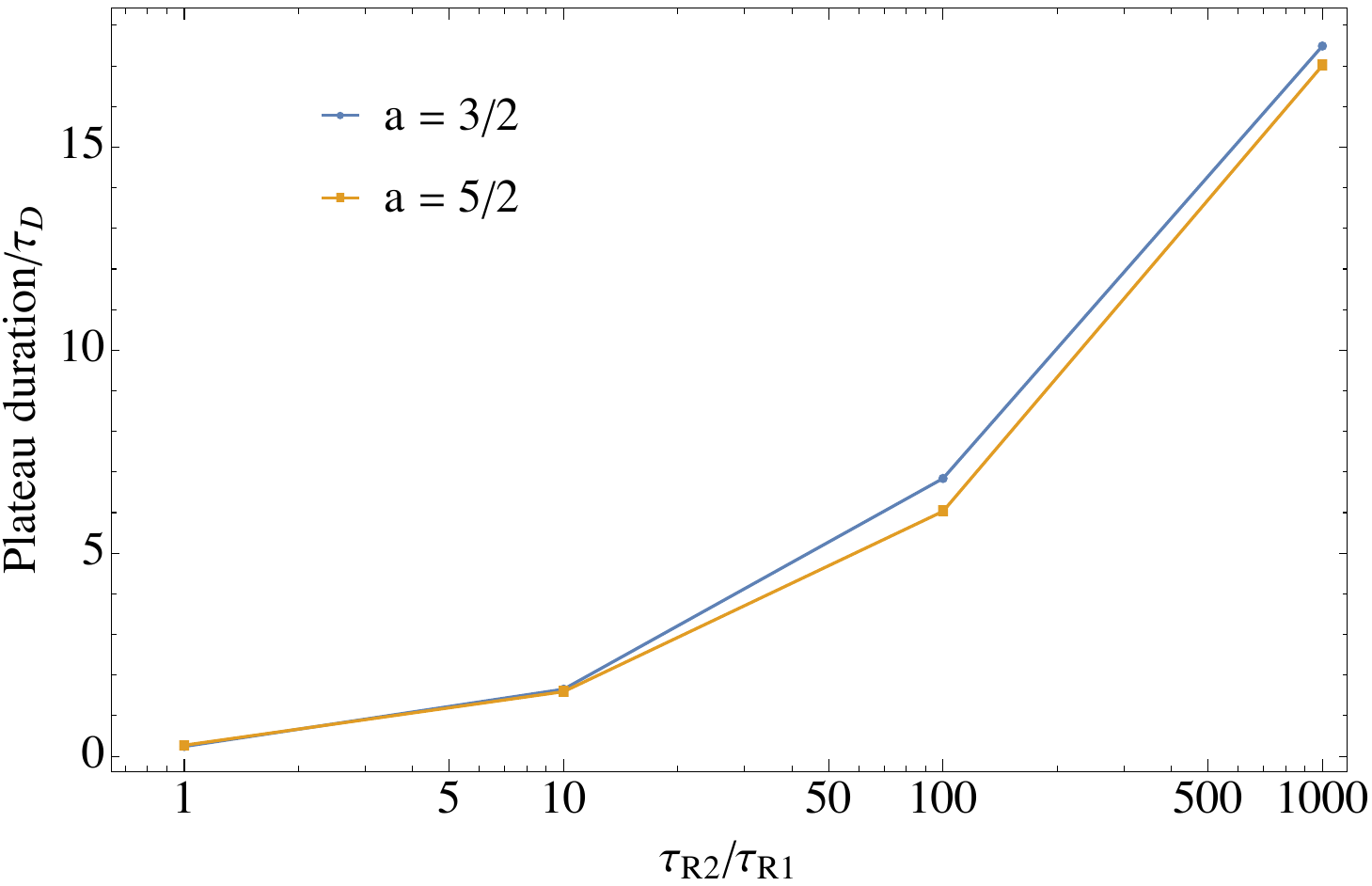}\\
\caption{The subsystem is that of Fig.~\ref{fig:mobility}: a quasi-particle band
($\omega\propto\varepsilon^{p-1}$, $p=5$) and a Hagedorn tail
$\Omega\propto E^{-a}e^{\beta_H E}$ truncated to
$E_0\le E\le E_{\max}$, with $E_0=\beta_H^{-1}$ and
$E_{\max}=150\,\beta_H^{-1}$, prepared in deliberately non-canonical initial data, with
$\beta_R=0.6\,\beta_H$ and $\tau_1=\tau_2=\tau_{R1}=\tau_D$. The reservoir
couples to the tail directly at rate $\tau_{R2}^{-1}$ in addition to the
cascade through the band; $\tau_{R2}=\infty$ is the two-sector dynamics of
Eqs.~\eqref{eq:beta2-dot}. (a) $\beta_S(t)$ for $a=3/2$. (b) The same for
$a=5/2$, the effective exponent of the dominant long-string sector in $d=3$
non-compact spatial dimensions; the behaviour is quantitatively almost
identical, and the inability of the quasi-canonical closure to reach this case
is an artifact of the boundary $\delta_2\to0^+$ of that closure. (c) Measured
mobility $\mu_{\rm eff}=(\tau_{R1}/4)\,\dot\beta_S/(\beta_R-\beta_S)$ for
$a=3/2$; dashed lines are the asymptotic floor $\tau_{R1}/\tau_{R2}$ to which
the direct term of Eq.~\eqref{eq:mu-floor} reduces as $\beta_S\to\beta_2$. Over
the plateau the factor $(\beta_R-\beta_2)/(\beta_R-\beta_S)$ differs from unity
by less than about $1\%$, so the distinction is not visible on this scale. The
collapse saturates at the floor for $\tau_{R2}\lesssim10^{2}\tau_{R1}$ and at
the truncation-limited value otherwise. Curves are shown up to the escape,
beyond which $\beta_R-\beta_S\to0$ and the measured mobility is not defined.
(d) Plateau duration, defined as the time at which $\beta_S$ crosses
$(\beta_H+\beta_R)/2$, versus $\tau_{R2}/\tau_{R1}$; dotted lines mark the
truncation-limited value attained as $\tau_{R2}\to\infty$.}
\label{fig:direct}
\end{figure}

Equation~\eqref{eq:tcross} is, however, the law for an untruncated tail, in
which the plateau is limited by the asymptotic freezing itself. The physically
relevant situation is the one of Fig.~\ref{fig:mobility}, where the available
long-string phase space is finite and the plateau is instead limited by its
exhaustion. Figure~\ref{fig:direct} therefore repeats the analysis at the level
of the full diagonal populations of both sectors. Three
features are worth noting.\\

First, the pinning survives the direct channel, with a duration set by whichever
of the two termination mechanisms acts first, and disappears entirely in the
isotropic limit $\tau_{R2}=\tau_{R1}$, where Eq.~\eqref{eq:mu-floor} returns
$\mu_{\rm eff}=O(1)$ and the no-go statement of
Sec.~\ref{sec:relationtoisolated} is recovered numerically. A reservoir coupling democratically to the whole string spectrum produces no slowing-down; this is contained in the present model as a limiting case.\\
 
Second, the hierarchy required is finite and is set by the available long-string phase space. With the truncation of
Fig.~\ref{fig:mobility}, roughly $70\%$ of the plateau survives already at
$\tau_{R2}=10^{3}\tau_{R1}$, and a plateau of several $\tau_D$ persists at
$\tau_{R2}=10^{2}\tau_{R1}$. What is required is that the direct channel be
slow compared with the rate at which the available long-string phase space is
exhausted, not compared with the asymptotic freezing rate of an infinite tail.\\
 
Third, the mechanism is not confined to $a<2$. The quasi-canonical closure
cannot describe $2<a<3$ beyond the finite time at which $\delta_2\to0^+$ and
the family is exhausted; that is a limitation of the closure and not of the
dynamics. At the level of the full populations the case $a_{\rm eff}=5/2$,
which is the effective exponent of the dominant long-string sector in $d=3$
non-compact directions and hence the case relevant to Sec.~\ref{sec:SDC},
behaves almost identically to $a=3/2$, as panel~(b) of
Fig.~\ref{fig:direct} shows.\\
 
It remains to ask whether the required hierarchy is realized microscopically. Two considerations compete. A state of energy $E$ deep in the Hagedorn tail can be reached directly only by processes of centre-of-mass energy $E$, so that the direct channel is kinematically a hard, multi-quantum process while the cascade proceeds by successive absorption of quanta of energy $\sim\beta_R^{-1}$; this favours the hierarchy. Against it, in the regime of interest the bath is hotter than the Hagedorn scale, and the Boltzmann suppression of the hard process competes with the exponential tail degeneracy. The efficient equilibration of a closed-string gas by an open-string gas reported in~\cite{Frey:2023xep,Frey:2024jqy} is plausibly a case in which the hierarchy fails, and Eq.~\eqref{eq:mu-floor} then predicts no pinning, consistently with the no-go result. We therefore do not claim that the bottleneck is generic. What the analysis provides is the condition under which it operates, Eqs.~\eqref{eq:mu-floor}--\eqref{eq:tcross}, in a form that can be confronted with a microscopic computation of the direct production rate in any given background.

\section{The Swampland Distance Conjecture and Thermodynamic Inertia}
\label{sec:SDC}

The open-system dynamics developed in Sec.~\ref{sec:open_system_Hagedorn} suggests a connection with
the Swampland Distance Conjecture (SDC). In its standard form, the SDC
states that when a scalar trajectory approaches an infinite-distance
boundary in moduli space, an infinite tower of states becomes
exponentially light, so that the original low-energy effective field
theory ceases to be valid \cite{Vafa:2005string,Grimm2018infinite,vanBeest:2021lhn,Agmon:2022thq,Brennan:2017rbf,Montero:2024gentle}. If $\Delta$ denotes the
proper field distance to the asymptotic boundary, the characteristic
mass scale of the tower behaves as
\begin{equation}
m_{\rm tower}(\Delta) \;\sim\; m_0\, e^{-\alpha \Delta},
\qquad \alpha > 0 ,
\label{eq:sdc-tower}
\end{equation}
in Planck units. This is a geometric statement about the asymptotic
structure of quantum-gravity moduli space, not a thermodynamic one. The
Hagedorn phenomenon, by contrast, is a statement at fixed string
tension: it concerns the thermodynamic consequences of the exponential growth of oscillator states,
\begin{equation}
\Omega(E) \;\sim\; E^{-a}\, e^{\beta_H E},
\qquad a > 0 .
\label{eq:sdc-hagedorn}
\end{equation}
The two statements are therefore not equivalent. In a large class of
infinite-distance limits, however, the light tower is realized by an
emergent, asymptotically tensionless critical string \cite{lee2022emergent}. In such
limits, Eqs.~(\ref{eq:sdc-tower}) and (\ref{eq:sdc-hagedorn}) describe
the same microscopic tower, the first through the lightness of its
states, the second through the exponential proliferation of its density
of states, and the tools developed in Sec.~\ref{sec:open_system_Hagedorn} allow the relation between them to be made quantitative rather than merely structural.

\subsection{Emergent-string limits and the moduli-dependent Hagedorn
scale}
\label{sec:SDC-realization}

Along an emergent-string trajectory the Hagedorn temperature is tied to
the string scale, $T_H = c\, m_s$ with $c$ an $\mathcal{O}(1)$
model-dependent constant. Identifying the tower scale
(\ref{eq:sdc-tower}) with the emergent string scale,
$m_s(\Delta) \sim m_0\, e^{-\alpha\Delta}$, the inverse Hagedorn
temperature acquires an exponential moduli dependence,
\begin{equation}
\beta_H(\Delta) \;=\; \beta_H(0)\, e^{\alpha \Delta} .
\label{eq:betaH-Delta}
\end{equation}
The open-system construction of Sec.~\ref{sec:open_system_Hagedorn} can now be evaluated along the trajectory. Consider a reservoir at fixed inverse temperature $\beta_R$
in Planck units, and let the subsystem be the emergent string gas, with
Hagedorn density of states given in Eq.~\eqref{eq:Omega_S_Hag} and Hagedorn scale
(\ref{eq:betaH-Delta}). As long as $\beta_R > \beta_H(\Delta)$, the
reservoir is colder than the Hagedorn scale and the dynamics possesses
the ordinary equilibrium fixed point of Sec.~\ref{sec:hagedornpinnin}, located at
\begin{equation}
\delta_{\rm eq}(\Delta) \;=\; \beta_R - \beta_H(\Delta) \;>\; 0 .
\label{eq:delta-eq}
\end{equation}
As the trajectory proceeds toward the infinite-distance boundary,
$\beta_H(\Delta)$ grows exponentially and $\delta_{\rm eq}(\Delta)$
shrinks, vanishing at the finite critical distance
\begin{equation}
\Delta_* \;=\; \frac{1}{\alpha}\,
\ln\!\left(\frac{\beta_R}{\beta_H(0)}\right) .
\label{eq:Delta-star}
\end{equation}
Beyond $\Delta_*$ the equilibrium fixed point no longer exists: any fixed
external temperature eventually exceeds the Hagedorn temperature of the
emergent string, and the reservoir attempts to drive the subsystem past the
Hagedorn boundary. This is precisely the regime analysed in
Secs.~\ref{sec:relationtoisolated}--\ref{sec:direct}. The exact form of the
reservoir contribution derived in Sec.~\ref{sec:relationtoisolated} shows that
the driven relaxation proceeds at the rate $(4\mu_R/\tau_S)(\beta_R-\beta_S)$,
with the mobility equal to unity on the hypoequilibrium family and suppressed
only when the fluctuation structure of the state departs from the canonical
form. The emergent-string tower can supply that departure, through the
mechanism of Sec.~\ref{sec:twosector}: when the energy reaches the Hagedorn sector
predominantly by cascading through the light modes, the mobility is throttled
by the ratio of the band variance to the diverging tail variance, the intensive
variable freezes, and the energy influx into the long-string sector remains
finite.
 
We stress that this requires a hierarchy of injection rates and not the absence
of a direct channel. Collisions of light quanta do produce highly excited
states, and that process is the channel $\tau_{R2}^{-1}$ of
Eq.~\eqref{eq:beta2-direct}. Its effect is quantified by
Eq.~\eqref{eq:mu-floor}: the mobility collapse saturates at
$\tau_{R1}/\tau_{R2}$, so that the pinning in the regime $\Delta>\Delta_*$ is a
long-lived transient whose duration is set by the direct production rate rather
than an asymptotic freezing. Whether that rate is slow enough in a given
background is a microscopic question which we do not settle here; what the
construction provides is the criterion. An explicit cosmological realization of
the geometry assumed here, in which a conifold modulus becomes dynamical after
inflation and lowers the Hagedorn temperature of a warped throat below the
temperature of a surrounding bath, has been studied
in~\cite{Frey:2005jk}, and provides a setting in which the crossing of
$\Delta_*$ is dynamically attained.

The strength of the effect near the critical distance follows from the
variance asymptotics given in Eq.~\eqref{eq:variance_asym}. Using
$\mathrm{Var}_S(H_S) \sim \delta_{\rm eq}^{\,a-3}$ for $1<a<3$ and
expanding $\delta_{\rm eq}(\Delta) \simeq \alpha\beta_R
(\Delta_*-\Delta)$ for $\Delta \to \Delta_*^{-}$, the thermodynamic
inertia of the subsystem diverges as a power law at finite proper
distance,
\begin{equation}
\mathrm{Var}_S(H_S)\Big|_{\delta_{\rm eq}(\Delta)}
\;\sim\;
\big[\alpha\,\beta_R\,(\Delta_* - \Delta)\big]^{\,a-3},
\qquad 1<a < 3 .
\label{eq:inertia-Delta}
\end{equation}
The exponent is controlled by the algebraic prefactor of the
emergent-string Hagedorn tail. As discussed in Sec.~\ref{sec:hagedornpinnin}, the sector
that dominates energy absorption near the Hagedorn scale is the single
long string, with effective exponent $a_{\rm eff} = 1 + d/2$ in $d$
non-compact spatial dimensions; for $d = 3$ this gives
$a_{\rm eff} = 5/2$ and
$\mathrm{Var}_S \sim (\Delta_* - \Delta)^{-1/2}$,   while for $d\ge5$, and for ensembles held at fixed winding, the inertia remains finite though possibly large.\\

Two consequences of this picture are worth stating explicitly. First,
in the pinned regime $\Delta > \Delta_*$ the subsystem inverse
temperature tracks the moduli-dependent Hagedorn scale,
$\beta_S(\Delta) \simeq \beta_H(\Delta)$, so that the effective
temperature of the string gas inherits the exponential behavior of the
tower,
\begin{equation}
T_S(\Delta) \;\simeq\; c\, m_s(\Delta)
\;=\; T_S(\Delta_*)\, e^{-\alpha(\Delta - \Delta_*)} .
\label{eq:TS-tracks-tower}
\end{equation}
The intensive thermodynamic variable itself decays exponentially in
proper field distance, with the same rate $\alpha$ that governs the
tower: Eq.~(\ref{eq:TS-tracks-tower}) is, in this precise sense, the
thermodynamic avatar of the SDC scaling (\ref{eq:sdc-tower}). Second,
in this regime the Hagedorn sector plays the role of an emergent
internal reservoir. The reservoir-dominance condition given in Eq.~\eqref{eq:reservoir_dominance} is a
statement about the mobility-weighted variance, and near the Hagedorn
boundary it is the long-string sector whose variance, and hence heat capacity via $\text{Var}(\hat H)=k_B T^2C_V$, diverges. The sector therefore behaves
exactly as a thermal bath pinned at its own temperature
$\beta_H(\Delta)$: it absorbs the incoming energy without a
corresponding change of its intensive state, which is the
nonequilibrium realization of the standard long-string condensation
picture.

\subsection{A thermodynamic discriminant of infinite-distance limits}
\label{sec:SDC-discriminant}

The mechanism described above is specific to towers with Hagedorn-type
degeneracies, and this specificity is itself informative. In a pure
decompactification limit the light tower is of Kaluza--Klein type, and
the associated density of states grows only sub-exponentially,
\begin{equation}
\Omega_{\rm KK}(E) \;\sim\;
\exp\!\big(\gamma\, E^{\frac{D-1}{D}}\big),
\label{eq:KK-dos}
\end{equation}
as for an ordinary particle gas in $D$ effective spacetime dimensions.
The canonical partition function then converges for every $\beta > 0$,
the energy variance remains finite at any finite reservoir temperature,
and no critical distance analogous to Eq.~(\ref{eq:Delta-star}) exists:
the subsystem temperature simply follows the reservoir, with mobility
$\mu_R$ of order unity, however deep into the limit the trajectory
proceeds. Within the present framework, the divergence of the
thermodynamic inertia at finite proper distance is therefore a
signature of an emergent-string limit, as opposed to a
decompactification limit. The nonequilibrium response of the intensive
variable thus provides a thermodynamic discriminant between the two
classes of infinite-distance limits distinguished by the Emergent
String Conjecture \cite{lee2022emergent}: heated toward its critical scale, a stringy
tower pins the temperature and condenses the energy into long strings,
whereas a Kaluza--Klein tower merely thermalizes.\\

It is worth noting that this discriminant is not weakened by the transient
character of the pinning established in Sec.~\ref{sec:direct}. The contrast
with a Kaluza--Klein tower is not a contrast between a long plateau and a
longer one: for a sub-exponential density of states the energy variance remains
finite at any finite reservoir temperature, so the mobility is of order unity
and there is no plateau at all, for any value of $\tau_{R2}$ and however deep
into the limit the trajectory proceeds. The divergence of the thermodynamic
inertia at finite proper field distance therefore remains a qualitative
signature of an emergent-string limit; what Sec.~\ref{sec:direct} adds is that
the resulting pinning is a transient of computable duration rather than a
permanent freezing.

\section{Conclusions}

In this work, we have reconsidered the Hagedorn regime of string theory from a nonequilibrium thermodynamic perspective. Starting from the standard microcanonical construction of the string density of states, we emphasized that the Hagedorn scale is encoded in the asymptotic growth
$\Omega(E)\sim E^{-a}e^{\beta_H E}$, which simultaneously determines the convergence boundary of the canonical partition function and the asymptotic approach of the microcanonical inverse temperature to $\beta_H$. In this sense, the Hagedorn phenomenon is already present at the level of
the microscopic density of states and is not an artifact of any particular thermodynamic ensemble.\\

We then formulated the problem within the framework of Steepest-Entropy-Ascent Quantum Thermodynamics (SEAQT), in which the effective inverse temperature is promoted to an instantaneous state-dependent quantity,
\begin{equation}
\beta(t)=\frac{1}{k_B}\frac{\mathrm{Cov}(\hat H,\hat S)}{\mathrm{Var}(\hat H)}.
\end{equation}
Within this framework, we derived the exact scalar evolution equation for $\beta$, showing that its dynamics is governed by the fluctuation structure of the nonequilibrium state. In the commuting limit, the evolution reduces to a classical-like expression controlled by higher-order mixed fluctuation moments. This makes clear that the Hagedorn problem can be reformulated as a dynamical question
about the response of an intensive variable on the state manifold, rather than solely as the breakdown of a canonical partition function.\\

The open-system extension developed here sharpens this interpretation. By introducing a system--reservoir splitting in the SEAQT metric structure, we obtained a reduced subsystem dynamics in which the reservoir drives the subsystem while the latter remains characterized by an internal
nonequilibrium inverse temperature $\beta_S$. In this setting, the Hagedorn regime can emerge not as an ordinary equilibrium fixed point, but as a mobility-induced bottleneck: the subsystem may continue to absorb energy from the reservoir, yet its inverse-temperature response can become increasingly weak near $\beta_H$. This provides a dynamical realization of the standard long-string intuition: incoming energy is redirected into the exponentially dense string sector rather than into a substantial further increase of temperature.\\

To make this mechanism explicit, we also evaluated a diagonal coarse-grained open-system model in which the subsystem density of states takes the Hagedorn form
\begin{equation}
\Omega_S(E)=A E^{-a}e^{\beta_H E}\Theta(E-E_0).
\end{equation}
Within a quasi-canonical ansatz, this allowed us to compute analytically the subsystem energy, variance, effective inverse temperature, and the driven evolution of the parameter $\delta=\lambda-\beta_H$. The resulting scaling relations show that the strength of the bottleneck is
controlled not only by the exponential Hagedorn growth, but also by the algebraic exponent $a$. In particular, the explicit diagonal model shows that literal divergence of the variance is not universal, but depends on the effective algebraic exponent of the Hagedorn tail. Thus, the density of states fixes the thermodynamic inertia of the diagonal open-system response; whether that inertia produces slowing-down depends in addition on how the energy is delivered (Sec.~\ref{sec:twosector}), while the genuinely quantum corrections require microscopic information beyond $\Omega(E)$.\\

The exact form of the reservoir contribution to the $\beta_S$-dynamics further shows that a divergent variance is necessary but not sufficient for the bottleneck. The inverse-temperature mobility equals unity identically on the quasi-canonical family, so a reservoir that drives the Hagedorn sector directly produces no slowing-down, however broad the tail. Suppression requires instead that energy reach the long-string sector predominantly through degrees of freedom of finite variance. We first isolated this route in a two-sector construction in which the reservoir couples only to the light quasi-particle modes and the Hagedorn tail is populated by internal exchange. In that limit the mobility collapses as the ratio of the band variance to the diverging tail variance. The intensive variable then freezes at $\beta_H$ while the energy flux into the long-string sector remains finite, and the light modes equilibrate at a stationary temperature strictly between the reservoir and Hagedorn scales, fixed by the mobility ratio.  This is a testable prediction of the mechanism, confirmed in the closure integration; the mobility collapse itself has been verified at the level of the full diagonal populations. Reinstating the direct production of highly excited strings by the reservoir, the mobility collapse saturates at a floor set by the ratio $\tau_{R1}/\tau_{R2}$ of the two injection times. Since the direct channel becomes relatively more important as the Hagedorn scale is approached, the pinning is a long-lived transient rather than a permanent freezing. Its duration is set by the direct production rate or by the exhaustion of the available long-string phase space, whichever acts first, and in the isotropic limit the plateau disappears altogether, recovering the no-go result. The bottleneck is therefore not generic. What the analysis provides is a quantitative criterion,
Eqs.~\eqref{eq:mu-floor}--\eqref{eq:tcross}, that can be confronted with a microscopic computation of the direct production rate in any given
background; the open-string gas of~\cite{Frey:2023xep,Frey:2024jqy} is plausibly a case in which it is not met.\\

This dynamical structure acquires a broader significance in connection with the Swampland Distance Conjecture, in the class of infinite-distance limits governed by an emergent critical string. Promoting the Hagedorn scale to a moduli-dependent quantity, $\beta_H(\Delta)=\beta_H(0)e^{\alpha\Delta}$, we showed that for any fixed reservoir temperature there exists a finite critical distance $\Delta_*$ beyond which the ordinary equilibrium fixed point disappears and the pinning mechanism activates, with the thermodynamic inertia diverging as $(\Delta_*-\Delta)^{a-3}$ for $1<a<3$; in the pinned regime the subsystem temperature tracks the string scale, inheriting the exponential moduli dependence of the tower. A cosmological setting in which such a crossing is dynamically attained, a conifold modulus lowering the Hagedorn temperature of a warped throat below that of a surrounding bath, was studied in~\cite{Frey:2005jk}. Since Kaluza--Klein towers, with sub-exponential densities of states, produce no such critical distance, the divergence of the thermodynamic inertia at finite proper distance acts as a thermodynamic discriminant between emergent-string and decompactification limits. The relation proposed here is not an equivalence and should not be read as a theorem: it is exhibited within the diagonal coarse-grained model and, beyond
$\Delta_*$, requires the hierarchy of injection rates identified above, so that the pinning in that regime is a long-lived transient. Within that class, however, it is a calculable statement rather than a structural analogy, with a definite critical distance and definite scaling exponents; the common theme of the two phenomena, an overwhelmingly large stringy sector that obstructs naive continuation of the original description, appears in the present thermodynamic language as an inertia of the intensive variable that diverges at finite distance in field space.\\

Several extensions suggest themselves. First, it would be valuable to compute the full noncommutative corrections to the $\beta$-dynamics for a more microscopic string-inspired eigenstructure, going beyond the diagonal coarse-grained treatment used here. Second, the present open-system construction could be generalized to include explicit interaction terms between subsystem and reservoir, thereby clarifying the robustness of the slowing-down mechanism away from the block-diagonal approximation. Third, it would be interesting to explore whether similar dynamical bottlenecks arise in other stringy regimes characterized by rapidly proliferating towers of states, including settings more directly tied to asymptotic tensionless-string limits.  Fourth, it would also be interesting to ask how the pinning mechanism interacts with gravitational backreaction: the long-string sector at $T_H$ is precisely the regime of the string/black-hole transition~\cite{HorowitzPolchinski1997, HorowitzPolchinski1998,ChenMaldacenaWitten2023,BalthazarChuKutasov2022}, and the finite condensation rate of Eq.~\eqref{eq:suppression} suggests a dynamical sharpening of that correspondence. Fifth, it would be valuable to test the mechanism in a setting where the Hagedorn regime is under independent control on both sides of a duality. Non-commutative open string theory provides such an arena: it exhibits a Hagedorn transition in a theory of open strings without gravity, and it admits a holographic description in which both the boundary theory and the bulk are
weakly coupled strings with Hagedorn densities of high-energy states~\cite{ferko2026holography}. Because gravitational backreaction is absent on the
open-string side, the two-sector construction of Sec.~\ref{sec:twosector} could be applied there without the complications raised in the preceding paragraph, and the resulting picture of the high-temperature behaviour would translate directly into a statement about the bulk.\\

\begin{center}
    {\bf Acknowledgments}
\end{center}
  C.D. is supported by SECIHTI Frontier Science CF-2023-I-682 while O. L-B is supported by CIIC-UG-DAIP No. 236/2022 and CBF-2025-I-2708.

\appendix
\section{\label{app:phi} Determination of the entropy and constraint gradients}
\subsection*{Determination of $|\Phi)$}
Let
\begin{equation}
\widetilde S(\hat{\gamma})=-k_B\,\mathrm{Tr}(\hat \rho\ln\hat \rho),
\qquad
\hat \rho=\hat{\gamma}\hat{\gamma}^\dagger .
\end{equation}
For an arbitrary variation $\delta\hat{\gamma}$, the induced variation of $\hat \rho$ is
\begin{equation}
\delta\hat \rho=\delta\hat{\gamma}\,\hat{\gamma}^\dagger+\hat{\gamma}\,\delta\hat{\gamma}^\dagger .
\end{equation}
Assuming $\hat \rho$ is positive definite, the Fr\'echet derivative of $\widetilde S$ at $\hat{\gamma}$ in the direction $\delta\hat{\gamma}$ is
\begin{equation}
D\widetilde S(\hat{\gamma})[\delta\hat{\gamma}]
=
-k_B\,\mathrm{Tr}\!\big[(\ln\hat \rho+I)\,\delta\hat \rho\big].
\end{equation}
Substituting $\delta\hat \rho$ gives
\begin{equation}
D\widetilde S(\hat{\gamma})[\delta\hat{\gamma}]
=
-k_B\,\mathrm{Tr}\!\big[(\ln\hat \rho+I)(\delta\hat{\gamma}\,\hat{\gamma}^\dagger+\hat{\gamma}\,\delta\hat{\gamma}^\dagger)\big].
\end{equation}
Using cyclicity of the trace,
\begin{equation}
D\widetilde S(\hat{\gamma})[\delta\hat{\gamma}]
=
-2k_B\,\mathrm{Re}\,\mathrm{Tr}\!\big[\delta\hat{\gamma}^\dagger(\ln\hat \rho+I)\hat{\gamma}\big].
\end{equation}
With the real inner product
\begin{equation}
\label{eq:realinner}
(A|B)=\frac12\mathrm{Tr}(A^\dagger B+B^\dagger A)=\text{Re}\,\mathrm{Tr}(A^\dagger B),
\end{equation}
the Riesz representation of the Fr\'echet derivative is therefore
\begin{equation}
D\widetilde S(\hat{\gamma})[\delta\hat{\gamma}]
=
\left(\frac{\delta \widetilde S}{\delta\hat{\gamma}}\,\middle|\,\delta\hat{\gamma}\right),
\end{equation}
with
\begin{equation} \label{eq:deltasgamma}
\frac{\delta \widetilde S}{\delta\hat{\gamma}}
=
-2k_B\,(\ln\hat \rho+I)\hat{\gamma}.
\end{equation}
Hence,
\begin{equation}
\frac{\delta \widetilde S}{\delta\hat{\gamma}}
=
-2k_B\,(\ln(\hat{\gamma}\hat{\gamma}^\dagger)+I)\hat{\gamma}
\end{equation}
\subsection*{Determination of $|\Psi_H)$ and $|\Psi_I)$}
Let
\begin{equation}
\widetilde H(\hat{\gamma})=\mathrm{Tr}(\hat \rho H),
\qquad
\widetilde I(\hat{\gamma})=\mathrm{Tr}(\hat \rho),
\qquad
\hat \rho=\hat{\gamma}\hat{\gamma}^\dagger .
\end{equation}
For an arbitrary variation $\delta\hat{\gamma}$, the induced variation of $\hat \rho$ is
\begin{equation}
\delta\hat \rho=\delta\hat{\gamma}\,\hat{\gamma}^\dagger+\hat{\gamma}\,\delta\hat{\gamma}^\dagger.
\end{equation}

For the energy functional,
\begin{equation}
D\widetilde H(\hat{\gamma})[\delta\hat{\gamma}]
=
\mathrm{Tr}(\delta\hat \rho\,H).
\end{equation}
Substituting $\delta\hat \rho$ gives
\begin{equation}
D\widetilde H(\hat{\gamma})[\delta\hat{\gamma}]
=
\mathrm{Tr}\!\big[(\delta\hat{\gamma}\,\hat{\gamma}^\dagger+\hat{\gamma}\,\delta\hat{\gamma}^\dagger)H\big].
\end{equation}
Using cyclicity of the trace and the self-adjointness of $H$,
\begin{equation}
D\widetilde H(\hat{\gamma})[\delta\hat{\gamma}]
=
2\,\text{Re}\,\mathrm{Tr}(\delta\hat{\gamma}^\dagger H\hat{\gamma}).
\end{equation}
With the real inner product
\begin{equation}
(A|B)=\frac12\mathrm{Tr}(A^\dagger B+B^\dagger A)=\text{Re}\,\mathrm{Tr}(A^\dagger B),
\end{equation}
this can be written as
\begin{equation}
D\widetilde H(\hat{\gamma})[\delta\hat{\gamma}]
=
(2H\hat{\gamma}|\delta\hat{\gamma}).
\end{equation}
Therefore, by the Riesz representation theorem,
\begin{equation}
\frac{\delta \widetilde H}{\delta\hat{\gamma}}=2H\hat{\gamma}.
\end{equation}

Similarly, for the normalization functional,
\begin{equation}
D\widetilde I(\hat{\gamma})[\delta\hat{\gamma}]
=
\mathrm{Tr}(\delta\hat \rho).
\end{equation}
Substituting $\delta\hat \rho$ yields
\begin{equation}
D\widetilde I(\hat{\gamma})[\delta\hat{\gamma}]
=
\mathrm{Tr}(\delta\hat{\gamma}\,\hat{\gamma}^\dagger+\hat{\gamma}\,\delta\hat{\gamma}^\dagger)
=
2\,\Re\,\mathrm{Tr}(\delta\hat{\gamma}^\dagger\hat{\gamma}).
\end{equation}
Hence,
\begin{equation}
D\widetilde I(\hat{\gamma})[\delta\hat{\gamma}]
=
(2\hat{\gamma}|\delta\hat{\gamma}),
\end{equation}
and therefore
\begin{equation}
\frac{\delta \widetilde I}{\delta\hat{\gamma}}=2\hat{\gamma}.
\end{equation}

\section{\label{app:reservoir} The reservoir contribution \texorpdfstring{$J_R$}{JR} and the explicit open-system \texorpdfstring{$\beta_S$}{betaS}-dynamics}
\subsection*{Derivation of the reservoir contribution $\mathcal{J}_R$
and of the explicit open-system $\beta_S$-dynamics}

The derivation of Sec.~\ref{sec:derivation} invoked energy conservation, Eq.~\eqref{eq:dHdt_zero}, at
two points; in the open system $\langle \hat H_S\rangle$ is not
conserved, so these steps must be revisited. Both survive unchanged.
For the variance,
\begin{equation}
\frac{d}{dt}\mathrm{Var}_S(\hat H)
= \mathrm{Tr}\!\left[(\Delta\hat H)^2 \frac{d\hat\rho_S}{dt}\right]
- 2\,\frac{d\langle\hat H\rangle}{dt}\,\langle \Delta\hat H\rangle
= \mathrm{Tr}\!\left[(\Delta\hat H)^2 \frac{d\hat\rho_S}{dt}\right],
\label{eq:app2-var}
\end{equation}
since $\langle \Delta\hat H\rangle = 0$; similarly, in the analogue of
Eq.~\eqref{eq:dAes_dt_before_frechet} the terms generated by the time dependence of the means
multiply $\langle \Delta\hat H\rangle$ or $\langle \Delta\hat S\rangle$
and vanish,
\begin{equation}
\frac{d}{dt}\mathrm{Cov}_S(\hat H,\hat S)
= \frac{1}{2}\,\mathrm{Tr}\!\left[\frac{d\hat\rho_S}{dt}
\{\Delta\hat H, \Delta\hat S\}\right]
+ \frac{1}{2}\,\mathrm{Tr}\!\left[\hat\rho_S
\left\{\Delta\hat H, \frac{d\hat S}{dt}\right\}\right].
\label{eq:app2-cov}
\end{equation}
The reversible contributions cancel between the two terms of
(\ref{eq:app2-cov}) exactly as in Eq.~\eqref{eq:comm_iden}, for any dissipator.
Consequently, at fixed state the map
$d\hat\rho_S/dt \mapsto d\beta_S/dt$ obtained from Eq.~\eqref{eq:dbetadt} is
linear, and the two terms of the split dissipator
(\ref{eq:dissipator-split}) may be evaluated independently. The first
term reproduces the isolated computation of Secs.~\ref{sec:derivation}--\ref{sec:structure_nonequilibrium} verbatim,
with $\beta \to \beta_S$ and $\tau_D \to \tau_S$, yielding the
polynomial $C_2\beta_S^2 + C_1\beta_S + C_0$ of Eq.~\eqref{eq:beta_quad}. It remains
to evaluate the drive term.

The drive term is derived first by writing $\Delta \beta := \beta_R - \beta_S$ and
\begin{equation}
\mathcal{D}_{\Delta \beta}[\hat\rho_S]
:= -\frac{2\Delta \beta}{\tau_S}\,\{\Delta\hat H, \hat\rho_S\} .
\label{eq:app2-drive}
\end{equation}
Its matrix elements in the instantaneous eigenbasis of $\hat\rho_S$ are
$(\mathcal{D}_{\Delta \beta})_{nm} = -\tfrac{2\Delta \beta}{\tau_S}(p_n+p_m)\Delta H_{nm}$.
The three channels of Sec.~\ref{sec:derivation} give:

\emph{(i) Variance channel.} Using (\ref{eq:app2-var}) and the trace
identity $\mathrm{Tr}(\hat A\{\hat B,\hat\rho\})
= \langle\{\hat A,\hat B\}\rangle$,
\begin{equation}
\frac{d}{dt}\mathrm{Var}_S(\hat H)\bigg|_{\Delta \beta}
= -\frac{2\Delta \beta}{\tau_S}\,
\big\langle \{(\Delta\hat H)^2, \Delta\hat H\}\big\rangle
= -\frac{4\Delta \beta}{\tau_S}\,\big\langle (\Delta\hat H)^3\big\rangle .
\label{eq:app2-varnu}
\end{equation}

\emph{(ii) Anticommutator channel.} The first term of
(\ref{eq:app2-cov}) gives
\begin{equation}
\frac{1}{2}\,\mathrm{Tr}\!\left[\mathcal{D}_{\Delta \beta}
\{\Delta\hat H,\Delta\hat S\}\right]
= -\frac{\Delta \beta}{\tau_S}\,
\big\langle \{\{\Delta\hat H, \Delta\hat S\}, \Delta\hat H\}
\big\rangle .
\label{eq:app2-anti}
\end{equation}

\emph{(iii) Fr\'echet channel.} From Eq.~\eqref{eq:dsdtnn}, the drive contribution
to $d\hat S/dt = -k_B\, D(\ln\hat\rho_S)[\mathcal{D}_{\Delta \beta}]$ reads
\begin{equation}
\left(\frac{d\hat S}{dt}\right)_{nm}^{(\Delta \beta)}
=
\begin{cases}
\displaystyle
\frac{2 k_B \Delta \beta}{\tau_S}\,
\frac{p_n+p_m}{p_n-p_m}\,\ln\!\frac{p_n}{p_m}\;\Delta H_{nm},
& n \neq m ,\\[10pt]
\displaystyle
\frac{4 k_B \Delta \beta}{\tau_S}\,\Delta H_{nn},
& n = m ,
\end{cases}
\label{eq:app2-frechet}
\end{equation}
where the diagonal line follows from
$(d\hat S/dt)_{nn} = -k_B\, p_n^{-1} (\mathcal{D}_{\Delta \beta})_{nn}$. Inserting
(\ref{eq:app2-frechet}) into the second term of (\ref{eq:app2-cov}),
written as $\tfrac{1}{2}\sum_{n,m}(p_n+p_m)\,
\Delta H_{nm}\,(d\hat S/dt)_{mn}$, and using the symmetry of
$(p_n-p_m)^{-1}\ln(p_n/p_m)$ under $n \leftrightarrow m$,
\begin{equation}
\frac{1}{2}\left\langle \left\{\Delta\hat H,
\frac{d\hat S}{dt}\right\}\right\rangle_{\!{\Delta \beta}}
=
\frac{k_B\Delta \beta}{\tau_S}\, Q_H[\hat\rho_S]
+ \frac{4 k_B \Delta \beta}{\tau_S}\,\mathrm{Var}_{\rm d}(\hat H_S) ,
\label{eq:app2-frechet-sum}
\end{equation}
with $Q_H$ and $\mathrm{Var}_{\rm d}$ as in Eq.~(\ref{eq:QH-Vard}).
Note that here, in contrast with the isolated case, the diagonal
Fr\'echet contribution survives unambiguously: it is proportional to
the manifestly positive quantity $\mathrm{Var}_{\rm d}$, not to a
covariance constrained by the multiplier.

Combining \eqref{eq:app2-varnu}--\eqref{eq:app2-frechet-sum} in
Eq.~\eqref{eq:Aee_dbeta_general}, the drive contribution to
$k_B \mathrm{Var}_S\, d\beta_S/dt$ is
\begin{equation}
\frac{\Delta \beta}{\tau_S}\Big[
4 k_B \beta_S \big\langle (\Delta\hat H)^3\big\rangle
- \big\langle \{\{\Delta\hat H, \Delta\hat S\}, \Delta\hat H\}
\big\rangle
+ k_B\, Q_H
+ 4 k_B\, \mathrm{Var}_{\rm d}
\Big]
\;\equiv\; -\,\mathcal{J}_R ,
\label{eq:app2-assembly}
\end{equation}
which, upon using $\Delta\hat S = k_B\beta_S(\Delta\hat H
- \Delta\hat F_S)$ (see Eq.~\eqref{eq:DeltaS_beta_relation} at the instantaneous $\beta_S$) to
eliminate the third moments,
\begin{equation}
\big\langle \{\{\Delta\hat H, \Delta\hat S\}, \Delta\hat H\}
\big\rangle
= 4 k_B \beta_S \big\langle (\Delta\hat H)^3\big\rangle
- k_B \beta_S \big\langle \{\{\Delta\hat H, \Delta\hat F_S\},
\Delta\hat H\}\big\rangle ,
\label{eq:app2-identity}
\end{equation}
yields the compact closed form quoted in Eq.~(\ref{eq:JR-exact}).

\section{\label{app:block} Block decomposition of the Lagrange multipliers}
In this appendix we derive the weighted-mean formula given in Eq.\eqref{eq:beta_eff_open} and its
hypoequilibrium reduction given in Eq.~\eqref{eq:beta_eff_weighted_average} directly from the general Lagrange
multiplier expression given in Eq.~\eqref{eq:beta2_general}, making explicit how the ansatz given in Eqs.~\eqref{eq:L_split}--\eqref{eq:L_open} leads to an exact separation of the covariances and
variances into system and reservoir contributions.

In the weak-coupling regime the composite square-root state may be taken
in product form,
\begin{equation}
\hat\gamma \;=\; \hat\gamma_S \otimes \hat\gamma_R ,
\qquad
\hat\rho \;=\; \hat\gamma\hat\gamma^\dagger
\;=\; \hat\rho_S \otimes \hat\rho_R ,
\label{eq:app-product}
\end{equation}
and the Hamiltonian is additive up to the neglected interaction term,
\begin{equation}
\hat H \;=\; \hat H_S \otimes \hat I_R + \hat I_S \otimes \hat H_R
\;+\; \hat H_{\rm int},
\qquad \hat H_{\rm int} \;\simeq\; 0 .
\label{eq:app-H-additive}
\end{equation}
For the entropy operator no further approximation is required: the
product form Eq.~\eqref{eq:app-product} implies
\begin{equation}
\hat S \;=\; -k_B \ln\big(\hat\rho_S \otimes \hat\rho_R\big)
\;=\; \hat S_S \otimes \hat I_R + \hat I_S \otimes \hat S_R ,
\qquad
\hat S_X := -k_B \ln \hat\rho_X ,
\label{eq:app-S-additive}
\end{equation}
so that the additivity of $\hat S$ in Eq.~\eqref{eq:H_S_split} is exact for product
states. In what follows we suppress the identity factors and write
$\hat A_S$ for $\hat A_S \otimes \hat I_R$, etc. For any sector
observable we define the fluctuation operator
$\Delta\hat A_X := \hat A_X - \langle \hat A_X\rangle_X$, with
$\langle \cdot\rangle_X = \mathrm{Tr}_X(\hat\rho_X\,\cdot\,)$.

The three vectors entering the constrained projection, Eqs.~\eqref{eq:basishi} and \eqref{eq:deltasgamma}, decompose as
\begin{align}
|\Psi_I) &= 2\hat\gamma ,
\label{eq:app-PsiI}\\
|\Psi_H) &= 2\hat H\hat\gamma
\;=\; 2\,(\Delta\hat H_S)\hat\gamma
    + 2\,(\Delta\hat H_R)\hat\gamma
    + 2\,\langle \hat H\rangle\,\hat\gamma ,
\label{eq:app-PsiH}\\
|\Phi) &= 2\big(\hat S - k_B\hat I\big)\hat\gamma
\;=\; 2\,(\Delta\hat S_S)\hat\gamma
    + 2\,(\Delta\hat S_R)\hat\gamma
    + 2\big(\langle \hat S\rangle - k_B\big)\hat\gamma ,
\label{eq:app-Phi}
\end{align}
with $\langle \hat H\rangle = \langle \hat H_S\rangle_S + \langle \hat
H_R\rangle_R$ and similarly for $\langle \hat S\rangle$. The key
kinematic property of the product state given in Eq.~\eqref{eq:app-product} is that
the three families of directions appearing above are mutually
orthogonal with respect to the real inner product given in Eq.~\eqref{eq:realinner}. Indeed, for
arbitrary Hermitian sector observables $\hat A_S$, $\hat B_R$,
\begin{equation}
\big((\Delta\hat A_S)\hat\gamma \,\big|\, (\Delta\hat B_R)\hat\gamma\big)
\;=\; \mathrm{Re}\,\mathrm{Tr}\!\left[\hat\rho\,
\Delta\hat A_S \otimes \Delta\hat B_R\right]
\;=\; \mathrm{Re}\,\Big[\langle \Delta\hat A_S\rangle_S\,
\langle \Delta\hat B_R\rangle_R\Big]
\;=\; 0 ,
\label{eq:app-cross-orthogonal}
\end{equation}
and likewise
$\big(\hat\gamma\,\big|\,(\Delta\hat A_X)\hat\gamma\big)
= \langle \Delta\hat A_X\rangle_X = 0$ for $X = S,R$. Eq.~
(\ref{eq:app-cross-orthogonal}) is the precise content of the direct-sum
ansatz given Eq.~\eqref{eq:L_split}: the product structure of the state induces an orthogonal
decomposition of the relevant tangent directions,
\begin{equation}
\mathcal{L} \;\supset\;
\mathrm{span}\{\hat\gamma\}
\;\oplus\; \mathcal{L}_S
\;\oplus\; \mathcal{L}_R ,
\qquad
\mathcal{L}_X := \big\{ (\Delta\hat A_X)\hat\gamma \big\},
\label{eq:app-decomposition}
\end{equation}
so that the vanishing of all cross-covariances between system and
reservoir fluctuations is not an additional assumption but a consequence
of (\ref{eq:app-product}). Within each sector, the same computation
gives the symmetrized second moments of Eq.~\eqref{eq:covHF},
\begin{equation}
\big((\Delta\hat A_X)\hat\gamma \,\big|\, (\Delta\hat C_X)\hat\gamma\big)
\;=\; \tfrac{1}{2}\big\langle \{\Delta\hat A_X, \Delta\hat C_X\}
\big\rangle_X
\;=\; \mathrm{Cov}_X(\hat A_X, \hat C_X) .
\label{eq:app-sector-cov}
\end{equation}
Genuine cross directions of the form
$(\Delta\hat A_S\,\Delta\hat B_R)\hat\gamma$ do exist in $\mathcal{L}$,
but for additive observables and product states they never appear in
(\ref{eq:app-PsiI})--(\ref{eq:app-Phi}) and are orthogonal to all
vectors that do; they would be sourced only by $\hat H_{\rm int}$ or by
classical/quantum correlations in $\hat\rho$, which is the precise sense
of the approximation sign in Eq.~\eqref{eq:H_S_split}.

Consistently with the decomposition (\ref{eq:app-decomposition}), the
dissipative operator given in Eq.~\eqref{eq:L_open} acts as
\begin{equation}
\hat L\,\big[(\Delta\hat A_X)\hat\gamma\big]
= \frac{1}{\tau_X}\,(\Delta\hat A_X)\hat\gamma ,
\qquad
\hat L\,\hat\gamma = \frac{1}{\tau_0}\,\hat\gamma ,
\label{eq:app-L-action}
\end{equation}
where $\tau_0$ denotes the (a priori unspecified) rate assigned to the
common normalization direction. Using
(\ref{eq:app-cross-orthogonal})--(\ref{eq:app-sector-cov}), the five
inner products entering Eqs.~\eqref{eq:beta1_general}--\eqref{eq:beta2_general} separate exactly:
\begin{align}
(\Psi_H|\hat L|\Phi) &= 4\Big[
\tau_S^{-1}\,\mathrm{Cov}_S(\hat H_S,\hat S_S)
+ \tau_R^{-1}\,\mathrm{Cov}_R(\hat H_R,\hat S_R)
+ \tau_0^{-1}\,\langle\hat H\rangle\big(\langle\hat S\rangle - k_B\big)
\Big],
\nonumber\\
(\Psi_H|\hat L|\Psi_H) &= 4\Big[
\tau_S^{-1}\,\mathrm{Var}_S(\hat H_S)
+ \tau_R^{-1}\,\mathrm{Var}_R(\hat H_R)
+ \tau_0^{-1}\,\langle\hat H\rangle^2
\Big],
\nonumber\\
(\Psi_H|\hat L|\Psi_I) &= 4\,\tau_0^{-1}\,\langle\hat H\rangle ,
\qquad
(\Psi_I|\hat L|\Psi_I) = 4\,\tau_0^{-1} ,
\qquad
(\Psi_I|\hat L|\Phi) = 4\,\tau_0^{-1}\big(\langle\hat S\rangle - k_B\big).
\label{eq:app-inner-products}
\end{align}
Note that in the isotropic limit $\tau_S = \tau_R = \tau_0 = \tau_D$
these expressions reduce term by term to the inner products of
Sec.~\ref{sec:special_case}, e.g.
$(\Psi_H|\hat L|\Psi_H) = \tfrac{4}{\tau_D}\big[\mathrm{Var}(\hat H)
+ \langle\hat H\rangle^2\big] = \tfrac{4}{\tau_D}\langle\hat H^2\rangle$,
which provides a check of (\ref{eq:app-inner-products}).

Substituting (\ref{eq:app-inner-products}) into the general formula
\eqref{eq:beta2_general} and writing $h := \langle\hat H\rangle$,
$\bar s := \langle\hat S\rangle - k_B$ for brevity,
\begin{align}
k_B\beta_{\rm eff}
&= \frac{(\Psi_H|\hat L|\Phi)(\Psi_I|\hat L|\Psi_I)
       - (\Psi_I|\hat L|\Phi)(\Psi_H|\hat L|\Psi_I)}
        {(\Psi_H|\hat L|\Psi_H)(\Psi_I|\hat L|\Psi_I)
       - (\Psi_H|\hat L|\Psi_I)^2}
\nonumber\\[4pt]
&= \frac{16\,\tau_0^{-1}\Big[\sum_X \tau_X^{-1}\mathrm{Cov}_X
   + \tau_0^{-1} h\bar s\Big]
   - 16\,\tau_0^{-2}\,\bar s\, h}
        {16\,\tau_0^{-1}\Big[\sum_X \tau_X^{-1}\mathrm{Var}_X
   + \tau_0^{-1} h^2\Big]
   - 16\,\tau_0^{-2}\, h^2}
\;=\;
\frac{\displaystyle\sum_{X=S,R} \tau_X^{-1}\,
      \mathrm{Cov}_X(\hat H_X,\hat S_X)}
     {\displaystyle\sum_{X=S,R} \tau_X^{-1}\,
      \mathrm{Var}_X(\hat H_X)} .
\label{eq:app-beta-eff}
\end{align}
All terms involving the means $h$ and $\bar s$ cancel identically
between the two products, and the overall factor $\tau_0^{-1}$ drops
out. This yields Eq.~\eqref{eq:beta_eff_open} as an exact consequence of
(\ref{eq:app-product})--(\ref{eq:app-S-additive}) and
(\ref{eq:app-L-action}): the only approximations are those made
\emph{before} the projection, namely the neglect of
$\hat H_{\rm int}$ and of correlations between the sectors, both of
which are subleading at weak coupling. Two remarks are in order. First,
$\beta_{\rm eff}$ is independent of the rate $\tau_0$ assigned to the
normalization direction; this reflects the fact that Eq.~\eqref{eq:beta2_general} is
precisely the projection of the entropy gradient orthogonal to the
constraint directions, and the normalization constraint is enforced
exactly for any $\tau_0$. The same computation applied to Eq.~\eqref{eq:beta1_general}
gives the normalization multiplier
$k_B\zeta_1 = \langle\hat S\rangle - k_B\beta_{\rm eff}\langle\hat
H\rangle - k_B$, which is the block analogue of Eq.~\eqref{eq:beta1_sh} and is
likewise $\tau_0$-independent. Second, in the isotropic limit
$\tau_S = \tau_R$ the weights cancel and (\ref{eq:app-beta-eff})
reduces to
$k_B\beta = (\mathrm{Cov}_S + \mathrm{Cov}_R)/(\mathrm{Var}_S +
\mathrm{Var}_R)$, i.e.\ to the global definition \eqref{eq:beta} evaluated on the
product state, for which the total covariance and variance are additive
by (\ref{eq:app-cross-orthogonal}). The anisotropic metric therefore
does not create the separation, the separation is kinematic, but
it replaces the plain additivity by the mobility weighting
$\tau_X^{-1}$.

\subsection*{Hypoequilibrium reduction}
If each sector is internally canonical (or, more generally, in a
hypoequilibrium state within its sector \cite{li2016generalized,LiVonSpakovsky2016Chem}),
$\hat\rho_X = Z_X^{-1} e^{-\beta_X \hat H_X}$, then
\begin{equation}
\hat S_X = -k_B \ln\hat\rho_X
= k_B\big(\beta_X \hat H_X + \ln Z_X\big)
\quad\Longrightarrow\quad
\Delta\hat S_X = k_B\,\beta_X\, \Delta\hat H_X ,
\label{eq:app-hypo}
\end{equation}
so that
$\mathrm{Cov}_X(\hat H_X,\hat S_X) = k_B\beta_X\,
\mathrm{Var}_X(\hat H_X)$, which is Eq.~\eqref{eq:sector_canonical_relation}. Substituting into
(\ref{eq:app-beta-eff}) yields Eq.~\eqref{eq:beta_eff_weighted_average},
\begin{equation}
\beta_{\rm eff}
= \frac{\tau_S^{-1}\beta_S\,\mathrm{Var}_S(\hat H_S)
      + \tau_R^{-1}\beta_R\,\mathrm{Var}_R(\hat H_R)}
       {\tau_S^{-1}\,\mathrm{Var}_S(\hat H_S)
      + \tau_R^{-1}\,\mathrm{Var}_R(\hat H_R)} .
\label{eq:app-46}
\end{equation}
We note that for the Hagedorn sector of Sec.~\ref{sec:hagedornpinnin} the identification
(\ref{eq:app-hypo}) holds with $\beta_X = \lambda$ even in the presence
of the algebraic prefactor $E^{-a}$, since, as shown there, the
prefactor belongs to the shell measure and not to the microscopic
probability. Expanding (\ref{eq:app-46}) in the reservoir-dominated
regime \eqref{eq:reservoir_dominance}, with
$\epsilon := \tau_S^{-1}\mathrm{Var}_S / \tau_R^{-1}\mathrm{Var}_R \ll 1$,
\begin{equation}
\beta_{\rm eff} = \beta_R + \epsilon\,(\beta_S - \beta_R)
+ \mathcal{O}(\epsilon^2),
\label{eq:app-reservoir-dominated}
\end{equation}
which quantifies the approach to Eq.~\eqref{eq:beta_eff_to_betaR}.

The weighting in (\ref{eq:app-beta-eff}) admits a direct physical
reading. Substituting the block decomposition into Eq.~\eqref{eq:pi_gamma}, the
sector components of the dissipative evolution take the isolated-system
form of Eq.~\eqref{eq:pi_gamma2} with the common multiplier $\beta_{\rm eff}$,
\begin{equation}
\Pi_\gamma\big|_X
= -\frac{2}{\tau_X}\left[\beta_{\rm eff}\,\Delta\hat H_X
- \frac{1}{k_B}\,\Delta\hat S_X\right]\hat\gamma ,
\label{eq:app-sector-dissipator}
\end{equation}
and repeating the steps leading to Eq.~\eqref{eq:E_open_balance_1} sector by sector, using
$(\Psi_I|\Pi_\gamma) = 0$ and (\ref{eq:app-sector-cov}), gives the
sector energy balances
\begin{equation}
\frac{d}{dt}\langle \hat H_X\rangle
= -\frac{4}{\tau_X}\,\mathrm{Var}_X(\hat H_X)
\big(\beta_{\rm eff} - \beta_X\big),
\qquad X = S, R ,
\label{eq:app-sector-balance}
\end{equation}
with $\beta_X := \mathrm{Cov}_X / (k_B \mathrm{Var}_X)$ as in
Eq.~\eqref{eq:beta_S_definition}. Summing over the sectors, total energy conservation
$\frac{d}{dt}\langle\hat H\rangle = 0$ holds if and only if
$\beta_{\rm eff}$ takes the value (\ref{eq:app-beta-eff}). The
weighted-mean formula given in Eq.~\eqref{eq:beta_eff_open} is therefore nothing but the statement that
the common Lagrange multiplier is fixed by energy conservation in the
presence of anisotropic mobilities: each sector couples to the
dissipative flow with an effective thermal conductance
$\tau_X^{-1}\mathrm{Var}_X(\hat H_X)$, and $\beta_{\rm eff}$ is the
conductance-weighted mean of the sector inverse temperatures.
Equation (\ref{eq:app-sector-balance}) is also the starting point of the
two-sector analysis of Sec.~\ref{sec:twosector}, where the same
construction is applied within the subsystem itself.

\bibliographystyle{JHEP}
%\bibliography{References_z.bib}

\providecommand{\href}[2]{#2}\begingroup\raggedright\endgroup

\end{document}